\documentclass[11pt,a4paper]{article}
\usepackage[utf8]{inputenc}
\usepackage{jheppub}
\usepackage{amsmath}
\usepackage{amsfonts}
\usepackage{amssymb}
\usepackage{graphicx} 
\usepackage{epsfig}
\usepackage{color}
\usepackage{amssymb}
\usepackage{bbold}
\usepackage{hhline}
\usepackage{tabu}
\usepackage{mathtools}

\newcommand{\dd}{\mathrm{d}}

\newcommand{\ee}[1]{\begin{equation}#1\end{equation}}
\newcommand{\ea}[1]{\begin{align}#1\end{align}}

\providecommand{\f}[2]{\frac{{#1}}{{#2}}}
\def\hmax{\varphi_{\rm bar}}
\def\MP{M_{\rm P}}

\title{Cosmological Aspects of Higgs Vacuum Metastability}
 
\author[a]{Tommi Markkanen}
\author[b]{Arttu Rajantie}
\author[c]{Stephen Stopyra}
\affiliation[a,b,c]{Department of Physics, Imperial College London, SW7 2AZ, UK}
\affiliation[c]{Department of Physics and Astronomy, University College London, WC1E 6BT, UK
}

\abstract{The current central experimental values of the parameters of the Standard Model give rise to a striking conclusion: metastability of the electroweak vacuum is favoured over absolute stability. A metastable vacuum for the Higgs boson implies that it is possible, and in fact inevitable, that a vacuum decay takes place with catastrophic consequences for the Universe. The metastability of the Higgs vacuum is especially significant for cosmology, because there are many mechanisms that could have triggered the decay of the electroweak vacuum in the early Universe. We present a comprehensive review of the implications from Higgs vacuum metastability for cosmology along with a pedagogical discussion of the related theoretical topics, including renormalization group improvement, quantum field theory in curved spacetime and vacuum decay in field theory.}

\emailAdd{t.markkanen@imperial.ac.uk}
\emailAdd{a.rajantie@imperial.ac.uk}
\emailAdd{stephen.stopyra09@imperial.ac.uk}
\begin{document}
\begin{flushleft}
	\hfill		  IMPERIAL/TP/2018/TM/05
\end{flushleft}
\maketitle

\makeatletter{}\section{Introduction}
\begin{figure}[ht!]
	\begin{center}
		\hspace{-18mm}		\includegraphics[width=0.65\textwidth]{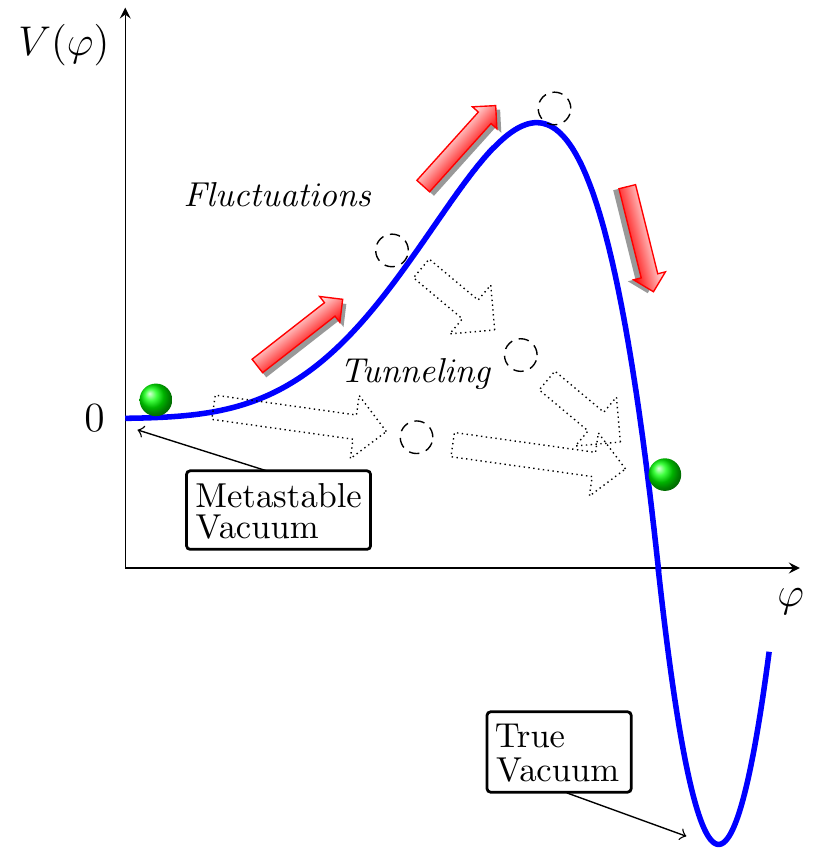}
	\end{center}
	\caption{\label{fig:ill} Illustration of vacuum decay for a potential with a metastable vacuum at the origin.}
\end{figure}

One of the most striking results of the discovery of Higgs boson \cite{Chatrchyan:2012xdj,Aad:2012tfa} has been that its mass lies in a regime that predicts the current vacuum state to be a false vacuum, that is, there is a lower energy vacuum state available to which the electroweak vacuum can decay into~\cite{Degrassi:2012ry,Buttazzo:2013uya}. That this was a possibility in the Standard Model (SM) has been known for a long time \cite{Hung:1979dn,Isidori:2001bm,Sher:1993mf,Ellis:2009tp,Casas:1996aq,EliasMiro:2011aa}. The precise behavior of the Higgs potential is sensitive to the experimental inputs, in particular the physical masses for the Higgs and the top quark and also physics beyond the SM. The current best estimates of the Higgs and top quark masses~\cite{Tanabashi:2018oca}, 
\begin{equation}
\label{equ:masses}
M_h = 125.18\pm0.16{\rm \,GeV},\quad M_t = 173.1\pm0.9{\rm \,GeV},
\end{equation}  
place the Standard Model squarely in the metastable region.

As in any quantum system, there are three main ways in which the vacuum decay can happen.
They are illustrated in Fig. \ref{fig:ill}. If the system is initially in the false vacuum state, the transition would take place through quantum tunneling. On the other hand, if there is sufficient energy available, for example in a thermal equilibrium state, it may be possible for the system to move classically over the barrier. The third way consists of quantum tunneling from an excited initial state. This is often the dominant process if the temperature is too low for the fully classical process. All three mechanisms can be relevant for the decay of the electroweak vacuum state, and their rates depending on the conditions. In each of them, the transition happens initially locally in a small volume, nucleating a small bubble of the true vacuum. The bubble then starts to expand, reaching the speed of light very quickly, any destroying everything in its way.

If the Universe was infinitely old, even an arbitrarily low vacuum decay rate would be incompatible with our existence. The implications of vacuum metastability can therefore only be considered in the cosmological context, taking into account the finite age and the cosmological history of the Universe.
Although the vacuum decay rate is extremely slow in the present day, that was not necessarily the case in the early Universe. High Hubble rates during inflation and high temperatures afterwards could have potentially increased the rate significantly. Therefore the fact that we still observe the Universe in its electroweak vacuum state allows us to place constraints on the cosmological history, for example the reheat temperature and the scale of inflation, and on Standard Model parameters, such as particle masses and the coupling between the Higgs field and spacetime curvature.

In this review we discuss the implications of Higgs vacuum metastability in early Universe cosmology and describe the current state of the literature. We also discuss all the theoretical frameworks, with detailed derivations, that are needed for the final results. 
This article complements earlier comprehensive reviews of electroweak vacuum metastability~\cite{Sher:1988mj,Schrempp:1996fb}, which focus on the particle physics aspects rather than the cosmological context, and the recent introductory review~\cite{Moss:2015fma} that explores the role of the Higgs field in cosmology more generally.

In Section \ref{sec:effpot} we present renormalization group improvement in flat space by using the Yukawa theory as an example before discussing the full SM.
Section \ref{sec:FLRW} contains an overview of quantum field theory on curved backgrounds relevant for our purposes, including the modifications to the SM. In Section \ref{sec:tunn} we go through the various ways vacuum decay can occur. In Section \ref{sec:Cosmology} we discuss the connection to cosmology and in Section \ref{sec:con} we present our concluding remarks.

Our sign conventions for the metric and curvature tensors are $(-,-,-)$ in the classification of  \cite{Misner:1974qy} and throughout we will use units where the reduced Planck constant, the Boltzmann constant and the speed of light are set to unity, $\hbar\equiv k_B\equiv c\equiv 1$. The reduced Planck mass is given by Newton's constant as
\begin{equation}
\label{equ:Planckmass}
M_{\rm P}\equiv(8\pi G)^{-1/2}\approx 2.435\times 10^{18}~{\rm GeV}.
\end{equation}
We will use $\varphi$ for the vacuum expectation value (VEV) of a spectator field (usually the Higgs), $\phi$ for the inflaton and $\Phi$ for the SM Higgs doublet. 
The inflaton potential is $U(\phi)$ and the
Higgs potential $V(\varphi)$. The physical Higgs and top masses read $M_h$ and $M_t$.

\makeatletter{}\section{Effective Potential in Flat Spacetime}
\label{sec:effpot}
\subsection{Example: Yukawa theory}
\label{sec:yukawa}
The possibility of quantum corrections destabilizing a classically stable vacuum has been known for quite some time~\cite{Krive:1976sg,Krasnikov:1978pu,Maiani:1977cg,Politzer:1978ic,Hung:1979dn,Cabibbo:1979ay}. Although our focus will be strictly on the SM, one should keep in mind that the instability that potentially arises in the SM is only a specific example of a more general phenomenon that could manifest in a variety of other theories of elementary particles. For this reason all the essential features of the vacuum instability in the SM can be illustrated with the simple Yukawa theory, which we will now discuss before moving on to the full Standard Model in Section~\ref{sec:effpSM}.

The action containing a massless, quartically self-interacting scalar field $\varphi$ Yukawa-coupled to a massless Dirac fermion $\psi$ is
\ee{S =\int d^4x\bigg[\f{1}{2}\partial_\mu\varphi\partial^\mu\varphi-\f{\lambda}{4}\varphi^4+\bar \psi\partial\!\!\!/\psi-g\varphi\bar\psi\psi\bigg]\,.\label{eq:yu}}
Classically, the potential for the scalar field is simply
\ee{V_{\rm cl}(\varphi)=\f{\lambda}{4}\varphi^4\,,}
which quite trivially has a well-defined state of lowest energy at the origin. 

When quantized the potential for the field $\varphi$ becomes modified by quantum corrections
\ee{V(\varphi)=V_{\rm cl}(\varphi)+ \text{\textit{quantum corrections}}\,,} which may be investigated within the usual framework of quantum field theory \cite{Peskin:1995ev}. Importantly, it has been for a long time understood that in some instances predictions in a quantum theory can deviate significantly from those of the classical case. A prime example of such behaviour is radiatively induced symmetry breaking \cite{Coleman:1973jx}. 

In the one-loop approximation the result for the quantum corrected or \textit{effective} potential for the Yukawa model has the form (see, for example, Ref.~\cite{Markkanen:2018bfx})
\ea{V_{{\rm eff}}(\varphi)&=\f{\lambda(\mu)}{4}\varphi^4(\mu)\nonumber \\&+\f{1}{64\pi^2}\bigg[{M}^4_\varphi(\mu)\bigg(\log\f{{M}^2_\varphi(\mu)}{\mu^2}
 - \f{3}{2} \bigg)-4{M}^4_\psi(\mu)\bigg(\log\f{{M}^2_\psi(\mu)}{\mu^2}
 - \f{3}{2} \bigg)\bigg]+\cdots
\label{eff_pot_scalar}
\,,}
with 
\ee{{M}^2_\varphi(\mu)\equiv 3\lambda(\mu)\varphi^2(\mu)\,;\qquad {M}^2_\psi(\mu)\equiv g^2(\mu)\varphi^2(\mu)\,.\label{eq:effm}}
In the above we have explicitly denoted the dependence on the {\it renormalization scale} $\mu$, which is an arbitrary energy scale, which one needs to choose in order to define the renormalised parameters of the theory.
There is also a similar dependence in 
$\varphi(\mu)$ which now refers to the renormalized one-point function of the quantized field, which is related to the bare field via the field renormalization constant \cite{Peskin:1995ev}
\ee{{\varphi}_{\rm bare}=\sqrt{Z(\mu)}{\varphi}(\mu)\,.\label{eq:fren}}

In the one-loop effective potential~(\ref{eff_pot_scalar}), 
the contribution from the fermion $\psi$ comes with a minus sign. For sufficient high values of $g$, it can overtake the classical contribution and lead to a region with negative potential energy. In the limit of  large field values $\varphi\rightarrow\infty$, one may write the potential as
\ee{V_{{\rm eff}}(\varphi\rightarrow\infty)\rightarrow	\varphi^4\f{9\lambda^2-4g^4}{32\pi^2}\log\bigg(\f{\varphi}{\mu}\bigg)+\cdots
\label{eff_pot_scalar2}
\,,}
implying that if
\ee{\lambda<\lambda_{\rm cr}\equiv \f{2 g^2}{3},
		\label{eq:infty}}
the potential has a barrier and starts to decrease without bound at high field values~\cite{Krive:1976sg}.
When $\lambda$ is larger than the critical threshold $\lambda_{\rm cr}$ the quantum correction approaches $+\infty$ indicating that an arbitrary small deviation from $\lambda_{\rm cr}$ leads either to $+\infty$ or $-\infty$ at large enough field values. 

Hence we have seen that in the Yukawa theory the low-field vacuum will be separated by a barrier from an infinitely deep well on the other side. Even if the barrier is very robust, after a sufficiently long time the system initialized in the classical vacuum must eventually make a transition to the other side of the barrier and evolve towards the state of minimum energy. 

A potential unbounded from below is a problematic concept and it is often assumed that, perhaps due to non-perturbative physics invisible to a loop expansion, some mechanism reverses the behavior of the potential at very high energies. This means that the minimum energy is in fact bounded from below, and the effect of the quantum corrections is to generate second local minimum beyond the barrier as depicted in Fig. \ref{fig:ill}. In theories containing $U(1)$ gauge fields, such as the SM, the reversal of the potential can be shown to happen and the issue of an infinitely deep well does not arise. In the effective theory framework, which arguably is the correct way of viewing the SM, this issue is also not present as one will always encounter a finite scale beyond which the calculation becomes unreliable. Indeed, gravitational corrections are a prime example of a modification that is expected to become significant at large field values.

From a practical point of view, whether or not the potential is infinitely or deep of has a second or more accurately a \textit{true} minimum beyond the barrier is not important for the generic prediction that the vacuum at the origin should eventually decay if the potential possesses regions with lower energy than at the origin.

However, conclusions based on the behaviour of the perturbative one-loop result (\ref{eff_pot_scalar}) may be premature. This is because for very large field values the logarithms become non-pertubatively large making the loop expansion invalid: generically one would expect higher powers of the logarithmic contributions in the square brackets of Eq.~(\ref{eff_pot_scalar2}) to be generated by higher orders in the expansion, as for example is evident in the results of Ref.~\cite{Chung:1999gi}. Concretely, for our Yukawa theory (\ref{eq:yu}) this requirement means that we can only draw conclusions in the region where
\ee{\f{4g^4}{64\pi^2}\log\bigg(\f{g^2\varphi^2}{\mu^2}\bigg)\lesssim 1\qquad \text{and}\qquad \f{9\lambda^2}{64\pi^2}\log\bigg(\f{3\lambda\varphi^2}{\mu^2}\bigg)\lesssim 1\,.\label{eq:logs}}
In principle, the smaller the logarithms the more accurate the result. 

\subsection{Renormalization group improvement}
\label{sec:rg}
By making use of renormalization group (RG) techniques it is possible to improve the accuracy of an existing perturbative expression such that the issue of large logarithms may be avoided~\cite{Kastening:1991gv,Bando:1992wy,Bando:1992np,Ford:1992mv}.

Demanding that the effective potential (\ref{eff_pot_scalar}) does not depend on the renormalisation scale $\mu$ gives rise to the Callan-Symazik equation~\cite{Callan:1970yg,Symanzik:1970rt,Symanzik:1971vw}
\begin{equation}
\frac{\dd}{\dd \mu}V_{{\rm eff}}(\varphi) =0\quad \Leftrightarrow\quad \bigg\{\mu \frac{\partial}{\partial \mu}+\beta_\lambda \frac{\partial}{\partial \lambda}+\beta_g \frac{\partial}{\partial g}-\gamma\varphi\f{\partial}{\partial\varphi}\bigg\}V_{{\rm eff}}(\varphi) =0\,,\label{eq:RGI}
\end{equation}
where we have defined the beta functions and the anomalous dimension in the usual manner
\ee{\beta_{c_i}\equiv \mu\f{\partial c_i}{\partial\mu}\,,\qquad\gamma\equiv \mu\f{\partial \log \sqrt{Z}}{\partial\mu}\,,}
with $\gamma$ from the field renormalization constant in (\ref{eq:fren}), which has a dependence on the renormalization scale $Z\equiv Z(\mu)$. Deriving the  beta functions and the anomalous dimension for the Yukawa theory is a well-known calculation (see for example Ref.~\cite{Bando:1992wy}) and here we simply state the results
\begin{align}
16\pi^2 \beta_{m^2} &= m^2\left(6\lambda + 4g^2\right),\label{eq:la0}  \\
16\pi^2 \beta_{\lambda} &= 18 \lambda ^2+8 g^2 \lambda-8 g^4,\label{eq:la} \\
\ 16\pi^2 \beta_{g} &=5g^3,\label{eq:g}\\
\ 16\pi^2 \gamma &=2g^2\label{eq:gamma}
\,,
\end{align}
where for completeness we have included the beta function also for a mass parameter of the scalar field.

The beta functions tell us how the values of the renormalised parameters ``run'', i.e., depend on the scale choice $\mu$.  For example, 
assuming renormalised coupling value $g(\mu_0)$ at some scale choice $\mu_0$,
one may solve the running of the Yukawa coupling $g(\mu)$ from Eq.~(\ref{eq:g}),
\ee{g^2(\mu)=\f{g^2(\mu_0)}{1-\f{5g^2(\mu_0)}{8\pi^2}\log({\mu}/{\mu_0})}\,.\label{eq:grun}}
This shows that increasing $\mu$ leads to a larger $g(\mu)$, and that the coupling $g(\mu)$ 
appears to diverge at scale 
\begin{equation}
\mu=\mu_0 \exp\left(\frac{8\pi^2}{5g^2(\mu_0)}\right),
\end{equation}
which is known as the Landau pole~\cite{pauli1955niels}.
However, well before the Landau pole is reached, the loop expansion ceases to be valid.
For more information on the effect of running couplings we refer the reader to more or less any textbook on quantum field theory (for example Refs.~\cite{Peskin:1995ev,Cheng:1985bj}).

Even though the full effective potential $V_{\rm eff}(\varphi)$ has to be independent of the scale choice $\mu$, for any finite-order perturbative result that is only true up to neglected higher-order terms. This means that some scale choices will work better than the others, and by a judicious choice, one can improve the accuracy of the perturbative result.
In general, one would choose the scale $\mu$ to optimise the perturbative expansion in such a way that the loop corrections are small as indicated in Eq.~(\ref{eq:logs}). However, for the effective potential (\ref{eff_pot_scalar2}), the loop corrections depend on the field value $\varphi$. Therefore each given choice of scale would only work well over a relatively narrow range of field values.

To ensure that Eq.~(\ref{eq:logs}) remains satisfied at any field values, one can take this approach further and make the renormalisation scale a function of the field $\varphi$, 
\begin{equation}
\label{equ:mustar}
\mu=\mu_*(\varphi),
\end{equation}
so that the expansion is optimised at all field values.
This procedure is generically called \textit{renormalization group improvement} (RGI).\footnote{In our work the improvement is understood to come from the specific step of optimizing the expansion via a particular choice of $\mu$. In some works, it simply means making use of running couplings.}
This way one can define the 
renormalization group improved (RGI) effective potential as 
\begin{equation}
V_{\rm RGI}(\varphi)\equiv V_{\rm eff,RG}(\varphi,\mu_*(\varphi)).
\label{eq:RGIfull}
\end{equation}
One should note that in this expression $\varphi$ refers to the field defined at the field-dependent renormalisation scale,
$\varphi=\varphi(\mu_*(\varphi))$ (for more discussion, see Ref.~\cite{Markkanen:2018bfx}),
and that at any finite order in perturbation theory the resulting function $V_{\rm RGI}(\varphi)$ depends on the choice of the function $\mu_*(\varphi)$.

In principle, one could choose $\mu_*$ in such a way that the loop correction vanishes exactly. For the one-loop potential (\ref{eff_pot_scalar2}) in the Yukawa theory, this would give
\begin{equation}
\mu_*^{\rm exact}(\varphi)=e^{-3/4}
\left(\frac{3\lambda}{g^2}\right)^{\frac{9\lambda^2}{18\lambda^2-8g^4}}g\varphi,
\label{equ:exactscale}
\end{equation}
where both the couplings $g$ and $\lambda$ and the field $\varphi$ are renormalised at scale $\mu_*^{\rm exact}(\varphi)$, and therefore the equation defines the scale $\mu_*(\varphi)$ implicitly.
With this choice, the RGI effective potential $V_{\rm RGI}(\varphi)$ is given simply by the tree-level potential with $\varphi$-dependent couplings,
\begin{equation}
\label{equ:RGIexact}
V_{\rm RGI}(\varphi)=\frac{1}{4}\lambda(\mu_*^{\rm exact}(\varphi))\varphi^4.
\end{equation}

In more realistic theories it is often impractical to choose $\mu_*(\varphi)$ that cancels the loop correction exactly~\cite{Markkanen:2018bfx}. Instead, one chooses some simpler function that keeps the loop correction sufficiently small.
The most common choice in the literature is simply 
\ee{\mu_*(\varphi) = \varphi\,.\label{eq:scale}} 
Because the loop correction in Eq.~(\ref{eff_pot_scalar2}) does not vanish for this scale choice it should still be included in the effective potential.
It is nevertheless, fairly common to make the further approximation of dropping it, and writing the tree-level RGI effective potential simply as 
\ee{V^{\text{tree}}_{\text{RGI}}(\varphi)=\f{\lambda(\varphi)}{4}\varphi^4\,.\label{eq:RGI0}}
For weak couplings this is not a good approximation, though. 
Eq.~(\ref{equ:exactscale}) shows that the loop correction vanishes for
$\mu_*\approx g\varphi$, and therefore a good approximation to RGI effective potential is
\begin{equation}
\label{equ:RGIerror}
V_{\rm RGI}(\varphi)\approx \frac{1}{4}\lambda(g\varphi)\varphi^4
=\frac{1}{4g^4}\lambda(g\varphi)(g\varphi)^4=\frac{1}{g^4}V_{\rm RGI}^{\rm tree}(g\varphi).
\end{equation}
From this we can see that the use of the tree-level RGI potential (\ref{eq:RGI0}) with the scale choice~(\ref{eq:scale}) gets the barrier position wrong by a factor of $g$ and the barrier height by a factor of $g^4$. 
Therefore one should either keep the one-loop correction, or use a more accurate scale choice.

From the beta function (\ref{eq:la}) for $\lambda$, we see that if $g^2\gg \lambda$, $\lambda$ can become negative at high scales $\mu$. 
It is conventional to define the \emph{instability scale} $\mu_\Lambda$ as the scale where this happens,
\ee{\lambda(\mu_\Lambda)=0\,.\label{eq:insta}} 
If $\mu_\Lambda<\infty$, the effective potential (\ref{equ:RGIexact}) becomes negative at high field values, too, implying an instability. Again, the root cause is a negative contribution from the fermions, this time in the beta functions. 

The solution for the running $\lambda(\mu)$ can be obtained analytically, but is unfortunately quite complicated (see e.g. Ref.~\cite{Bando:1992wy}). However, 
it is easy to see that the critical value of the coupling, below which the instability appears, is
\begin{equation}
\underline{\lambda}_{\rm cr}=\f{1+\sqrt{145}}{18}g^2.
\label{equ:RGIlambdacr}
\end{equation} 
Close to this critical value one may provide relatively simple analytical results. Suppose we have initial conditions given at some reference scale $\mu_0$ for the running parameters $g(\mu_0)$ and $\lambda(\mu_0)$ the latter of which we parametrize as a fixed value $\underline{\lambda}_{\rm cr}$ and a perturbation $\delta\lambda$ as
\ee{g(\mu_0)\,;\qquad\lambda(\mu_0)\equiv\underline{\lambda}_{\rm cr}-\delta\lambda	\,.\label{eq:cr2}}
By solving Eqs.~(\ref{eq:la})--(\ref{eq:g}) explicitly one may show that $\lambda(\mu)$ has the following expansion
\ee{\f{\lambda(\mu)}{g^2(\mu)}=\f{\underline{\lambda}_{\rm cr}}{g^2(\mu_0)}-\f{\delta\lambda}{g^2(\mu_0)}\bigg(\f{g^2(\mu)}{g^2(\mu_0)}\bigg)^{\sqrt{29/5}}+\mathcal{O}(\delta\lambda^2)\,.\label{eq:thresh}}
From Eq.~(\ref{eq:grun}) it is apparent that, because $g(\mu)$ is a monotonically increasing function of $\mu$, the RGI effective potential (\ref{equ:RGIexact}) is unbounded from below at large field values, for an arbitrarily small positive perturbation $\delta\lambda>0$. For comparison, the threshold (\ref{eq:infty}) in the unimproved case was $\lambda_{\rm cr}/g^2=2/{3}$, somewhat lower than the RGI result (\ref{equ:RGIlambdacr}).

The above makes apparent a very important generic feature: renormalization group improvement can lead to 
conclusions that are qualitatively different from the unimproved results.
In particular, sizes of couplings deemed as well-behaved and hence giving rise to a stable potential may in fact reveal to result in an instability by the RG improved results. This also implies that close to the critical value the higher loop corrections become quite important as even a small change may tilt the conclusion from stable to unstable, or vice versa. This is also suggested by the fact that the couplings run very gradually and the precise value of the instability scale is very sensitive to small corrections: even a tiny change in the initial values or the running may change $\mu_\Lambda$ by several orders of magnitude. These features are illustrated in the example below.

\begin{figure}
\begin{center}
		\hspace{-20mm}\includegraphics[width=0.8\textwidth]{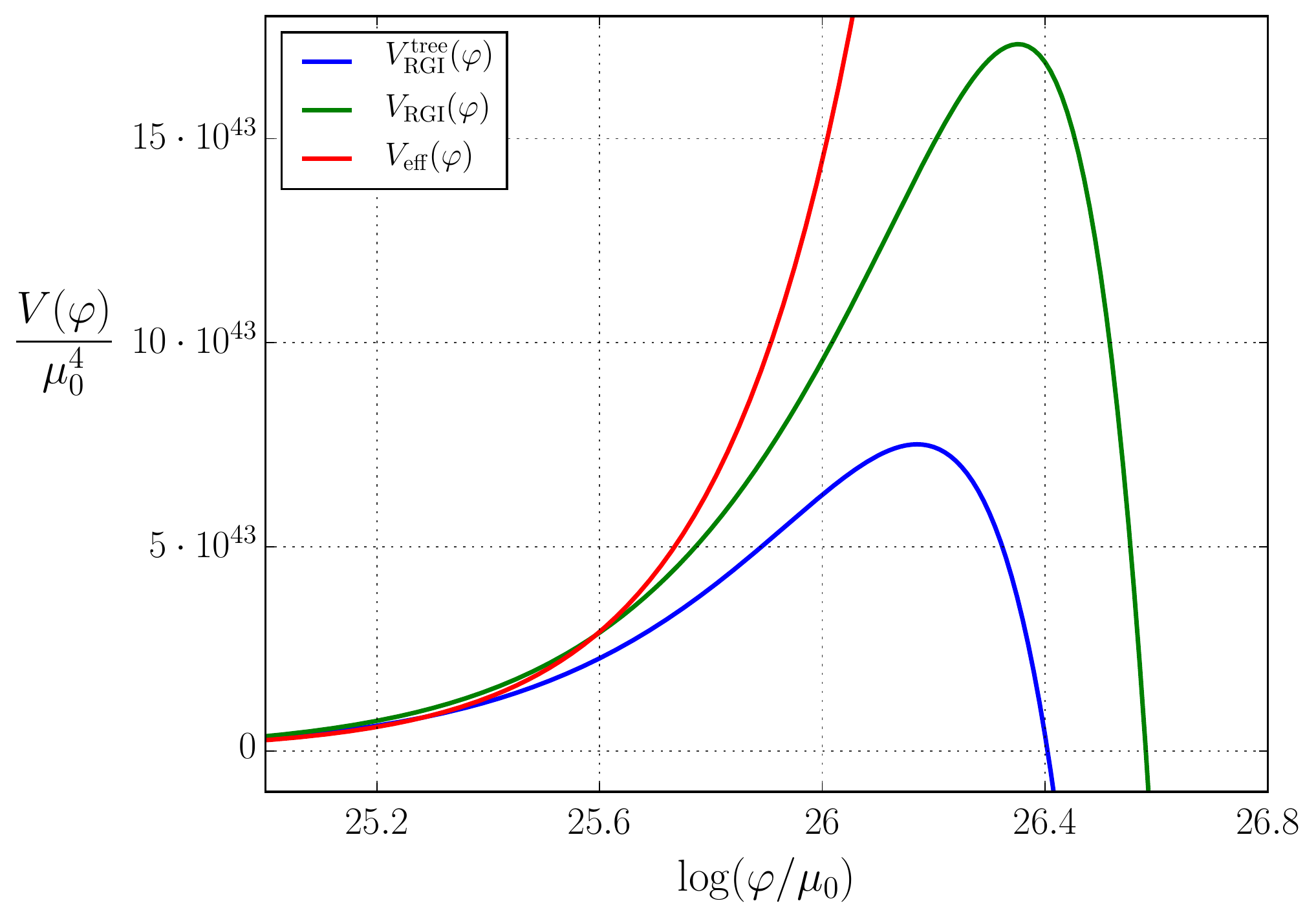}
	
	\end{center}
	\caption{\label{fig:sca}
	Behaviour of the 1-loop RGI effective potential (\ref{eq:RGIfull}) (green),
     the tree-level RGI effective potential (\ref{eq:RGI0}) (blue), and the non-improved result (\ref{eff_pot_scalar}) (red) with the choices (\ref{eq:inncond}) at the reference scale $\mu_0$. The RGI scale choice was $\mu_*(\varphi)=\varphi$.
     }
\end{figure}

For concreteness, let us consider a numerical example that highlights the importance of renormalization group improvement. Specifically, we choose the Yukawa theory with a negligible mass parameter and with the initial conditions defined at the renormalisation scale $\mu_0$ as
as
\ee{g(\mu_0)=\f{1}{\sqrt{2}}\,;\qquad \lambda(\mu_0)=\f{1+\sqrt{145}}{36}-10^{-2}\approx0.352\label{eq:inncond}	\,.}
which from (\ref{eq:cr2}) can be seen to correspond to a choice that is {\it below} the critical value by 
\ee{\delta \lambda = 10^{-2}\,.}

Since Eq.~(\ref{eq:inncond}) satisfies $\lambda(\mu_0)>\f{2}{3}g^2(\mu_0)$ the unimproved effective potential (\ref{eq:infty}) implies no instability. This is however not the case after renormalization group improvement as shown in Fig. \ref{fig:sca}. We must however make sure that the above scale is such that all parameters remain perturbative, in particular for the Yukawa theory we need to check that the $g$-coupling is sufficiently small. For our parametrization this can be loosely expressed as ${2g^2(\mu_\Lambda)}\lesssim 4\pi$ and perturbativity is easily demonstrated with the help of Eq.~(\ref{eq:grun}). This check is quite important since if $g(\mu)$ reaches a large value before $\mu_\Lambda$, it will render the entire derivation inconsistent. 

What is also apparent from Fig. \ref{fig:sca} that there is a clear difference between the tree-level RGI  approximation (\ref{eq:RGI0}) and the full RGI result (\ref{eq:RGIfull}), when using the simple non-exact scale choice (\ref{eq:scale}). In many applications this would result in a non-negligible inaccuracy, but as shown in Eq.~(\ref{equ:RGIerror}), it changes the barrier position by a factor $O(g)$ and height by $O(g^4)$, which can be important for vacuum stability.  This sensitivity to quantum corrections and the choice of $\mu_*$ comes from the fact that the instability occurs precisely at the point where the tree-level contribution vanishes.

\subsection{Effective potential in the Standard Model}
\label{sec:effpSM}
The SM has a far richer particle content than the simple Yukawa theory of Section \ref{sec:yukawa}, but the main reason for the possible vacuum instability remains the same: Quantum corrections from the fermions contribute with a minus sign and if significant enough can lead to the formation of regions with lower potential energy than the electroweak vacuum. In the SM the effect is mostly due to the top quark, because it is by far the heaviest and thus has the largest Yukawa coupling. As discussed in Section \ref{sec:tunn}, general field theory principles then dictate that after a sufficiently long time has passed the system should relax into the configuration with the lowest energy resulting in the decay of the electroweak vacuum.

Through increasing experimental accuracy and improved analytic estimates in recent years it has become apparent that the central values for the couplings of the SM allow extrapolation to energy scales close to the Planck scale and that they are in fact incompatible with the situation where the electroweak vacuum would be the state of lowest energy. Some important early works addressing the question of vacuum instability are Refs.~\cite{Krive:1976sg,Krasnikov:1978pu,Maiani:1977cg,Politzer:1978ic,Hung:1979dn,Cabibbo:1979ay}. The full body of work studying aspects of the vacuum instability is vast (to say the least) and includes Refs.~\cite{Linde:1979ny,Lindner:1985uk,Lindner:1988ww,Arnold:1989cb,Arnold:1991cv,Ellwanger:1992jp,Ford:1992mv,Sher:1993mf,Altarelli:1994rb,Casas:1994qy,Espinosa:1995se,Casas:1996aq,Hambye:1996wb,Nie:1998yn,Frampton:1999xi,Isidori:2001bm,Gonderinger:2009jp,Ellis:2009tp,Holthausen:2011aa,EliasMiro:2011aa,Chen:2012faa,EliasMiro:2012ay,Rodejohann:2012px,Bezrukov:2012sa,Datta:2012db,Alekhin:2012py,Chakrabortty:2012np,Anchordoqui:2012fq,Masina:2012tz,Chun:2012jw,Chung:2012vg,Gonderinger:2012rd,Degrassi:2012ry,Buttazzo:2013uya,Dev:2013ff,Nielsen:2012pu,Tang:2013bz,Klinkhamer:2013sos,He:2013tla,Chun:2013soa,Jegerlehner:2013cta,Branchina:2013jra,DiLuzio:2015iua,Martin:2013gka,Gies:2013fua,Branchina:2015nda,Eichhorn:2015kea,Antipin:2013sga,Chao:2012mx,Spencer-Smith:2014woa,Chetyrkin:2012rz,Chetyrkin:2013wya,Gabrielli:2013hma,Branchina:2014rva,Bednyakov:2015sca,Branchina:2014usa,Bednyakov:2012en,Bednyakov:2013eba,Bednyakov:2013cpa,Kobakhidze:2013pya,Salvio:2016mvj,Chigusa:2018uuj,Chigusa:2017dux,Garg:2017iva,Khan:2015ipa,Khan:2014kba,Liu:2012qua,Bambhaniya:2016rbb,Schrempp:1996fb,Sher:1988mj,Moss:2015fma}. 

The modern high precision era of vacuum instability investigations can be thought to have been initiated by the detailed analyses performed in Refs.~\cite{Degrassi:2012ry,Buttazzo:2013uya}, which presented the first complete next-to-next-to-leading order analysis
of the Standard Model Higgs potential and the running couplings.
\begin{figure}[ht!]
	\begin{center}
				\includegraphics[width=\textwidth]{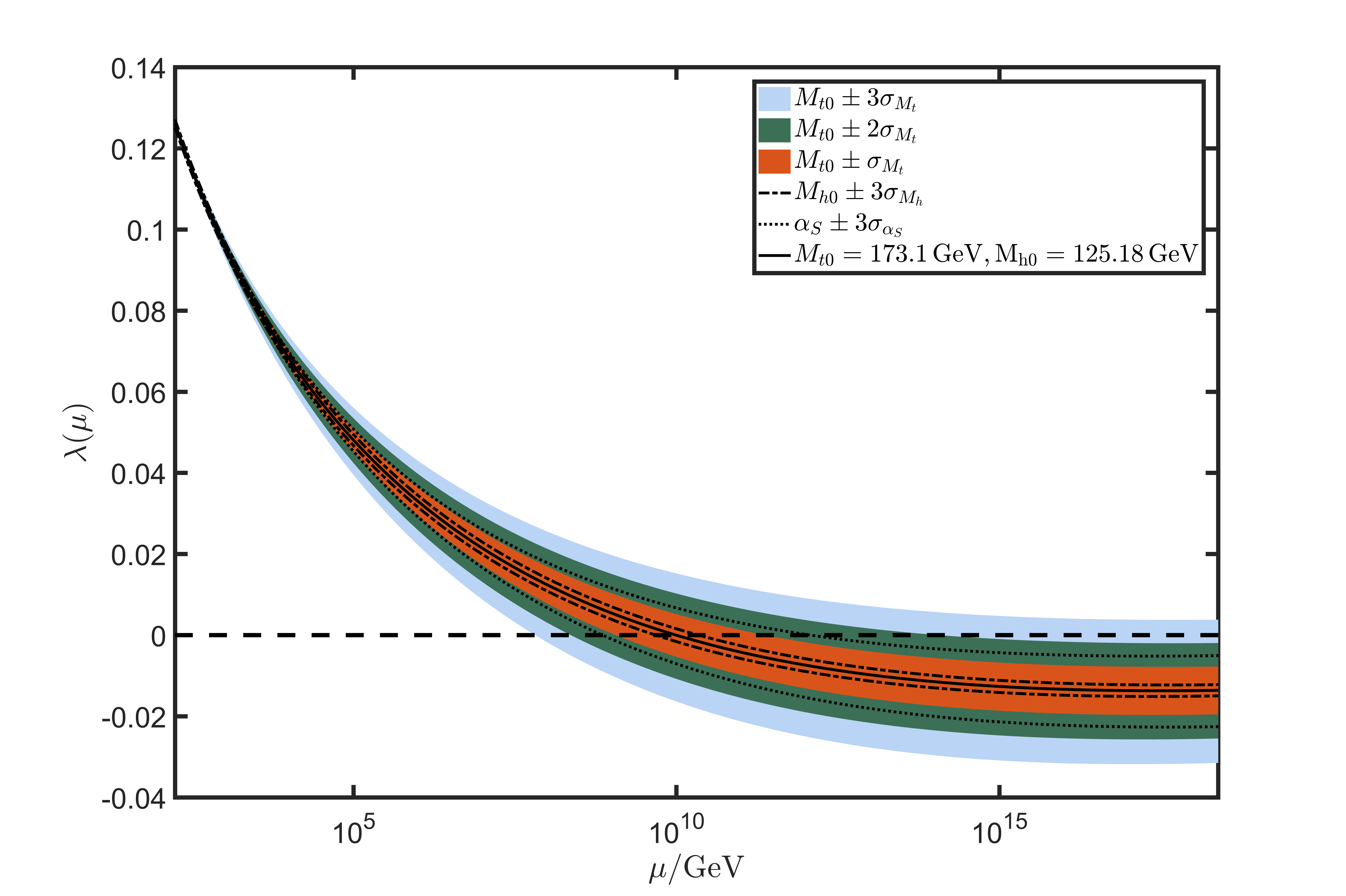}
	\end{center}
	\caption{\label{fig:run2} RG evolution of the Higgs four-point coupling. The bands represent uncertainties up to $3\sigma$ coming from the mass of the Higgs, the top quark and the strong coupling constant $M_h$, $M_t$ and $\alpha_S$, respectively, using central values~\cite{Tanabashi:2018oca} of $M_h = 125.18\pm 0.16\rm{\,GeV}$, $M_t = 173.1\pm 0.9\rm{\,GeV}$, $\alpha_S = 0.1181\pm 0.0011$.
    }
\end{figure}

The current state-of-the-art calculation for the running of Standard Model parameters uses two-loop matching conditions, three-loop RG evolution and pure QCD corrections to four-loop order~\cite{Bednyakov:2015sca}.
The running of the Higgs self-coupling $\lambda$ is shown in Fig.~\ref{fig:run2} for the central mass values (\ref{equ:masses}), together with bands showing the effects of the estimated errors in the parameter values.
For the central mass values (\ref{equ:masses}), the instability scale (\ref{eq:insta}), defined by $\lambda(\mu_\Lambda)=0$, is
\ee{\mu_{\Lambda} = 9.92\times 10^{9}\rm{\,GeV}.\label{eq:iscale2}}
This depends sensitively on the top and Higgs masses:
At $1\sigma$ the range is  $1.16\times 10^{9}\rm{\,GeV} < \mu_{\Lambda} < 2.37\times 10^{11}\rm{\,GeV}$, and the case in which $\lambda(\mu)$ is never negative at still
included within $3\sigma$ uncertainty. 
Using the three-loop running, and including the one-loop correction in the RGI effective potential with the scale choice $\mu_*(\varphi)=\varphi$,
the top of the potential barrier lies at
\begin{equation}
\hmax = 4.64\times 10^{10}{\rm{\,GeV}},
\label{equ:hbarrier}
\end{equation}
and the barrier height is
\begin{equation}
\Delta V(\hmax) = V(\hmax) - V(\varphi_{\rm{fv}}) = 
3.46\times 10^{38}{\rm{\,GeV^4}} = (4.31\times 10^9 {\rm{\,GeV}})^4
.
\label{equ:Vbarrier}
\end{equation}
For comparison, the tree-level RGI form (\ref{eq:RGI0}),
which means dropping the one-loop correction
and is common in the literature,
would give a significantly lower position fot the top of the potential barrier, 
$\hmax = 7.70\times 10^9{\rm{\,GeV}}$. 
Using the unimproved one-loop effective potential with parameters renormalised at the electroweak scale gives as even lower value $\hmax= 5.78\times 10^4~{\rm GeV}$.
This demonstrates that, as discussed in Section~\ref{sec:rg}, the use of renormalisation group improvement and the inclusion of at least the one-loop correction in the RGI effective potential are both crucial for accurate results.

A slightly more formal issue that must also be kept in mind is that the barrier position $\hmax$ is in fact gauge dependent and strictly speaking has limited physical significance~\cite{DiLuzio:2014bua, Andreassen:2014gha,Espinosa:2016nld,Espinosa:2016uaw}. The value of the potential at its extrema are however gauge independent as demanded by the famous Nielsen identity \cite{Nielsen:1975fs}. In the simplest approximation the probability of  vacuum decay involves only the values of the potential at the extrema and subtleties involving gauge dependence are evaded. Furthermore, more precise calculations of the rate of vacuum decay, since it is a physical process, can be expected to always be cast into a gauge-invariant form \cite{Plascencia:2015pga}.  
\makeatletter{}\section{Field Theory in Expanding Universe}
\label{sec:FLRW} 
\subsection{Spectator field on a curved background}
\label{sec:scal}
In the extreme conditions of the early Universe, gravity plays a significant role.
In order to investigate the consequences from Higgs metastability we must therefore make use of an approach that incorporates also gravitational effects. This can be achieved in the framework of quantum field theory in a curved spacetime. The study of quantum fields theory on curved backgrounds is hardly a recent endeavour. For a thorough discussion on the subject we refer the reader to the standard textbooks, such as Refs.~\cite{Birrell:1982ix,Mukhanov:2007zz,Parker:2009uva}.

As a representative model we choose an action consisting only of a self-interacting scalar field
\ee{S=\int d^4x\,\sqrt{|g|}\bigg[\f{1}{2}\nabla_\mu\varphi\nabla^\mu\varphi-\f{1}{2}m^2\varphi^2-\f{\xi}{2}R\varphi^2-\f{\lambda}{4}\varphi^4\bigg]\,,\label{eq:act}} 
where the curved background is visible in the metric dependence of integration measure, $\sqrt{|g|}$, the covariant derivative $\nabla_\mu$ and in the appearance of the non-minimal coupling $\xi$ that connects the field to the scalar curvature of gravity $R$. The necessity of an operator $\propto R\varphi^2$ was discovered already in Refs.~\cite{Tagirov:1972vv,Callan:1970ze,Chernikov:1968zm}, the reasons for which we will elaborate in Section \ref{eq:effpotc}. It will turn out to be a key ingredient for the implications of the vacuum (in)stability in the early Universe. 

Since our discussion assumes a classical curved background with no fluctuations of the metric $g^{\mu\nu}$ some effects visible in a complete quantum gravity approach are possibly missed. For energy scales below the Planck threshold and for spectator fields with a negligible effect on the evolution of the background modifications from quantum gravity are expected to be suppressed. For the case of a quasi de Sitter background this was verified in detail in Ref.~\cite{Markkanen:2017dlc} for the SM Higgs. The reason why quantum gravity is not relevant for a potential SM metastability can be understood from the simple fact that the instability scale (see Section~\ref{sec:effpSM}) is significantly lower than the Planck mass
\ee{\f{\mu_\Lambda}{M_{\rm P}}\approx 10^{-8}\,.}
\subsection{Homogeneous and isotropic spacetime}
\label{sec:hom}
From the cosmological point of view it is often sufficient to consider the special case of a homogeneous and isotropic spacetime with the Friedmann--Lema\^{i}tre--Robertson--Walker (FLRW) line-element given in cosmic time as
\ee{ds^2=dt^2-a(t)^2d\mathbf{x}^2\,,\label{eq:le}}
where $a(t)\equiv a$ is the scale factor describing cosmic acceleration. We will furthermore assume that the energy and pressure densities of the background, $\rho$ and $p$, are connected via the constant equation of state parameter $w$ as \ee{p=w\rho\,;\qquad \rho = T_{00}\,,~~p = T_{ii}/a(t)^2\,,}
where $T_{\mu\nu}$ is the energy-momentum tensor of the background. With the line-element (\ref{eq:le}) the Einstein equation reduces to the the Friedmann equations \ea{
\begin{dcases} \phantom{-(}3H^2M_{\rm P}^2&=\phantom{\bigg[]}3\bigg(\f{\dot{a}}{a}\bigg)^2M_{\rm P}^2\quad\qquad= \rho\\ -(3H^2+2\dot{H})M_{\rm P}^2 &=-\bigg[\bigg(\f{\dot{a}}{a}\bigg)^2+2\f{\ddot{a}}{a}\bigg]M_{\rm P}^2 =p=w\rho \end{dcases}\,,\label{eq:e}
}
which allow one to easily find expressions for the Hubble rate $H\equiv\dot{a}/a$ and the scale factor as functions of $w$\begin{eqnarray}
\label{eq:rehaH}
a&=&\bigg(\frac{t}{t_0}\bigg)^{\frac{2}{3(w+1)}}\,,~~ H=\frac{2}{3(w+1)t}\,,\quad {\rm for}~~ w\neq-1\,
\nonumber\\
a&=&e^{H t}\,,~~ H=H_0\,,\quad {\rm for}~~w=-1\,.
\end{eqnarray}
For the purposes of this discussion the most important quantity characterizing gravitational effects will be the scalar curvature of gravity $R$, which may be written as a function of the equation of state parameter and the Hubble rate\ee{R=6\bigg[\bigg(\frac{\dot{a}}{a}\bigg)^2+\frac{\ddot{a}}{a}\bigg]=3(1-3w)H^2\,.\label{eq:R} }
\subsection{Amplified fluctuations}
\label{sec:mod}

Let us then concentrate on a free quantum theory by setting $\lambda=0$. For this case the action~(\ref{eq:act})  leads to the equation of motion
\ee{\left(\Box+m^2+\xi R\right)\hat{\varphi}=0\,, \label{eq:eom}}
whose solutions, as usual, can be expressed as a mode expansion
\ee{\hat{\varphi}=\int \f{d^{3}\mathbf{k}\, e^{i\mathbf{k\cdot\mathbf{x}}}}{\sqrt{(2\pi )^{3}a^2}}\left[\hat{a}_\mathbf{k}^{\phantom{\dagger}}f^{\phantom{\dagger}}_{k}(\eta)+\hat{a}_{-\mathbf{k}}^\dagger f^*_k(\eta)\right]\,\label{eq:adsol2}\, ,}
with $[\hat{a}_{\mathbf{k}}^{\phantom{\dagger}},\hat{a}_{\mathbf{k}'}^\dagger]=\delta^{(3)}(\mathbf{k}-\mathbf{k}'),~~[\hat{a}_{\mathbf{k}}^{\phantom{\dagger}},\hat{a}_{\mathbf{k}'}^{\phantom{\dagger}}]=[\hat{a}_{\mathbf{k}}^{{\dagger}},\hat{a}_{\mathbf{k}'}^\dagger]=0$, where $\mathbf{k}$ is the co-moving momentum and $k\equiv|\mathbf{k}|$. In the above we have also made use of conformal time defined as 
\ee{\eta=\int^{t}\f{dt'}{a(t')}\qquad \Rightarrow\qquad ds^2=a^2(d\eta^2-d\mathbf{x}^2)\,.\label{eq:conft}}
From Eqs.~(\ref{eq:eom}) and (\ref{eq:adsol2}) we may write down the mode equation
\ee{f''_{k}(\eta)
+ \bigg[\mathbf{k}^2  +a^2\mathcal{M}^2\bigg] f_{k}(\eta) = 0 \,,
\label{eq:mode}}
where the primes denote derivatives with respect to conformal time and we have defined the effective mass
\ee{\mathcal{M}^2\equiv m^2+  \left(\xi-\frac{1}{6}\right)R\,.
\label{equ:curvaturemass}}

Equation (\ref{eq:mode}) may be interpreted as that of a harmonic oscillator with a time-dependent mass. The crucial point is that for many cosmologically relevant combinations of $m$, $\xi$ and $w$ the $\mathcal{M}^2$-contribution is in fact negative. A prime example would be a massless minimally coupled scalar field during cosmological inflation for which $m=0$, $\xi=0$ and $w=-1$ giving $\mathcal{M}^2=-2H^2$. If $\mathcal{M}^2<0$ it is a simple matter to show that the modes with $(\mathbf{k}/a)^2+\mathcal{M}^2<0$ contain an exponentially growing branch, which implies that a large field fluctuation can be generated. This effect coming from an imaginary mass-like contribution is sometimes called tachyonic or spinodal instability/amplification \cite{Felder:2001kt}. We note that even if no tachyonicity occurs, a large fluctuation can nonetheless be generated if there is a rapid i.e. a non-adiabatic change in $\mathcal{M}$.

A more precise way of understanding the generation of a large fluctuation is by calculating the infrared (IR) portion of the variance i.e. a loop with a low-momentum cut-off $\Lambda_{\rm IR}$. This shows that in many situations that can broadly be characterized as having $\mathcal{M}^2\lesssim 0$ the result diverges \cite{Markkanen:2017edu}\footnote{For example in de Sitter space one has $|f_{k \rightarrow0} (\eta)|^2\propto H^2 k^{-3}$ for $m=\xi=0$.}  
\ee{\langle\hat{\varphi}^2\rangle_{\Lambda_{\rm IR}}
	\propto \int^{\Lambda_{\rm IR}}_0dk\, k^2\,|f_k(\eta)|^2\overset{t\rightarrow\infty}{\longrightarrow\infty}\label{eq:IR}\,;\quad\text{for}\quad {\lambda=0}\,.}

When the theory is not free interactions will via backreaction prevent the generation of arbitrary large fluctuations. In practice one may understand this as the emergence of positive mass-like contributions 
from the interactions making the field heavy and thus preventing tachyonic or non-adiabatic amplification. The functioning of this mechanism usually allows a significant $\langle\hat{\varphi}^2\rangle$ term indicating that quite generally an IR divergence in the free theory implies a large fluctuation when interactions are included.

This rather simple discussion leads to an important implication in regards the vacuum instability problem in the cosmological setting: even if in flat space the decay of a metastable vacuum is enormously unlikely, this may not have been the case during the earlier cosmological epochs when a transition over the barrier can be induced by a large fluctuation generated by the dynamics on a curved background.

\subsection{Quantum theory in de Sitter space}
\label{sec:StaroYoko}
Even in the simple special case of a de Sitter background it is difficult to perform analytic calculations for an interacting quantum theory. This is mostly due to the non-trivial infrared behavior of quantum fields in de Sitter space \cite{Allen:1985ux,Sasaki:1992ux, Mukhanov:1996ak,Abramo:1997hu,Prokopec:2002uw,Enqvist:2008kt,Onemli:2004mb,Losic:2005vg}. A manifestation of this is the lack of a perturbative expansion based on a non-interacting propagator due to the infrared divergence as described in Eq.~(\ref{eq:IR}). The infrared properties of de Sitter space have attracted significant attention over the years and we refer the interested reader to the review \cite{Seery:2010kh} for more information and references.

One popular way forward is to use techniques based on the so-called two-particle-irreducible (2PI) diagrams, which are essentially non-perturbative resummations of distinct classes of Feynman diagrams. The 2PI approach is attractive in that it is derivable via first principles from quantum field theory without any approximations. Hence, in principle it can be used up to arbitrary accuracy. Unfortunately, only the leading terms that come by the Hartree approximation are analytically tractable. Applications of 2PI techniques to de Sitter space include Refs.~\cite{Riotto:2008mv,Tranberg:2008ae,Arai:2011dd,Serreau:2011fu,Garbrecht:2013coa,Herranen:2013raa,Gautier:2015pca}. 

A non-perturbative framework for calculating quantum correlators in de Sitter space was laid out in Refs.~\cite{Starobinsky:1986fx,Starobinsky:1994bd}. This technique is generally known as the stochastic formalism and is surprisingly straightforward calculationally. It is based on the insight that to a good approximation in de Sitter space one may neglect the quantum nature of the problem and devise a set-up in which the correlators may be calculated from a classical probability distribution ${P}(t,\varphi)$. 
If the scalar field $\hat\varphi$ is light, $m\ll H$,
coarse graining over horizon sized patches allows one to approximate its dynamics with a Langevin equation 
\ee{\dot{\varphi}=-\f{V'(\varphi)}{3H}+f(t)\label{eq:Lan}\,,}
where $V(\varphi)$ is the classical potential and $f(t)$ is a white noise term satisfying 
\ee{\langle f(t')f(t)\rangle =\f{H^3}{4\pi^2}\delta(t'-t)\,. \label{eq:WN}}
The reason why the 'hat' notation has been dropped from $\varphi$ is that Eq.~(\ref{eq:Lan}) contains only classical stochastic quantities i.e. the quantum features are no longer visible. 

The Langevin equation (\ref{eq:Lan}) can be cast in the form of a Fokker-Planck equation for the probability density ${P}(t,\varphi)$~\cite{Starobinsky:1994bd}
\begin{equation}
\dot{P}(t,\varphi)=\frac{1}{3H}\frac{\partial}{\partial \varphi}\big[ {P}(t,\varphi){V}'(\varphi)
\big]+\f{H^3}{8\pi^2}\frac{\partial^2}{\partial \varphi^2 }{P}(t,\varphi)\,.\label{eq:FP}
\end{equation}
After a sufficiently long time has passed one would expect that ${P}(t,\varphi)$ reaches a constant equilibrium distribution.
When $\dot{P}(t,\varphi)=0$, Eq.~(\ref{eq:FP}) has a simple analytic solution as 
\begin{equation}
P_{\rm eq}(\varphi)=N \exp\left\{-\f{8\pi^2V(\varphi)}{3H^4}\right \}\, ,\label{eq:p}
\end{equation}
where $N$ is a normalization factor. 

As an example, for a theory with only a quartic term $V(\varphi)=(\lambda/4)\varphi^4$, which in many cases is the relevant approximation for the SM Higgs in the early Universe, this results in
the equilibrium probability distribution
\begin{equation}
P_{\rm eq}(\varphi)=\left(\frac{32\pi^2\lambda}{3H^4}\right)^{1/4}\frac{1}{\Gamma(1/4)}
\exp\left\{
-\frac{2\pi^2 \lambda \varphi^4}{3H^4}
\right\}.
\end{equation}
The corresponding field variance becomes
\ee{
\label{hstar}
\langle\hat{\varphi}^2\rangle = \sqrt{\frac{3}{2\pi^2}}{\frac{\Gamma(\frac{3}{4})}{\Gamma(\frac{1}{4})}} \frac{H^2}{\sqrt{\lambda}}\approx 0.132 \frac{H^2}{\sqrt{\lambda}}\,.}
This means that the Higgs field develops a non-zero value $\varphi\sim \lambda^{-1/4}H$, which is sometimes called a condensate~\cite{Kunimitsu:2012xx,Enqvist:2013kaa,Enqvist:2014bua, Kusenko:2014lra,Enqvist:2015sua,Pearce:2015nga,Freese:2017ace,Hardwick:2018sck}. 

The  central assumption  that  leads  to  the  stochastic description is 
that the effect of the ultraviolet physics on the infrared behaviour can be described as a white noise term in the Langevin equation~(\ref{eq:Lan}). 
The ultraviolet modes also contribute to the effective potential $V(\varphi)$ in the Fokker-Planck equation (\ref{eq:FP}), as was discussed in Section \ref{sec:rg} in flat space. 
These are two separate effects, which both need to be included in the calculation~\cite{Markkanen:2018bfx}.
Especially when investigating the vacuum stability of the SM it is therefore imperative that the quantum corrections are incorporated in the stochastic approach, for example by making use of the RGI effective potential as the input in Eq.~(\ref{eq:FP}). 

\subsection{Curvature corrections to the effective potential}
\label{eq:effpotc}
It is clearly evident from the derivations of Section \ref{sec:mod} that a scalar field in curved spacetime feels the curvature of the background. It then follows that also the effective potential must receive a contribution from curvature. In order to reliably investigate the implications from the SM metastability in the early Universe these contributions then must be included in a discussion of quantum corrections to the potential. 

Investigations of the effective potential on a curved background have been performed by a number of authors in a variety of models~\cite{Hu:1984js,Elizalde:1993ew,Buchbinder:1986yh,Elizalde:1993ee,Elizalde:1994im,Kirsten:1993jn,Toms:1982af,Toms:1983qr,Odintsov:1993rt,Czerwinska:2015xwa,Elizalde:1993qh,Odintsov:1990mt,Ford:1981xj,George:2012xs,Buchbinder:1992rb,Bounakis:2017fkv}. However, the derivation of the effective potential for the full SM in curved spacetime was only recently carried out it in Ref.~\cite{Markkanen:2018bfx}. 

Deriving the effective potential for a quantized scalar field on a curved background is naturally much more difficult than in flat space: for many backgrounds even the case of a free scalar field admits no closed form solutions for the mode equation \cite{Birrell:1982ix}. Another complication that arises is that choosing the boundary condition i.e. the specific quantum state in which the effective potential is calculated is far from obvious. This is due to the fact that in curved space the concept of a particle and hence the vacuum state is no longer well-defined globally, but depends on the specific dynamics and perceptions of a given particular observer  \cite{Gibbons:1977mu}.

However, even on an arbitrary curved background some things remain universal: renormalizability of a quantum field theory imposes the requirement that all quantum states should have coinciding divergences. From this it follows that it is possible to derive an effective potential retaining terms only originating from the very high ultraviolet (UV), which is a contribution that is always present irrespective of the quantum state one is interested in. Such an expression would then allow one to determine all the generated operators and their respective runnings, as RG effects are ultimately the result of UV physics.

Let us once more study the Yukawa theory of Section \ref{sec:yukawa} only this time in curved spacetime and without neglecting the mass parameter for the scalar. In curved spacetime the action reads
\ee{S =\int d^4x\sqrt{|g|}\,\bigg[\f{1}{2}\nabla_\mu\varphi\nabla^\mu\varphi-\f{1}{2}m^2\varphi^2-\f{\xi}{2}R\varphi^2-\f{\lambda}{4}\varphi^4+\bar \psi\nabla\!\!\!\!/\psi-g\varphi\bar\psi\psi\bigg]\,.\label{eq:yu2}}

The most convenient way of deriving the effective potential is the Heat Kernel method reviewed in Ref.~\cite{Avramidi:2000bm}, see also \cite{Buchbinder:1992rb}. This approach has been known for a long time, see Refs.~\cite{Schwinger:1951nm,DeWitt:1965jb,Seeley:1967ea,Gilkey:1975iq,Minakshisundaram:1949xg,Hadamard} for early work. We will make use of the resummed form of the Heat Kernel expansion derived in Refs.~\cite{Parker:1984dj,Jack:1985mw}, which for the action (\ref{eq:yu2}) gives (for details, see Ref.~\cite{Markkanen:2018bfx})

\ee{V_{\rm eff}(\varphi)=\f{1}{2}m^2\varphi^2+\f{\xi}{2}R\varphi^2+\f{\lambda}{4}\varphi^4+V^{(1)}_\varphi(\varphi)+V^{(1)}_\psi(\varphi)
\label{eq:curve2}\,,}
where the one-loop quantum corrections from the scalar and the fermion, $V^{(1)}_\varphi(\varphi)$ and $V^{(1)}_\psi(\varphi)$, read

\ee{V^{(1)}_\varphi(\varphi)=\f{\mathcal{M}_\varphi^4}{64\pi^2}\bigg[\log\bigg(\f{|\mathcal{M}_\varphi^2|}{\mu^2}\bigg)-\f{3}{2}\bigg]+\f{\f{1}{90}\left(R_{\mu\nu\delta\eta}R^{\mu\nu\delta\eta}-R_{\mu\nu}R^{\mu\nu}\right)}{64\pi^2}\log\bigg(\f{|\mathcal{M}_\varphi^2|}{\mu^2}\bigg)\label{eq:curve3}\,,}
and
\ea{V^{(1)}_\psi(\varphi)=-\f{4\mathcal{M}_\psi^4}{64\pi^2}\bigg[\log\bigg(\f{|\mathcal{M}_\psi^2|}{\mu^2}\bigg)-\f{3}{2}\bigg]+\f{\f{1}{90}\left(
(7/2)R_{\mu\nu\delta\eta}R^{\mu\nu\delta\eta}+4R_{\mu\nu}R^{\mu\nu}\right)}{64\pi^2}\log\bigg(\f{|\mathcal{M}_\psi^2|}{\mu^2}\bigg)\label{eq:curve4}\,,}
respectively, and the curved space effective masses $\mathcal{M}^2_\varphi$ and $\mathcal{M}^2_\psi$ are now
\ee{\mathcal{M}^2_\varphi\equiv m^2+3\lambda\varphi^2+\left(\xi- 1/6\right)R\,;\qquad \mathcal{M}^2_\psi\equiv g^2\varphi^2+R/12\,.\label{eq:effm2}}
The $R^{\mu\nu}$ and $R^{\mu\nu\alpha_\beta}$ are the Ricci and Riemann tensors, respectively. We have introduced the absolute values in the logarithms to ensure that the result is never complex. A complex effective potential in flat space can be interpreted as a finite lifetime of the quantum state~\cite{Weinberg:1987vp}, but this is ultimately an infrared effect and hence not correctly represented in an UV expansion. Therefore, the effective potential in curved space derived with the Heat Kernel expansion correctly represents the local physics and can for example be used to determine the running of parameters in curved space and the possible generation of new operators (see the next section), but in order to answer questions about vacuum decay one needs additional technology, which is discussed in Section \ref{sec:tunn}.

What the above clearly shows is that on a curved background a highly non-trivial dependence on the curvature emerges: A curved spacetime leads to the generation of additional operators that couple to the scalar field. Importantly, the non-minimal term $\propto R \varphi^2$ directly coupling $\varphi$ to $R$ is not the only one, but terms $\propto R^2$,  $R_{\mu\nu}R^{\mu\nu}$ and $R_{\mu\nu\delta\eta}R^{\mu\nu\delta\eta}$ are also unavoidable and they couple to the scalar field via the logarithmic loop contributions. These terms are not necessarily small, for example in de Sitter space with a constant Hubble rate $H$ the various curvature contributions may be written as
\ee{R=144H^4\,,\quad R_{\mu\nu}R^{\mu\nu}=36H^4\,,\quad R_{\mu\nu\delta\eta}R^{\mu\nu\delta\eta}=24H^4
\label{dSRiemann}\,,}
and in the early Universe the Hubble rate can be several orders of magnitude larger than any mass parameter of the SM. Simply put, since curvature is felt by the scalar field its inclusion in the calculation is vital for making robust predictions because the scale provided by $H$ often is the largest scale of the problem.

\subsection{Running couplings in curved space}
\label{sec:crun}
The basic principles laid out in the flat space analysis of Section \ref{sec:rg} remain unchanged  when the background in no longer flat: Demanding a result independent of the renormalization scale $\mu$ leads to the Callan-Symanzik equation from which the beta functions may be solved given the anomalous dimension $\gamma$. Since $\gamma$ is a dimensionless number it will receive no contributions from constants associated with the curvature of space such as $\xi$. Otherwise parameters only visible in the action when the background is curved would nonetheless influence the RG running of, say, $\lambda$. Similar arguments imply that all beta functions present in flat space remain unchanged when the background is curved.

As one may see from Eqs.~(\ref{eq:curve2}) -- (\ref{eq:curve4}) operators that are not present in the tree-level action (\ref{eq:yu2}) are generated by the loop correction. This means that even if one renormalizes these terms to zero, they may resurface via RG running. Ultimately, this is the reason behind the non-minimal term $\propto R\varphi^2$ already in (\ref{eq:yu2}). For the same reason in our theory we must include the following purely gravitational action \ee{S_g=-\int d^4x\,\sqrt{|g|}\bigg[V_{\Lambda}-\kappa R+\alpha_{1} R^2+\alpha_{2} R_{\mu\nu}R^{\mu\nu}+\alpha_{3} R_{\mu\nu\delta\eta}R^{\mu\nu\delta\eta}\bigg]\label{eq:treecurve}\,.}
A straightforward application of the Callan-Symanzik equation (\ref{eq:RGI}) with Eq.~(\ref{eq:gamma}) for Eqs.~(\ref{eq:curve2}) -- (\ref{eq:curve4}) gives the beta functions for the Yukawa theory
\ea{\beta_\xi&=\f{(\xi-1/6)}{16\pi^2}\left(6\lambda+4g^2\right)\,;&\beta_{V_\Lambda}&=\f{m^4/2}{16\pi^2}\,; &\beta_\kappa&=-\f{m^2(\xi-1/6)}{16\pi^2}\,;& \nonumber \\\beta_{\alpha_1}&=\f{(\xi-1/6)^2/2-1/72}{16\pi^2}\,; &\beta_{\alpha_2}&=\f{1/60}{16\pi^2}\,;& \beta_{\alpha_3}&=\f{1/40}{16\pi^2}\,. \label{eq:betaC}}
which along with Eqs.~(\ref{eq:la0})--(\ref{eq:gamma})  provide a complete set of RG equations for the Yukawa theory in curved spacetime.

A crucial difference to the flat space case arises when implementing renormalization group improvement. In Section \ref{sec:rg} we exploited the fact that the full quantum result must be independent of the renormalization scale $\mu$ in order to optimize the pertubative expansion. Namely, we made the choice (\ref{equ:mustar}) in order to keep the logarithms small also at large scales. 
In curved space the logarithms in the loop corrections (\ref{eq:curve3}) and (\ref{eq:curve4}) have dependence on the scalar curvature $R$, and therefore it must be included in the optimization. 
The one-loop calculation shows that the exact scale choice that would fully cancel the loop correction is not possible across the whole range of field values~\cite{Markkanen:2018bfx}. 
Instead, a sensible choice for the optimized scale $\mu_*$ is a linear combination of $\varphi^2$ and $R$ i.e.
\ee{\mu_*^2(\varphi,R)=a\varphi^2+bR\,,\label{eq:ccrun}}
where the parameters $a$ and $b$ are chosen in such a way that the logarithms remain under control.
 
Equation (\ref{eq:ccrun}) highlights an often neglected effect arising in curved spaces after renormalization group improvement: In a curved background the optimal scale choice depends significantly on curvature. This phenomenon may be characterized as \textit{curvature induced running} and was recently studied in detail for the full SM in Ref.~\cite{Markkanen:2018bfx}. In situations where the curvature of the background is significant it can give the dominant contribution to the scale. Considering the metastability of the SM in the early Universe this in fact is often the case as during and after inflation one may have a Hubble rate much larger than the instability scale, $H\gg \mu_\Lambda$.
\subsection{The Standard Model}
The Standard Model particle content can be expressed with the Lagrangian
\ee{
{\cal L}_{\rm SM} ={\cal L}_{\rm YM} + {\cal L}_{\rm F} + {\cal L}_{\Phi}+{\cal L}_{GF}+{\cal L}_{GH}\,.
\label{eq:lag}}
The first three terms in Eq.~(\ref{eq:lag}) describe the contributions coming from the gauge fields, the fermions and the Higgs doublet $\Phi$ whose one point function we write as
$\langle\hat{\Phi}\rangle\equiv\varphi$, 
from now on dropping the hats. The 'GF' and 'GH' are the gauge fixing and ghost Lagrangians, respectively. Here we show explicitly only the Higgs contribution (for the full result see Ref.~\cite{Markkanen:2018bfx})
\ea{
{\cal L}_{\Phi} &= \left(D_\mu \Phi \right)^\dagger \left(D^\mu \Phi \right) +m^2\Phi^\dagger \Phi-\xi R\Phi^\dagger \Phi- \lambda (\Phi^\dagger \Phi)^2\label{eq:higg1}
\,,}
with the SM covariant derivative
\ea{
D_\mu &= \nabla_\mu - ig \tau^a A^a_\mu - ig'Y A_\mu;\qquad \tau^a=\sigma^a/2\, ,
}
where $\nabla_\mu$ contains the connection appropriate for Einsteinian gravity, $g$ and $g'$ ($A^a_\mu$ and $A_\mu$) are the $SU(2)$ and $U(1)$ gauge couplings (fields),
$\tau$ and $Y$ the corresponding generators, and $\sigma^a$ the Pauli matrices.

As de Sitter space is the most important application of our results here we show the perturbative 1-loop correction for the SM in a spacetime with an equation of state $w=-1$ i.e. a constant Hubble rate $H$ (see Section \ref{sec:hom})
\ea{
	V_{\rm SM}^{(1)}(\varphi,\mu)& = 
	\f{1}{64\pi^2} \sum\limits_{i=1}^{31}\bigg\{ n_i\mathcal{M}_i^4\bigg[\log\left(\f{|\mathcal{M}_i^2|}{\mu^2}\right) - d_i \bigg] +{n'_i}H^4\log\left(\f{|\mathcal{M}_i^2|}{\mu^2}\right)\bigg\}\,,
	\label{potential}}
where the sum is over all degrees of freedom of the SM, which may be found in tables \ref{tab:contributions} and \ref{tab:contributions2}. 
\begin{table}
	\caption{\label{tab:contributions}The 1-loop effective potential (\ref{potential}) contributions with tree-level couplings to the Higgs. $\Psi$ stands for $W^{\pm}$ and $Z^0$ bosons, the 6 quarks q, the 3 charged leptons $l$, the Higgs $h$. The Goldstone bosons are $\chi_{W}$ and $\chi_{Z}$ and ghosts $c_{W}$ and $c_{Z}$. The masses may be found in Eq.~(\ref{eq:masses}).}
	\begin{center}
		{\tabulinesep=1.6mm
			\begin{tabu}{|c||ccccc|}
				\hline
				$\Psi$ & $~~i$ & $~~n_i$   & $~d_i$    &$~n'_i$      &$\hspace{-2.6cm}\mathcal{M}_i^2$    \\\hhline{|=#=====|}
				$~$ & $~~1$  & $~~2$       & $\quad{3}/{2}~~$        & $-34/15$        &  $\hspace{-2.8cm}m^2_W+H^2$      \\
				$~W^\pm$ & $~~2$  & $~~6$       & $\quad{5}/{6}~~$       & $-34/5$        &  $\hspace{-2.8cm}m^2_W+H^2$       \\
				$~$ & $~~3$  & $-2$      & $\quad{3}/{2}~~$         & $~~4/15$        & $\hspace{-2.7cm} ~m^2_W-2H^2$        \\\hline
				$~$ & $~~4$  & $~~1$        & $\quad{3}/{2}~~$ & $-17/15$        & $\hspace{-2.8cm} m^2_Z+H^2$      \\
				$Z^0$ & $~~5$  & $~~3$        & $\quad{5}/{6}~~$ & $-17/5$        & $\hspace{-2.8cm} m^2_Z+H^2$     \\
				$~$ & $~~6$  & $-1$      & $\quad{3}/{2}~~$ & $~~2/15$        & $\hspace{-2.7cm} ~m^2_Z-2H^2$     \\\hline
				q & $7-12$  & $-12$     & $\quad{3}/{2}~~$    & $~~38/5$        & $\hspace{-2.8cm} m^2_{q}+H^2$      \\\hline
				$l$ & $13-15$  & $-4$     & $\quad{3}/{2}~~$    & $~38/15$        & $\hspace{-2.8cm} m^2_{l}+H^2$      \\\hline
				$h$ & $~16$  & $~~1$       & $\quad{3}/{2}~~$          & $-2/15$      & $\hspace{-0.9cm}m_h^2+12(\xi-{1}/{6})H^2$  \\\hline
				${\chi}_W$ & $~17$  & $~~2$      & $\quad{3}/{2}~~$           & $-4/15$      & $\,\,\,~~\quad m_\chi^2\,+\zeta_W m^2_W+12(\xi-{1}/{6})H^2$   \\\hline
				${\chi}_Z$ & $~18$  & $~~1$       & $\quad{3}/{2}~~$           & $-2/15$      & $\,~~\quad m_\chi^2+\zeta_Z m^2_Z\,+12(\xi-{1}/{6})H^2$   
				\\\hline ${c}_W$ & $~19$  & $-2$       & $\quad{3}/{2}~~$           & $~~4/15$      & $\hspace{-1.9cm}\zeta_W m^2_W-2H^2$  
				\\\hline
				${c}_Z$ & $~20$  & $-1$      & $\quad{3}/{2}~~$            & $~~2/15$      & $\hspace{-2.1cm}\zeta_Z m^2_Z-2H^2$   
				\\\hline
			\end{tabu}
		}
	\end{center}
	
\end{table}
\begin{table}
	\caption{\label{tab:contributions2}Contributions to the effective potential (\ref{potential}) with no coupling to the Higgs at tree-level. The $\Psi$ include the photon $\gamma$, the 8 gluons $g$, the 3 neutrinos $\nu$ and the respective ghosts $c_\gamma$ and $c_g$. }
	\vspace{2mm}
	\begin{center}
		{\tabulinesep=1.6mm
			\begin{tabu}{|c||ccccc|}
				\hline
				$\Psi$ & $~~i$ & $~~n_i$   & $~d_i$    &$~n'_i$      &$\mathcal{M}_i^2$    \\\hhline{|=#=====|}
				$~$ & $~21$  & $~~1$       & $\quad{3}/{2}~~$        & $-17/15$        &  $H^2$      \\
				$~\gamma$ & $~22$  & $~~3$       & $\quad{5}/{6}~~$       & $-17/5$        &  $H^2$       \\
				$~$ & $~23$  & $-1$      & $\quad{3}/{2}~~$         & $~~2/15$        & $ -2H^2$        \\\hline
				$~$ & $~24$  & $~~8$        & $\quad{3}/{2}~~$ & $-136/15$        & $ H^2$      \\
				$g$ & $~25$  & $~~24$        & $\quad{5}/{6}~~$ & $-136/5$        & $ H^2$     \\
				$~$ & $~26$  & $-8$      & $\quad{3}/{2}~~$ & $~~16/15$        & $-2H^2$     \\\hline
				$\nu$ & $27-29$  & $-2$      & $\quad{3}/{2}~~$           & $~~19/15$      & $H^2$   \\\hline
								${c}_\gamma$ & $~30$  & $-1$       & $\quad{3}/{2}~~$           & $~~2/15$      & $-2H^2$  
				\\\hline
				${c}_g$ & $~31$  & $-8$      & $\quad{3}/{2}~~$            & $~~16/15$      & $-2H^2$   
				\\\hline
			\end{tabu}
		}
	\end{center}
\end{table}The masses are defined as
\ea{
m^2_h&=-m^2+3\lambda\varphi^2,\quad \,m^2_i=\f{y_i^2}{2}\varphi^2,\quad m^2_W = \f{g^2}{4}\varphi^2\,,\nonumber \\
m^2_Z &= \f{g^2 + (g')^2}{4}\varphi^2\,,\quad m^2_\chi=-m^2+\lambda\varphi^2
\,.\label{eq:masses}}
and the $\zeta_i$ are the gauge fixing parameters.

The flat space beta functions have of course been known for some time, see for example Refs.~\cite{Ford:1992mv,Buttazzo:2013uya}. The complete set of SM beta functions to 1-loop order was however first calculated only in Ref.~\cite{Markkanen:2018bfx}. The 1-loop SM beta functions for couplings associated with gravity coming from the action (\ref{eq:treecurve}), $\xi, V_{\Lambda}, \kappa, \alpha_1, \alpha_2$ and $\alpha_3$ can be solved from Eq.~(\ref{potential}) and read 
\ea{16\pi^2 \beta_{\xi} &= \label{eq:gravbetas}\bigg(\xi - \f{1}{6}\bigg)\bigg[12 \lambda +2Y_2-\frac{3 (g')^2}{2}-\frac{9 g^2}{2}\bigg]\,, \\ {16\pi^2}\beta_{V_\Lambda}&={2m^4}\,, \\{16\pi^2}\beta_\kappa&={4}{m^2\bigg(\xi - \f{1}{6}\bigg)}\,, \\ {16\pi^2}\beta_{\alpha_1}&=2 \xi ^2-\frac{2 \xi }{3}-\frac{277}{144}\,, \label{eq:gravbetas1}\\16\pi^2 \beta_{\alpha_2}&=\frac{571}{90}\,,\label{eq:gravbetas2} \\16\pi^2 \beta_{\alpha_3}&=-\frac{293}{720}\,,\label{eq:gravbetas3}}
where
\begin{align}
Y_2 &\equiv 3(y_u^2 + y_c^2 + y_t^2) + 3(y_d^2 + y_s^2 + y_b^2) + (y_e^2 + y_{\mu}^2 + y_{\tau}^2)\,,
\end{align}
with $y_i$ being a Yukawa coupling for a fermion type $i$.

Much like in the flat space case in Eq.~(\ref{eq:RGIfull}) we can write the RGI effective potential by choosing an optimized scale $\mu_*(\varphi,R)$ in such a way that the loop correction is small~\cite{Markkanen:2018bfx}. In curved space in addition to the Lagrangian from Eq.~(\ref{eq:higg1}) we must include the purely gravitational terms from Eq.~(\ref{eq:treecurve}) in addition to the one loop contributions (\ref{potential}) giving rise to
\ea{V^{\rm SM}_{\rm RGI}(\varphi)&=\f{\xi(\mu_*)}{2}R\varphi^2 + \f{\lambda(\mu_*)}{4}\varphi^4+\alpha_{1}(\mu_*) R^2+\alpha_{2}(\mu_*) R_{\mu\nu}R^{\mu\nu}+\alpha_{3}(\mu_*) R_{\mu\nu\delta\eta}R^{\mu\nu\delta\eta}\nonumber \\ &+ V^{(1),{\rm SM}}_{\rm RGI}(\varphi,\mu_*) \,,\label{eq:effiSM}}
where $\mu_*$ generally depends on both $\varphi$ and $R$, and 
we have assumed $|R|\gg |m_h^2|$, which is usually true for the SM Higgs in the early Universe. 

When the Hubble rate is above electroweak scales it is quite obvious that the highly non-trivial curvature dependence apparent in Eq.~(\ref{potential}) and also in Eq.~(\ref{eq:effiSM}) with the optimized scale (\ref{eq:ccrun}) cannot be neglected: it is just as, if not more, important as what would have been obtained by using only a flat space derivation. The most obvious difference is the emergence of the direct non-minimal coupling between the Higgs and the scalar curvature $R$. Due to the curvature dependence of the optimized renormalization scale in curved space~(\ref{eq:ccrun}), which can be traced back to the curvature dependence of the one-loop correction~(\ref{potential}), the generation of the non-minimal coupling in the current cosmological paradigm is unavoidable. It will be sourced by the changing Hubble rate $H$. Furthermore, as can be read from the beta function (\ref{eq:gravbetas}), $\xi=0$ is not a fixed point of the RG flow. Depending on the sign of $\xi R$, the non-minimal coupling can have a stabilizing or destabilizing effect, which can be very significant in the early Universe.

\begin{figure}[ht]
\centering 
\includegraphics[height = 0.4\textheight]{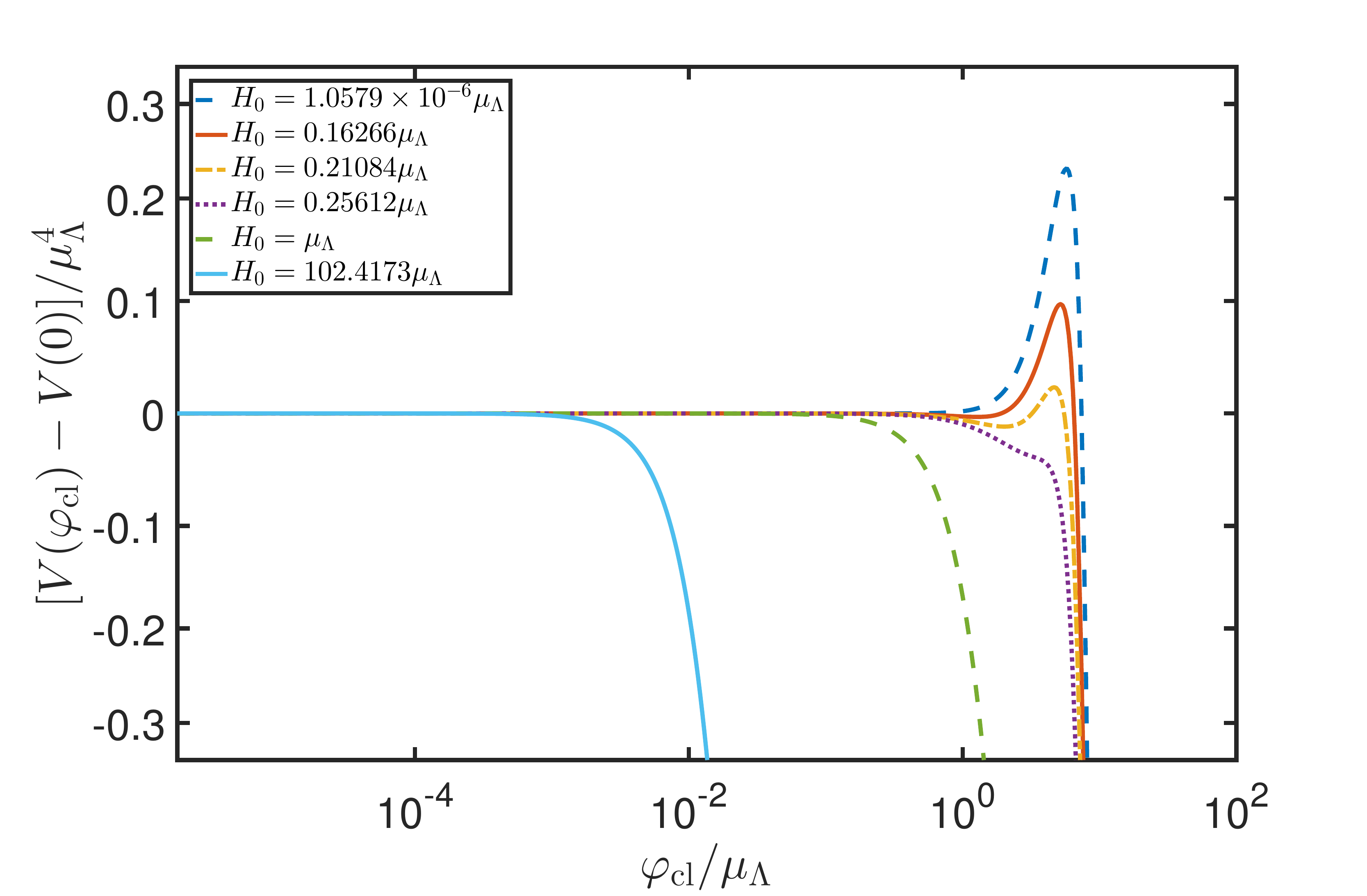}
    \caption{\label{fig:xi0}The one-loop RGI effective potential for the full SM in de Sitter space with $\xi=0$ at the electroweak scale, in units of the instability scale (\ref{eq:insta}), using the optimised scale choice~(\ref{eq:ccrun}). The $x$ axis is given by the field renormalized at the physical top mass, $\varphi_{\rm cl}\equiv\varphi(M_t)$. The disappearance of the potential barrier at large Hubble rates can be traced back to the RG running of the non-minimal coupling $\xi$.     	Figure taken from Ref.~\cite{Markkanen:2018bfx}.}
\end{figure}

In Fig.\,\ref{fig:xi0} we illustrate the behavior of effective potential for the full SM in de Sitter space including the one loop quantum correction (\ref{potential}). We have chosen to set the renormalised non-minimal coupling $\xi$ to zero at the electroweak scale. We use the field renormalized at the physical top mass 
\ee{\varphi_{\rm cl}\equiv\varphi(M_t)=\sqrt{\f{Z(\mu_*)}{Z(M_t)}}\varphi(\mu_*)\,,} as the x-axis. It is clearly evident that in curved space the potential may have drastically different predictions to flat space. 
As can be read off from the beta function for $\xi$ (\ref{eq:gravbetas}), if $\xi=0$ at some low scale, it will run to negative values at high scales. Furthermore, since in de Sitter space $R=12H^2>0$, a negative $\xi$ can prevent the emergence of a potential barrier, even if robustly present on a flat background, as visible in Fig.\,\ref{fig:xi0}. 
\makeatletter{}\section{Vacuum Decay}
\label{sec:tunn}

\subsection{Quantum tunnelling and bubble nucleation}
\label{sec:tunnintro}
The main mechanism behind vacuum decay in the Standard Model is essentially a direct extension of ordinary quantum tunnelling to quantum field theories. In ordinary quantum mechanics, the wave-function for particles trapped by a potential barrier can penetrate the classically forbidden region of the barrier, leading to a non-zero probability to be found on the other side. The transition rate for particles of energy $E$ incident on a barrier described by potential $W(x)$ can be estimated using the WKB method \cite{Coleman:1985rnk},
\begin{equation}
T = \exp\left(-2\int_{x_1(E)}^{x_2(E)}dx\sqrt{2(W(x) - E)}\right),\label{eq:TWKB}
\end{equation}
where $x_1,x_2$ are the turning points of the potential. As is clear from this expression, the tunnelling rate is suppressed by wide and tall barriers.

Although Eq. (\ref{eq:TWKB}) can in principle be evaluated directly, we will follow a different approach
that readily generalises to quantum field theories~\cite{Coleman:1977py,Brown:2007sd}.
The idea is to use the equation of motion,
\begin{equation}
\frac{d^2x}{dt^2} = -W'(x) \rightarrow \frac{1}{2}\left(\frac{dx}{dt}\right)^2 + W(x) = E.\label{eq:LorEq}
\end{equation}
The region $(x_1,x_2)$ is classically forbidden, since $W(x) - E > 0$ there. We can apply a trick, however, by analytically continuing time to an imaginary value: $\tau = it$, which gives a \emph{Euclidean} equation of motion,
\begin{equation}
\frac{d^2x}{ d\tau^2} = +W'(x) \implies \frac{1}{2}\left(\frac{dx}{d\tau}\right)^2 - W(x) = -E.\label{eq:EuclEq}
\end{equation}
The most notable feature of these equations is that the potential has effectively been inverted. This means that we can find a classical solution that rolls through the barrier between the turning points $x_1$ and $x_2$. If we can find this solution, it allows us to re-express the integral in Eq. (\ref{eq:TWKB}) as
\begin{align}
2\int_{x_1}^{x_2}dx\frac{dx}{d\tau} =& 2\int_{\tau_1}^{\tau_2} d\tau \left(\frac{dx}{d\tau}\right)^2 \nonumber\\
=&2\int_{\tau_1}^{\tau_2} d\tau\left[\frac{1}{2}\left(\frac{dx}{ d\tau}\right)^2 + W(x) - E\right] = S_{\rm{E}}[x_B(\tau)] - S_{\rm{E}}[x_{\rm{fv}}(\tau)],\label{eq:TunExpon}
\end{align}
where $S_{\rm{E}}$ is the \emph{Euclidean} action corresponding to Eq. (\ref{eq:EuclEq})
\begin{equation}
S_E[x(\tau)] =\int d\tau\left[\frac{1}{2}\left(\frac{dx}{ d\tau}\right)^2 + W(x)\right],\label{eq:ActionQM}
\end{equation}
while $x_B(\tau)$ is a bounce solution of the Euclidean equations of motion satisfying $x'(\tau_1) = x'(\tau_2) = 0$, and $x_{\rm{fv}}(\tau)$ is a constant solution, sitting in the false vacuum with energy $E$. The `bounce' solution is so named because we see, by energy conservation, that it starts at $x_1$, rolls down the inverted potential before `bouncing' off $x_2$ and rolling back. By finding this solution and evaluating its action, we can compute the rate for tunnelling through a barrier.

This argument generalised straightforwardly to many-body quantum systems, where we use the action
\begin{equation}
S_{\rm{E}}[q_i(\tau)]=\int d\tau\left[\sum_i\frac{1}{2}\left(\frac{dq_i}{ d\tau}\right)^2 + W(q_i)\right].\label{eq:manyBody}
\end{equation}
With more than one degree of freedom, however, there are actually an infinite number of paths that $q_i(\tau)$ could take when passing through the barrier, corresponding to an infinite number of solutions. However, since the decay rate is exponentially dependent on the action, $\Gamma \propto e^{-S_{\rm{E}}[q_i]}$, it is clear that only the solution with smallest Euclidean action will contribute significantly, as this will dominate the decay rate (in other words, the tunnelling takes the `path of least resistance').

The generalisation from a many body system, $q_i$, to a quantum field theory with scalar field $\varphi(x)$ is then straightforward,
\begin{equation}
S_{\rm{E}}[\varphi(x)] = \int d^4x\left[\frac{1}{2}\partial_{\mu}\varphi\partial^{\mu}\varphi + V(\varphi)\right].\label{eq:actionQFT}
\end{equation}
The integral here is over flat four-dimensional Euclidean space, and note that the opposite sign of the potential leads to an opposing sign in the equations of motion,
\begin{equation}
-\nabla_{\mu}\nabla^{\mu}\varphi+ V'(\varphi) = 0.\label{eq:BounceEofMPDE}
\end{equation}
Although it is tempting to interpret $V(\phi)$ as the potential to be tunneled through, this is only somewhat true. The analogue of $W(q_i)$ in Eq.~(\ref{eq:manyBody}) is a functional of the field configuration $\varphi(x)$, given by an integral over three-dimensional space space,
\begin{equation}
U[\varphi(x)] = \int d^3x\left[\frac{1}{2}(\nabla\varphi)^2 + V(\varphi)\right],\label{eq:QFTPot}
\end{equation}
where $\nabla\varphi$ represents the \emph{spatial} derivative of the field. 
In the analogy with quantum mechanics, this term should be considered part of the potential, as its many body equivalent is a nearest-neighbour interaction between adjacent degrees of freedom, $q_i,q_{i\pm 1}$. This means, in particular, 
while in quantum mechanics, the particle emerges after tunneling at a point $x_2$ that has the same potential energy, $W(x_1) = W(x_2)$, in quantum field theory, 
the field emerges \emph{lower} down the potential $V$.

In a field theory, the analogue of $x_2$ is a field configuration, $\varphi(x)$, given by slicing the bounce solution at its mid-way point. This is a nucleated `true-vacuum' bubble, whose decay rate is determined by the Euclidean action of the bounce solution, $\varphi_B$. As we will see in Section \ref{sec:BubbleEvolution}, the dominant Euclidean solutions have $O(4)$ symmetry, which means that the bubble nucleates with $O(3,1)$ symmetry. This causes it to expand at near the speed of light, resulting in the space around a nucleation point being converted to the true vacuum, releasing energy into the bubble wall. Apart from the destruction that this would unleash, and the different masses of fundamental particles in the bubble interior, the result is also gravitational collapse of the bubble  \cite{Coleman:1980aw}, making its nucleation in our past light-cone completely incompatible with the trivial observation that the vacuum has not decayed (yet).

In cosmological applications, but also other areas, it is also important to consider the effect of thermally induced fluctuations over the barrier. Brown and Weinberg  \cite{Brown:2007sd} describe how thermal effects can be included in the above argument. At non-zero temperature, we must integrate over the possible excited states, and the decay exponent which depends on energy,
\begin{equation}
T \propto \int dE e^{-\beta E}e^{-B(E)},
\end{equation}
where $B(E)$ is the (energy dependent) difference in Euclidean action between the bounce solution and the excited state of energy $E$. This integral is dominated by the energy that minimises the exponent $\beta E + B(E)$, which is easily shown to satisfy
\begin{equation}
\beta = 2(\tau_2(E) - \tau_1(E)),
\end{equation}
where $\tau_1,\tau_2$ are the initial and final values in imaginary time of the (energy dependent) bounce solution. In other words, the bounce solution is periodic in imaginary time, with period controlled by the temperature.

In quantum field theory, the decay rate per unit volume and time of a metastable vacuum decays was first discussed by Coleman \cite{Coleman:1977py,Callan:1977pt}, and is given by
\begin{equation}
\Gamma = 
A\exp\left(-B\right),\quad
A=
\left(\frac{B}{2\pi}\right)^2 \left|\frac{\det '(S''[\varphi_{\rm B}])}{\det (S''[\varphi_{\rm{fv}}])}\right|^{-\frac{1}{2}}
,\label{eq:tun:decay_rate}
\end{equation}
where
\begin{equation}
B = S[\varphi_{\rm B}] - S[\varphi_{\rm{fv}}]\label{eq:B}
\end{equation}
is the difference between the Euclidean action of a so called bounce solution $\varphi_{\rm B}$ of the Euclidean (Wick rotated) equations of motion, and the action of the constant solution $\varphi_{\rm fv}$ which sits in the false vacuum. $S''$ denotes the second functional derivative of the Euclidean action of a given solution, and $\det'$ denotes the functional determinant after extracting the four zero-mode fluctuations which correspond to translations of the bounce (these are responsible for the formula giving a decay rate per unit volume). 
Precise calculations of the pre-factor $A$ in the Standard Model were performed in \cite{Isidori:2001bm}, and involve computing the fluctuations around the bounce solution of all fields that couple to the Higgs. This requires renormalising the loop corrections, and also to avoid double-counting, expanding around the tree-level bounce, rather than the bounce in the loop corrected potential.

In the gravitational case, the prefactor $A$ is harder to compute. The main issue is that it includes both Higgs and \emph{gravitational} fluctuations, and without a way of renormalizing the resulting graviton loops, the calculation becomes much harder. Various attempts have been made to do this using the fluctuations discussed in Section \ref{sec:bounce_eigenvalues} (see Refs.~\cite{Dunne:2006bt,Lee:2014uza,Koehn:2015hga} for example), but a full description, especially for the Standard Model case, is not yet available.

In most cases, it is reasonable to estimate  the prefactor $A$ using dimensional analysis. 
Because $A$ has dimension four, one would expect 
\begin{equation}
\label{equ:prefactor}
A\sim\mu^4, 
\end{equation}
where $\mu$ the characteristic energy scale of the instanton solution.
Due to the exponential dependence on the decay exponent, $B$, this will not lead to large errors,
and therefore we will use this result in the absence of more accurate estimates.

\subsection{Asymptotically flat spacetime at zero temperature}
\label{sec:tun:flat}

In flat Minkowski space, the bounce solution corresponds to a saddle point of the Euclidean action,
\begin{equation}
S[\varphi] = \int d^4 x\left[\frac{1}{2}\partial_{\mu}\varphi\partial^{\mu}\varphi + V(\varphi)\right],\label{eq:ScalarEuclACtion}
\end{equation}
with one negative eigenvalue (see Section~\ref{sec:bounce_eigenvalues}).
Since Eq. (\ref{eq:tun:decay_rate}) depends exponentially on the bounce action, only the lowest action bounce solutions will contribute. 
In flat space, it is always the case that the lowest action solution has $O(4)$ symmetry~\cite{Coleman:1977th}. This means that the equations of motion for the bounce can be reduced to
\begin{equation}
\ddot{\varphi} + \frac{3}{r}\dot{\varphi} - V'(\varphi) = 0,\label{eq:O4symFlat}
\end{equation}
subject to the boundary conditions $\dot{\varphi}(0) = 0$ and $\varphi(r\rightarrow\infty)\rightarrow \varphi_{\rm{fv}}$. These ensure that the bounce action is finite and thus gives non-zero contribution to the decay rate.
There are always trivial solutions corresponding to the minima of the potential $V(\varphi)$, but they
do not contribute to vacuum decay because they have no negative eigenvalues.

For example, in a theory with a constant negative quartic coupling, that is,
\begin{equation}
V(\varphi) = -|\lambda|\frac{\varphi^4}{4},\label{eq:tun:constLambda}
\end{equation}
there exists the Lee-Weinberg or Fubini bounce \cite{Lee:1985uv,Fubini:1976jm}. This is a solution of the form:
\begin{equation}
\varphi_{\rm{LW}}(r) = \sqrt{\frac{2}{|\lambda|}}\frac{2r_B}{r_B^2 + r^2},\label{eq:LWbounce}
\end{equation}
where the arbitrary parameter $r_B$ characterises the size of the bounce (and thus the nucleated bubble). This arbitrary parameter appears in the theory because the potential Eq. (\ref{eq:tun:constLambda}) is conformally invariant, and thus bounces of all scales contribute equally with action
\begin{equation}
S[\varphi_{\rm{LW}}] = \frac{8\pi^2}{3|\lambda|}.\label{eq:LWaction}
\end{equation}
In fact, similar bounces contribute approximately in the Standard Model, where the running of the couplings breaks this approximate conformal symmetry, so that bounces of order the scale at which $\lambda$ is most negative (which is the minimum of the $\lambda(\mu)$ running curve) dominate the decay rate \cite{Isidori:2001bm}. 

The complete calculation
would also include gravity, and would therefore involve finding the corresponding saddle point of the 
action 
\begin{equation}
S[\varphi,g_{\mu\nu}] = \int d^4x\left[\frac{1}{2}\nabla_{\mu}\varphi\nabla^{\mu}\varphi + V(\varphi) - \frac{M_{\rm{P}}^2}{2}R\right],
\label{eq:CdLAction}
\end{equation}
where $R$ is the Ricci scalar.
The leading gravitational correction to Eq.~(\ref{eq:LWaction}) is~\cite{Isidori:2007vm}
\begin{equation}
\Delta S_{\rm gravity}=\frac{256\pi^3}{45(r_BM_{\rm P}\lambda)^2}.
\end{equation}
Another approach is to solve the bounce equations numerically, which makes it possible to use the exact field and Einstein equations and the full effective potential. The difference is a second order correction \cite{Isidori:2007vm}. 
Using the tree-level RGI effective potential (\ref{eq:RGI0}), the full numerical result 
including gravitational effects
for $M_t = 173.34{\rm{\,GeV}}$, $M_h = 125.15{\rm{\,GeV}}$, $\alpha_S(M_z) = 0.1184$ and minimal coupling $\xi=0$
is \cite{Rajantie:2016hkj}
\begin{equation}
B_{\rm{grav}} = 1808.3.\label{eq:BgravNumerical}
\end{equation}

\begin{figure}[t]
	\centering
	\includegraphics[width=\textwidth]{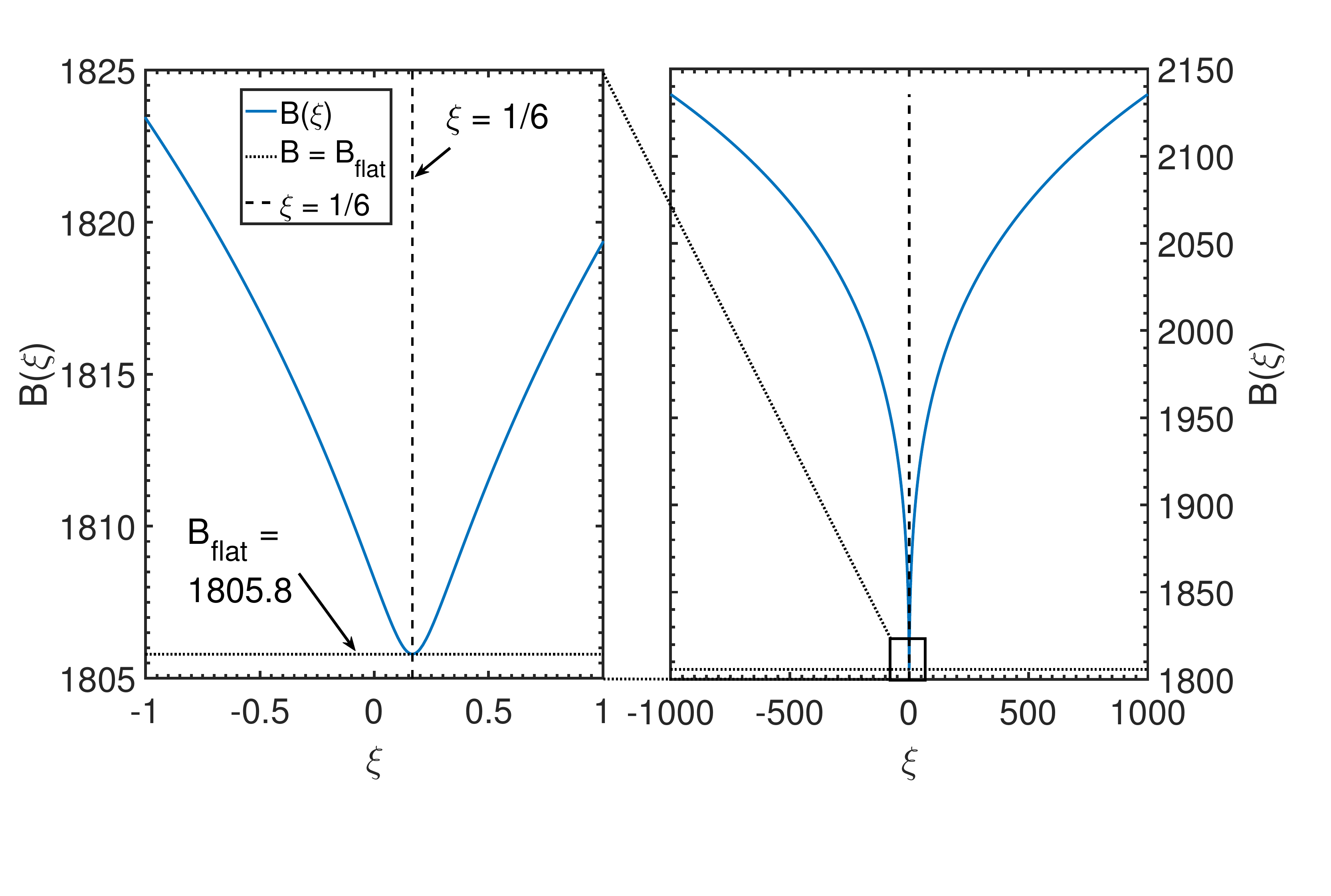}
	\caption{\label{fig:nonminB}Plot of the decay rate for a flat false vacuum for different values of the non-minimal coupling, $\xi$. The minimal action is obtained close to the conformal value $\xi = 1/6$, and agrees well with the flat space result (\ref{eq:BflatNumerical}). Originally published in Ref.~\cite{Rajantie:2016hkj}.}
\end{figure}

A non-minimal value of the Higgs curvature coupling $\xi$
changes the action and the shape of the bounce solution (and thus the scale that dominates tunneling) \cite{Isidori:2007vm,Czerwinska:2016fky,Czerwinska:2017pgb,Rajantie:2016hkj,Salvio:2016mvj}.
Fig.~\ref{fig:nonminB} shows the bounce action $B$ as a function of $\xi$, computed numerically in Ref.~\cite{Rajantie:2016hkj}.
As the plot shows, the action is smallest near the conformal value $\xi=1/6$.
For $\xi\approx 1/6$, the result agrees well with the perturbative calculation~\cite{Salvio:2016mvj}),
\begin{equation}
\Delta S_{\rm gravity}=\frac{32\pi^2(1-6\xi)^2}{45(r_BM_{\rm P}\lambda)^2}.
\end{equation}

For comparison, for the same parameters, the numerically computed decay exponent in flat space is \cite{Rajantie:2016hkj}
\begin{equation}
B_{\rm{flat}} = 1805.8,\label{eq:BflatNumerical}
\end{equation}
which is very close to the full gravitational result with the conformal coupling $\xi=1/6$.
The analytical approximation (\ref{eq:LWaction}) using $\mu_{\rm{min}} = 2.79\times 10^{17}{\rm{\,GeV}}$ gives
\begin{equation}
\label{eq:Bflatanalytica}
S[\varphi_{LW}] = 1804.5.
\end{equation}

Calculations of the prefactor $A$ show that the decay rate (\ref{eq:tun:decay_rate}) is well approximated by \cite{Isidori:2001bm}
\begin{equation}
\label{equ:Gamma0}
\Gamma \sim \mu_{\rm{min}}^4e^{-B}\sim 10^{-716}~{\rm GeV}^4,
\end{equation}
where the numerical value corresponds to the action (\ref{eq:BgravNumerical}).
This agrees with the estimate from dimensional analysis (\ref{equ:prefactor}).
Note, however, that the rate is very sensitive to the top quark and Higgs boson masses, and also to higher-dimensional operators~\cite{Branchina:2013jra,Branchina:2014rva}. 

The presence of a small black hole can catalyse vacuum decay and make it significantly faster~\cite{Gregory:2013hja,Burda:2015isa,Burda:2015yfa,Burda:2016mou,Tetradis:2016vqb}. 
The action of the vacuum decay instanton in the presence of a seed black hole is given by
\begin{equation}
B=\frac{M_{\rm seed}^2-M_{\rm remnant}^2}{2\MP ^2},
\end{equation}
where $M_{\rm seed}$ and $M_{\rm remnant}$ are the masses of the seed black hole and the left over remnant black hole.
For black holes of mass $M_{\rm seed}\lesssim 10^5 \MP \approx 1{\rm g}$ the vacuum decay rate becomes unsuppressed. This can be interpreted~\cite{Tetradis:2016vqb,Mukaida:2017bgd} as a thermal effect due to the black hole temperature $T_{\rm seed}=\MP ^2/M_{\rm seed}$.
The catalysis of vacuum decay does not necessarily rule out cosmological scenarios with primordial black holes, because 
positive values of non-minimal coupling $\xi$ would suppress the vacuum decay in the presence of a black hole~\cite{Canko:2017ebb}.

\subsection{Non-zero temperature}
\label{sec:finiteTdecay}
The presence of a heat bath with non-zero temperature has a significant impact on the vacuum decay rate $\Gamma$ \cite{Anderson:1990aa,Arnold:1991cv}. On one hand, 
the thermal bath modifies the effective potential of the Higgs field, and on the other hand, as discussed in Section~\ref{sec:tunnintro}, it modifies the process itself because it can start from an excited state rather than the vacuum state.

At one-loop level, the finite-temperature effective potential can be written
as \cite{Arnold:1991cv}
\begin{equation}
\label{equ:thermalpot}
V_{\rm eff}(T,\varphi)=
V_{\rm eff}(\varphi)
+T\sum_i n_i\int \frac{d^3k}{(2\pi)^3}\ln\left[
1\mp e^{-\sqrt{k^2+{\cal M}_i^2}/T}
\right],
\end{equation}
where $n_i$ and ${\cal M}_i^2$ are given in Table~\ref{tab:contributions} (taking $H=0$).
In the high-temperature limit, $T\gg M_h$, this can be approximated by
\begin{equation}
V_{\rm eff}(T,\varphi)\approx V_{\rm eff}(0,\varphi)
+\frac{1}{2}\gamma^2 T^2\varphi^2,
\end{equation}
where 
\begin{equation}
\gamma^2\approx\frac{1}{12}\left(
\frac{3}{4}g^2+\frac{9}{4}{g'}^2+3y_t^2+6\lambda
\right).
\label{equ:thermalmass}
\end{equation}
Therefore the thermal fluctuations give rise to a positive contribution to the quadratic term. This raises the height of the potential barrier, and therefore would appear to suppress the decay rate.

At non-zero temperatures the decay process is described by a periodic instanton solution with period $\beta$ in the Euclidean time direction. In the high-temperature limit, the solution becomes independent of the Euclidean time, and has the interpretation of a classical sphaleron configuration. The instanton action is therefore given by 
\begin{equation}
B(T)=E_{\rm sph}(T)/T,
\end{equation}
where $E_{\rm sph}$ is the energy of the sphaleron, which is the three-dimensional saddle point configuration analogous to the Coleman bounce (\ref{eq:O4symFlat}),
and satisfies the equation
\begin{equation}
\ddot\varphi+\frac{2}{r}\dot\varphi-V'(\varphi)=0.
\end{equation}

Using the approximation of constant negative $\lambda$, the action is \cite{Arnold:1991cv}
\begin{equation}
B(T)=\frac{E_{\rm sph}(T)}{T}\approx 18.9  \frac{\gamma}{|\lambda|}.
\end{equation}
Because $\gamma\ll 1$, this is smaller than the zero-temperature action (\ref{eq:LWaction}). 
Therefore the net effect of the non-zero temperature is to increase the vacuum decay rate compared to the zero-temperature case.

More accurately, the sphaleron solutions have been calculated numerically in Ref. \cite{Rose:2015lna,Salvio:2016mvj}. At high temperatures $T\gtrsim 10^{16}~{\rm GeV}$, the action is roughly 
\begin{equation}
B(T\gtrsim 10^{16}~{\rm GeV} )\sim 300.
\label{equ:thermalBnumerical}
\end{equation}
When the temperature decreases, the action increases, so that
$B(10^{14}~{\rm GeV})\sim 400$.

Salvio et al. \cite{Salvio:2016mvj} obtained fully four-dimensional instanton solutions numerically, without assuming independence on the Euclidean time, and found that the three-dimensional sphaleron solutions have always the lowest action and are therefore the dominant solutions. They also showed that including the two-loop corrections to the quadratic term (\ref{equ:thermalmass}) or the one-loop correction to the Higgs kinetic term gives only small correction to the action.

Taking also the prefactor into account, the vacuum decay rate at non-zero temperature is \cite{Espinosa:2007qp,Rose:2015lna}
\begin{equation}
\label{equ:Gammathermal}
\Gamma(T)\approx T^4\left(\frac{B(T)}{2\pi}\right)^{3/2}e^{-B(T)}
.
\end{equation}

\subsection{Vacuum decay in de Sitter space}
\label{sec:CdL}

In extending from flat space to curved space, the theorem~\cite{Coleman:1977th} that guarantees $O(4)$ symmetry of the bounce no longer applies. 
There is some evidence, however, that in background metrics that do respect this symmetry, $O(4)$ symmetric solutions should still dominate \cite{Masoumi:2012yy}. This would include the special case of particular interest in this review - an inflationary, or de Sitter background.\footnote{In principle, inflation is not exact de Sitter, and so the background does not respect exact $O(4)$ symmetry if Euclideanised, but for slow roll inflation models, it is a reasonable approximation to make.} 
A Wick rotated metric can be placed in a co-ordinate system that makes the $O(4)$ symmetry of the bounce immediately manifest,
\begin{equation}
ds^2 =  d\chi^2 + a^2(\chi) d\Omega_3^2,\label{eq:O4metric}
\end{equation}
where $\chi$ is a radial variable, $ d\Omega_3^2$ is the 3-sphere metric, and $a^2(\chi)$ is a scale factor that physically describes the radius of curvature of a surface at constant $\chi$. The bounce equations of motion then take the form \cite{Coleman:1980aw}
\begin{align}
&\ddot{\varphi} + \frac{3\dot{a}}{a}\dot{\varphi} - V'(\varphi) = 0\label{eq:tun:bounceGrav}\\
&\dot{a}^2 = 1-\frac{a^2}{3M_{\rm{P}}^2}\left(-\frac{\dot{\varphi^2}}{2} + V(\varphi)\right).\label{eq:tun:aGrav}
\end{align}
We will consider the case in which the false vacuum has a positive energy density, $V(\varphi_{\rm fv})>0$, and therefore non-zero Hubble rate
\begin{equation}
H^2=\frac{V(\varphi_{\rm fv})}{3M_{\rm P}^2}.
\end{equation}

The boundary conditions the bounce solution must satisfy require special attention: $a(0) = 0$ is required because of the definition of $a(\chi)$ as a radius of curvature of a surface of constant $\chi$, while we require $\dot{\varphi}(0) = \dot{\varphi}(\chi_{\rm{max}}) = 0$, where $\chi_{\rm{max}} > 0$ is defined by $a(\chi_{\rm{max}}) = 0$. These boundary conditions avoid the co-ordinate singularities at $\chi = 0,\chi_{\rm{max}}$ giving infinite results, but allow for the peculiar property that the bounces are compact, and do not approach the false vacuum anywhere.

One way of understanding this peculiar feature was discussed by Brown and Weinberg~\cite{Brown:2007sd}. They considered vacuum decay in de Sitter space, specifically the static patch co-ordinates where the metric takes the form
\begin{equation}
dS_n^2 = -\left(1- H^2r^2\right)dt^2 + (1-H^2r^2)^{-1}dr^2 + r^2 d\Omega_{n-2}^2,\label{eq:staticMetric}
\end{equation}
where $ d\Omega_{n-2}^2$ is the $n-2$-sphere metric (in this case, $n = 4$). The important feature of these co-ordinates is that they are valid only up to the horizon at $r = 1/H$. The Euclidean action can then be re-written as
\begin{align}
S_E =& \int_{-\frac{\pi}{H}}^{\frac{\pi}{H}} d\tau\int d^3x\sqrt{\det h}\left[\frac{1}{2}(1-H^2r^2)^{-\frac{1}{2}}\left(\frac{ d\varphi}{ d\tau}\right)^2 + \frac{1}{2}(1-H^2r^2)^{\frac{1}{2}}h^{ij}\partial_i\varphi\partial_j\varphi\right.\nonumber\\&\left. + (1-H^2r^2)^{\frac{1}{2}}V(\varphi)\right],\label{eq:dSThermalAction}
\end{align}
where $h_{ij}$ is the remaining spatial metric. Brown and Weinberg interpreted this to mean that tunneling takes place on a \emph{compact} Euclidean space, with a curved three-dimensional geometry. This compactness condition is reflected in the boundary conditions $\dot{\varphi}(0) = \dot{\varphi}(\chi_{\rm{max}})$, which inevitably produce a compact bounce solution. They observed that the same effect would be seen in considering a spatially curved universe with this same spatial geometry, but with a non-zero temperature, 
\begin{equation}
\label{equ:GibbonsHawking}
T_{\rm GH} = \frac{H}{2\pi}.
\end{equation}
This corresponds to the Gibbons-Hawking temperature of de Sitter space \cite{Gibbons:1977mu}, and implies that bounces in de Sitter space may have a thermal interpretation.

The simplest solution of Eqs.~(\ref{eq:tun:bounceGrav}) and (\ref{eq:tun:aGrav}) is the Hawking-Moss solution \cite{Hawking:1981fz}. This is a constant solution, for which $\varphi = \varphi_{\rm{bar}}$ sits at the top of the barrier for the entire Euclidean period, and the scale factor is given by
\begin{equation}
a(\chi) = \frac{1}{H_{\rm{HM}}}\sin(H_{\rm{HM}}\chi),
\quad H_{\rm{HM}}^2 = \frac{V(\varphi_{\rm{bar}})}{3M_{\rm{P}}^2}.
\label{eq:HMa}
\end{equation}
Hence $\chi_{\rm{max}} = {\pi}/{H_{\rm{HM}}}$. The action difference of Eq.~(\ref{eq:B}) is then easily computed analytically to be
\begin{equation}
B_{\rm HM}
 = 24\pi^2M_{\rm{P}}^4\left(\frac{1}{V(\varphi_{\rm{fv}})} - \frac{1}{V(\varphi_{\rm{bar}})}\right).\label{eq:HMaction}
\end{equation}
A particularly important limit is that in which $\Delta V(\varphi_{\rm{bar}}) = V(\varphi_{\rm{bar}}) - V(\varphi_{\rm{fv}}) \ll V(\varphi_{\rm{fv}})$. In that case, Eq. (\ref{eq:HMaction}) is approximately
\begin{equation}
B_{\rm HM} = \frac{8\pi^2\Delta V(\varphi_{\rm{bar}})}{3H^4},\label{eq:HMactionFixedBackground}
\end{equation}
where $H^2 = V(\varphi_{\rm{fv}})/{3M_{\rm{P}}^2}$ is the background Hubble rate. 
The prefactor (\ref{equ:prefactor}) in the decay rate can be expected to be at the scale of the Hubble, and therefore the vacuum decay rate due to the Hawking-Moss instanton can be approximated by
\begin{equation}
\Gamma(H)\sim H^4 e^{-B_{\rm HM}(H)}
\label{equ:GammaHM}
\end{equation}

Eq. (\ref{eq:HMactionFixedBackground}) has a simple thermal interpretation: It is the ratio of the energy required to excite an entire Hubble volume, ${4\pi}/{3H^3}$ from the false vacuum to the top of the barrier, divided by the background Gibbons-Hawking temperature (\ref{equ:GibbonsHawking}).
Therefore it can be understood as Boltzmann suppression in classical statistical physics.

\begin{figure}
	\centering
	\includegraphics[width=\textwidth]{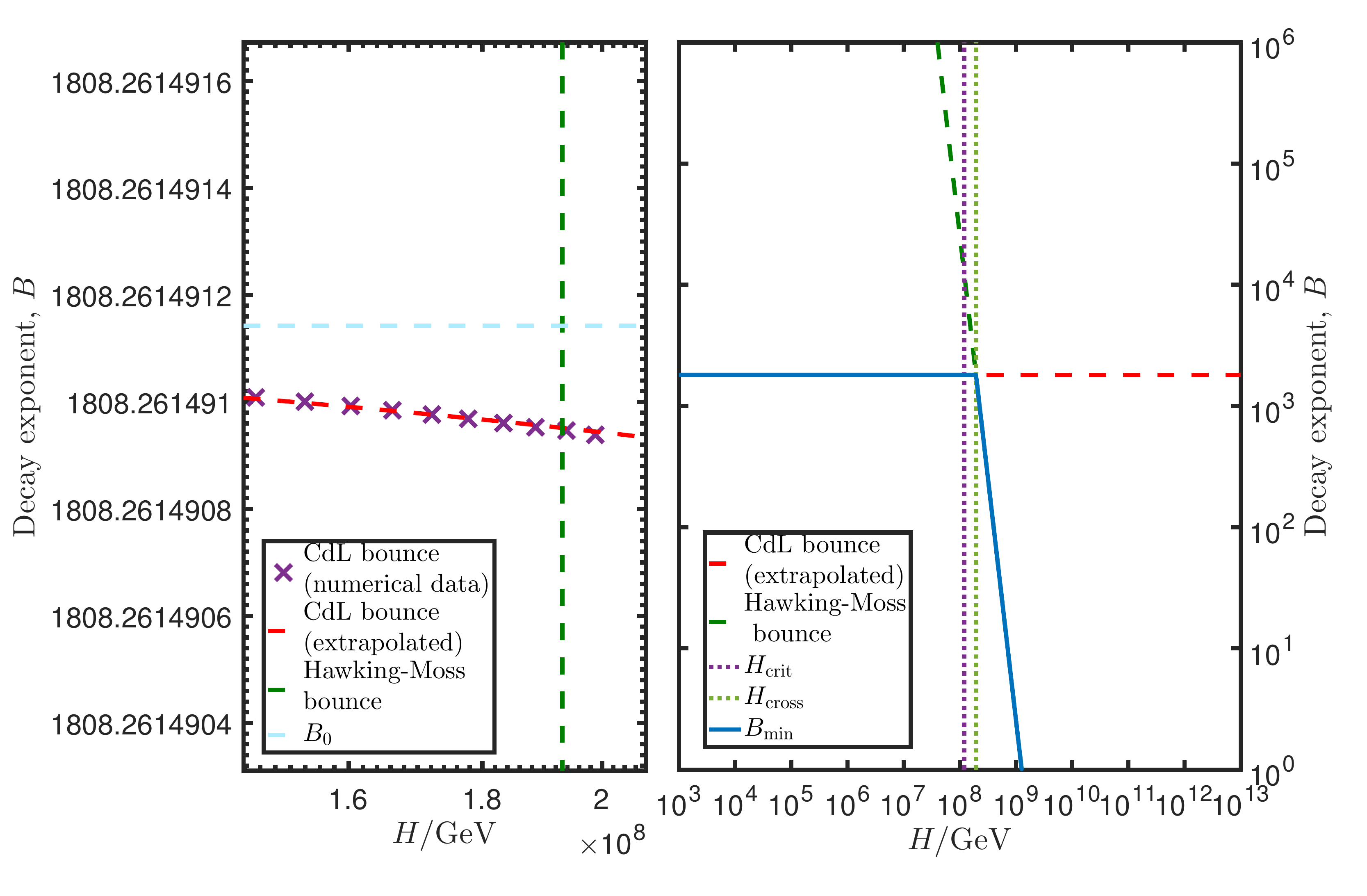}
	\caption{\label{fig:SM_bounce_action}CdL bounce decay exponent plotted against the Hawking-Moss solution in the Standard Model with $M_t = 173.34{\rm{\,GeV}},M_h = 125.15{\rm{\,GeV}},\alpha_S(M_Z) = 0.1184$. The critical values $H_{\rm{crit}} = 1.193\times 10^8{\rm{\,GeV}}$, $H_{\rm{cross}} = 1.931\times 10^{8}{\rm{\,GeV}}$ are also plotted, along with $B_0$, the bounce action obtained at $H = 0$.}
\end{figure}

The bounce equations (\ref{eq:tun:bounceGrav}) and (\ref{eq:tun:aGrav}) also often have Coleman-de Luccia (CdL) instantons, in which the field increases monotonically from $\varphi(0)<\varphi_{\rm bar}$ to $\varphi(\chi_{\rm min})>\varphi_{\rm bar}$.
For low false vacuum Hubble rates, $H\ll \mu_{\rm min}$,
a CdL solution can be found as a perturbative correction to Eq.~(\ref{eq:LWbounce}), with the action~\cite{Shkerin:2015exa}
\begin{equation}
B_{\rm CdL}\approx \frac{8\pi^2}{3|\lambda(\mu_{\rm min})|}\left[
1+36\left(\xi-\frac{1}{6}\right)\frac{H^2}{\mu_{\rm min}^2}\ln\frac{\mu_{\rm min}}{H},
\right].
\end{equation}

Numerical HM and CdL bounce solutions in the Standard Model were found in Ref.~\cite{Rajantie:2017ajw} and the corresponding actions are shown in Fig.~\ref{fig:SM_bounce_action}, for the parameters $M_h = 125.15{\rm{\,GeV}}$, $M_t = 173.34{\rm{\,GeV}}$, $\alpha_S = 0.1184$. 
We can see that at low Hubble rates, the CdL solution has a lower action than the HM solution.
For example, for the case of background Hubble rate $H = 1.1937\times 10^8{\rm{\,GeV}}$, the numerical result is $B_{\rm CdL} = 1805.8$ in a fixed de Sitter background metric, and $B_{\rm CdL} = 1808.26$ including gravitational back-reaction. 
The CdL action is also almost independent of the Hubble rate.

On the other hand, the Hawking-Moss action (\ref{eq:HMaction}) decreases rapidly as the Hubble rate increases.
It
crosses below $B_{\rm{CdL}}$ at Hubble rate \cite{Rajantie:2017ajw}
\begin{equation}
\label{equ:Hcross}
H_{\rm{cross}} = 1.931\times 10^{8}{\rm{\,GeV}}.
\end{equation} 
At Hubble rates below this, $H>H_{\rm cross}$ vacuum decay is dominated by the Coleman-de Luccia instanton, which describes quantum tunneling through the potential barriers, whereas
above this, $H>H_{\rm cross}$, the dominant process is the Hawking-Moss instanton.
This is discussed further in Section \ref{sec:HMtransition}.

In addition to the HM and CdL solutions, one may also find oscillating solutions~\cite{Weinberg:2005af,Hackworth:2004xb,Lee:2017byw,Lee:2014ula}, which cross the top of the barrier $\varphi_{\rm bar}$ multiple times between $\chi=0$ and $\chi=\chi_{\rm max}$, and additional CdL-like solutions with higher action~\cite{Hackworth:2004xb,Rajantie:2017ajw}. The latter were found numerically in the Standard Model in Ref.~\cite{Rajantie:2017ajw}. Because these solution have a higher action than the HM and CdL solutions, they are highly subdominant as vacuum decay channels.
Oscillating solutions also have more than one negative eigenvalues~\cite{Lavrelashvili:2006cv,Dunne:2006bt}.

\subsection{Negative eigenvalues\label{sec:bounce_eigenvalues}}
In order for a stationary point of the action to describe vacuum decay, it has to have precisely one negative 
eigenvalue.
The reason is that the decay rate of a metastable vacuum is determined by the imaginary part of the energy as computed by the effective action~\cite{Callan:1977pt}, and thus only solutions that contribute an imaginary part to the vacuum energy will contribute to metastability.

This requirement comes in via the functional determinant which encodes the quantum corrections to the bounce solution. This functional determinant is given by a product over the eigenvalues for fluctuations around the relevant bounce solution. In flat space, these all satisfy \cite{Callan:1977pt}
\begin{equation}
-\nabla_{\mu}\nabla^{\mu}\delta\varphi + V''(\varphi_B)\delta\varphi = \lambda\delta\varphi,\label{eq:fluct}
\end{equation}
where $\varphi_B$ is the solution expanded around. The $O(4)$ symmetric bounce solutions in flat space can be shown to have at least one negative eigenvalue, since they possess zero modes corresponding to translations of the bounce around the space-time. In fact, there must only be one such eigenvalue. Solutions with more negative eigenvalues do not contribute to tunneling rates, because while they are stationary points of the Euclidean action, they are not minima of the barrier penetration integral (\ref{eq:TWKB}) obtained from the WKB approximation~\cite{Coleman:1985rnk}.

The situation is somewhat different in the gravitational case, however, due to the fact that in addition to the scalar field, we can also consider metric fluctuations about a bounce solution. A quadratic action for fluctuations about a bounce in curved space was first derived by \cite{Lavrelashvili1985} and has been considered by several authors \cite{Lee:2014uza,Lavrelashvili:2006cv,Koehn:2015hga}. This takes the gauge invariant form
\begin{equation}
\mathcal{L}^{(2)}(\zeta_l,\dot{\zeta}_l) = \frac{a^3(\chi)\left(1-\frac{1}{3}l(l+2)\right)}{2\left(Q - \frac{1}{3}\dot{a}^2l(l+2)\right)}\left[\dot{\zeta}_l^2(\chi) + f(a,\phi)\zeta_l^2(\chi) \right],\label{eq:quadratic_action}
\end{equation}
where
\begin{equation}
Q = 1 - \frac{a^2(\chi)V(\varphi)}{3M_{\rm{P}}^2}\label{eq:Q},
\end{equation}
and $f$ is a complicated function of $a$ and $\varphi$ which can be found in Refs.~\cite{Lee:2014uza,Lavrelashvili:2006cv,Koehn:2015hga}. The analysis of this Lagrangian is complicated, but some conclusions can be drawn. To begin with, it is possible to argue that expanded around a CdL bounce solution, this action \emph{always} has an infinite number of negative eigenvalues. This is the so called `negative mode problem' \cite{Lee:2014uza,Lavrelashvili:2006cv,Koehn:2015hga}. The argument, as expressed in Ref.~\cite{Lee:2014uza}, is that we can re-write $Q$ using Eq. (\ref{eq:tun:aGrav}) as
\begin{equation}
Q = \dot{a}^2 - \frac{a^2\dot{\varphi}^2}{3M_{\rm{P}}^2}.\label{eq:Qrewritten}
\end{equation}
Note that the bounce always has a point satisfying $\dot{a} = 0$, which is the largest value obtained by $a(\chi)$. Consequently, there is always a region where $Q$ is negative, so for the $l = 0$ modes it is possible to construct a negative kinetic term in Eq. (\ref{eq:quadratic_action}). This means that sufficiently rapidly varying fluctuations will have their action unbounded below, so there is an infinite tower of high frequency, rapidly oscillating fluctuations that all have negative eigenvalues. Note that for $l = 1$ the quadratic Lagrangian is zero (these are the zero-modes associated to translations of the bounce), and for $l > 1$, both numerator and denominator in Eq. (\ref{eq:quadratic_action}) are negative, thus the kinetic terms are always positive. Since $Q = 1$ in flat space (obtained by taking the $M_{\rm{P}}\rightarrow \infty$ limit), it is clear that these `rapidly oscillating' modes are somehow associated to the gravitational sector.

At first, this seems concerning, however, it was pointed out in Ref.~\cite{Lee:2014uza} that these high frequency oscillations are inherently associated with quantum gravity contributions, and thus may not affect tunneling. If we focus on the `slowly varying' modes, the structure of these is much more similar to the analogous flat space bounces. The conclusion we should draw then, is that a solution is relevant only if there is a single \emph{slowly varying} negative eigenvalue.

\subsection{Hawking-Moss/Coleman-de Luccia transition\label{sec:HMtransition}}
As discussed in Section~\ref{sec:CdL},  there are two types of solutions that contribute to vacuum decay in de Sitter space. The first is the Hawking-Moss solution (\ref{eq:HMa}), and the second is the Coleman-de Luccia solution, which crosses the barrier once. By considering the negative eigenvalues of the HM solution, one gains insight into which solutions exist and contribute to vacuum decay at a given Hubble rate.

The eigenvalues of the Hawking-Moss solution are~\cite{Lee:2014uza}
\begin{equation}
\lambda_N = \frac{V''(\varphi_{\rm{bar}})}{H_{\rm{HM}}^2} + N(N+3),\label{eq:HMeigen}
\end{equation}
and their degeneracy is \cite{Rubin:1983be}
\begin{equation}
D_N(4,0) = \frac{(N+1)(N+2)(2N+3)}{6}. \label{eq:HMdegen}
\end{equation}
Because $V''(\varphi_{\rm{bar}}) < 0$,
the $N = 0$ mode is self evidently negative, and has degeneracy 1.
Higher modes will all be positive if and only if
\begin{equation}
\lambda_1 = \frac{V''(\varphi_{\rm{bar}})}{H_{\rm{HM}}^2} + 4 > 0.\label{eq:lambda1}
\end{equation}
This imposes a lower bound on $H_{\rm{HM}}$, below which the Hawking-Moss solution has multiple negative eigenvalues. Hence, it cannot contribute to vacuum decay for Hubble rates below the critical threshold~\cite{Brown:2007sd,Coleman:1985rnk}. An alternative way of expressing this is in terms of a critical Hubble rate. If we define $H^2 = {V(\varphi_{\rm{fv}})}/{3M_{\rm{P}}^2}$ to be the background Hubble rate in the false vacuum, then the condition for Hawking-Moss solutions to contribute to vacuum decay is $H > H_{\rm{crit}}$ where
\begin{equation}
H_{\rm{crit}}^2  = -\frac{V''(\varphi_{\rm{bar}})}{4} - \frac{\Delta V(\varphi_{\rm{bar}})}{3M_{\rm{P}}^2}.\label{eq:Hcrit}
\end{equation}
Here, $\Delta V(\varphi)\equiv V(\varphi) - V(\varphi_{\rm{fv}})$. However, the second term generally only contributes significantly if the difference in height between the top of the barrier and the false vacuum is comparable to the Planck Mass. For most potentials, only the second derivative at the top of the barrier matters.

At low Hubble rates, $H < H_{\rm{crit}}$, the Hawking-Moss solution does not contribute to vacuum decay,
but on the other hand, 
a CdL solution is \emph{guaranteed} to exist~\cite{Balek:2003uu}. 
In most potentials, the CdL solution smoothly merges into the Hawking-Moss solution as the Hubble rate approached $H_{\rm crit}$ from below, and and the Hawking-Moss solution becomes relevant \cite{Balek:2004sd,Hackworth:2004xb}.
Close to the critical Hubble rate,
$H \sim H_{\rm{crit}}$, one can define the quantity~\cite{Tanaka:1992zw,Balek:2004sd,Joti:2017fwe}
\begin{equation}
\Delta \equiv -\frac{1}{14}\left[V^{(4)}(\varphi_{\rm{bar}}) - \frac{(V^{(3)}(\varphi_{\rm{HM}}))^2}{3V^{(2)}(\varphi_{\rm{bar}})} - \frac{8V^{(2)}(\varphi_{\rm{HM}})}{3M_{\rm{P}}^2}\right],\label{eq:Delta}
\end{equation}
which divides potentials into two classes \cite{Balek:2004sd,Rajantie:2017ajw}. Those with $\Delta < 0$ are `typical' potentials, for which the perturbative solution only exists for $H < H_{\rm{crit}}$ \cite{Balek:2004sd}, while those with $\Delta > 0$ only have perturbative solutions for $H > H_{\rm{crit}}$. 
When a perturbative solution exists, its action is given by \cite{Balek:2004sd}
\begin{equation}
B_{\rm CdL} = B_{\rm{HM}} + \frac{2\pi^2(\varphi_0 - \varphi_{\rm{HM}})^4\Delta}{15H_{\rm{HM}}^4},\label{eq:pertAction}
\end{equation}
where $\varphi_0$ is the true vacuum side value of the bounce (which approaches $\varphi_{\rm{HM}}$ in the $H\rightarrow H_{\rm{crit}}$ limit) and $\varphi_{\rm{bar}}$ is the top of the barrier. 

Hence one can see that if $\Delta<0$, a CdL solution with lower action, $B_{\rm CdL}<B_{\rm HM}$, exists for $H<H_{\rm crit}$, and approaches the Hawking-Moss solution smoothly as $H\rightarrow H_{\rm{crit}}$, until it vanishes at $H_{\rm{crit}}$. At the same point, the second eigenvalue of the HM solution turns positive, and therefore the HM solution starts to contribute to vacuum decay.

On the other hand, if $\Delta>0$, 
which is the case for the Standard Model Higgs potential~\cite{Rajantie:2017ajw},
the perturbative CdL solution exists only for $H>H_{\rm crit}$.
Below $H_{\rm crit}$, the HM solution has two negative eigenvalues, which means that it does not contribute to vacuum decay. Instead, the relevant solution is the CdL solution, which also has a lower action (see Fig.~\ref{fig:SM_bounce_action}). When the Hubble rate is increased, a second, perturbative CdL solution appears smoothly at $H=H{\rm crit}$, at the same as the second eigenvalue of the HM solution becomes positive. At $H>H_{\rm crit}$ there are, therefore, at least two distinct CdL solutions, and in fact, numerical calculations indicate that there are at least four~\cite{Rajantie:2017ajw}.
For the parameters used in Fig.~\ref{fig:SM_bounce_action}, the critical Hubble rate is $H_{\rm{crit}} = 1.193\times 10^{8}{\rm{\,GeV}}$.

\subsection{Evolution of bubbles after nucleation}
\label{sec:BubbleEvolution}

The bounce solution $\varphi_B$ determines the field configuration to which the vacuum state tunnels \cite{Callan:1977pt,Brown:2007sd}, and therefore sets the initial conditions for its later evolution.
It is the equivalent of the second turning point on the true vacuum side, $x_2$, appearing in Eq. (\ref{eq:TWKB}). In ordinary quantum mechanics, a particle with energy $E$ emerges on the true vacuum side of the barrier at $x_2(E)$ after tunnelling. This is related to the bounce solution, which starts at $x_1$, rolls until reaching $x_2$, and then bounces back to $x_1$, thus $x_2$ represents a slice of the bounce solution half way through.

In complete analogy, the field emerges at a configuration corresponding to a slice half way through the bounce solution (in Euclidean time). In flat space tunnelling, the bounce is $\varphi_B(\chi)$ where $\chi^2 = \tau^2 + r^2$, and thus touches the false vacuum at $\tau\rightarrow \pm \infty$. Hence the mid-way points occurs at $\tau = 0$ and the solution emerges with $\phi(x,0) = \varphi_B(r)$. One can then use this as an initial condition at $t = 0$ for the Lorentzian field equations,
\begin{equation}
\nabla_{\mu}\nabla^{\mu}\varphi + V'(\varphi) = 0\label{eq:phiLorenzt}.
\end{equation}

However, this is not really necessary, as the $O(4)$ symmetry of the bounce solution carries over into $O(3,1)$ solution \cite{Callan:1977pt}, and thus the solution can be read off as 
\begin{equation}
\varphi(x,t) = \varphi_B(\sqrt{r^2 - t^2})\quad\mbox{for}\quad r>t.
\end{equation}
From this one can see that the bubble wall is moving outwards asymptotically at the speed of light.
The inside of the light cone corresponds to an anti-de Sitter spacetime collapsing into a singularity~\cite{Espinosa:2007qp,Espinosa:2015qea,Burda:2016mou,East:2016anr}.

The situation in de Sitter space is considerably more complicated, 
but the conclusion is the same~\cite{Brown:2007sd}. First, 
de Sitter bounces can be thought of as bounces at finite temperature on a curved \emph{spatial} background described by constant time slices of the static patch of de Sitter space,
\begin{equation}
ds^2 = -(1-H^2r^2)dt^2 + (1-H^2r^2)^{-1}dr^2 + r^2 d\Omega_2^2.\label{eq:StaticMetric}
\end{equation}
The temperature in this case is the Gibbons-Hawking temperature (\ref{equ:GibbonsHawking}) of de Sitter space.
Bounces at finite temperature $\beta = 1/k_BT$ correspond to periodic bounces in Euclidean space \cite{Brown:2007sd}, with period $\tau_{\rm{period}} = \beta$. In this case, the bounce starts at the false vacuum at $\tau = -\pi/H$, hits its mid-point at $\tau = 0$, and returns to the false vacuum side at $\tau = \pi/H$. Thus, the $\tau = 0$ hypersurface describes the final state of the field after tunnelling. 

Analytic continuation of the metric back to real space can be performed using the approach of \cite{Burda:2016mou}. The $O(4)$ symmetric Euclidean metric is of the form
\begin{equation}
ds^2 =  d\chi^2 + a^2(\chi)[ d\psi^2 + \sin^2\psi d\Omega_2^2],\label{eq:psiChiCoords}
\end{equation}
where in the de Sitter case,
\begin{equation}
a(\chi) = \frac{1}{H}\sin(H\chi).\label{eq:fixedBacka}
\end{equation}
Since it is straightforward to analytically continue the flat space metric back to real space via the transformation $\tau = it$, then the same thing can be done with any conformally flat metric, by changing variables to $\tilde{\tau},\tilde{r}$ such that
\begin{equation}
ds^2 = \frac{a^2(\chi)}{f^2(\chi)}[ d\tilde{\tau}^2 +  d\tilde{r}^2 + \tilde{r}^2 d\Omega_2^2],\label{eq:conformallyFlatForm}
\end{equation}
which is achieved by choosing $f(\chi)$ such that $f'(\chi) = f/a, f(0) = 0$. In the de Sitter case, this means
\begin{equation}
f(\chi) = C\frac{\sin(H\chi)}{1+\cos(H\chi)} = C\tan(H\chi/2),\label{eq:fdS}
\end{equation}
where $C$ is an arbitrary constant - we can choose it to be $1$. This co-ordinate system is obtained from the $O(4)$ symmetric co-ordinates via
\begin{align}
\tilde{\tau} &= f(\chi)\cos(\psi),\label{eq:tauCon}\\
\tilde{r} &= f(\chi)\sin(\psi).\label{eq:rCon}
\end{align}
One then transforms back to real space exactly as in flat space, via $\tilde{\tau} = it$. The co-ordinate $\chi$ is then related to $\tilde{t}$ and $\tilde{r}$ via
\begin{equation}
\chi = f^{-1}(\sqrt{\tilde{r}^2 - \tilde{t}^2}).\label{eq:Chirt}
\end{equation}
It should be noted that $\tilde{t},\tilde{r}$ as defined only cover the $\tilde{r} > \tilde{t}$ portion of de Sitter space. Because the metric is manifestly conformally flat in these co-ordinates, we can see that this corresponds to the portion of de Sitter space \emph{outside} the light-cone, which lies at $\tilde{r}=\pm\tilde{t}$.

Doing this for de Sitter yields the real space metric
\begin{equation}
ds^2 = \frac{4}{H^2[1 + \tilde{r}^2 - \tilde{t}^2]^2}[- d\tilde{t}^2 + d\tilde{r}^2 + \tilde{r}^2 d\Omega_2^2],\label{eq:dSConformallyFlat}
\end{equation}
which at first glance, is not obviously de Sitter space. However, the transformation
\begin{align}
t &= \frac{1}{2H}\log\left|\frac{1 - \tilde{r}^2 + 2\tilde{t} + \tilde{t}^2}{1 - \tilde{r}^2 - 2\tilde{t} + \tilde{t}^2}\right|,\label{eq:ConToStatict}\\
r &= \frac{2\tilde{r}}{H(1 + \tilde{r}^2 - \tilde{t}^2)},\label{eq:ConToStaticr}
\end{align}
can be readily shown to yield Eq. (\ref{eq:StaticMetric}), thus this is indeed a valid analytic continuation of the Euclidean 4-sphere back to de Sitter space.

To describe the subsequent evolution of the bubble, it is argued in Ref.~\cite{Burda:2016mou} that $\phi(r,t) = \phi_B(\chi(r,t))$ matches the symmetry of the $O(4)$ symmetric bounce, just as in flat space, with $\chi(r,t)$ defined by Eq. (\ref{eq:Chirt}). As mentioned before, this describes only the evolution of the scalar field outside the light-cone. For $\tilde{r} < \tilde{t}$, it is necessary to solve the Euclidean equations directly. That calculation demonstrates explicitly that the formation of a singularity in the negative-potential region is inevitable~\cite{Burda:2016mou}, confirming previous calculations using the thin wall approximation in Ref.~\cite{Coleman:1980aw}. 

As for the evolution outside the light-cone, it can be seen that, much as in flat space, a point of constant field value $\varphi_0$ corresponding to $\chi_0$ where $\varphi_0 = \varphi(\chi_0)$, satisfies
\begin{equation}
\tilde{r}(\tilde{t}) = \sqrt{\tilde{t}^2 + f^2(\chi_0(\phi_0))},\label{eq:BubblePosition}
\end{equation}
which means that it rapidly approaches the speed of light as $\tilde{t}\rightarrow\infty$. 
Thus, just as in flat space, the bubble expands outwards at the speed of light.

Even if the bubble wall moves outward at the speed of light, it does not necessarily grow to fill the whole Universe, if it is trapped behind an event horizon.
Scenarios in which bubbles of true vacuum form primordial black holes have been discussed~\cite{Hook:2014uia,Kearney:2015vba,Espinosa:2017sgp,Espinosa:2018euj}. This can happen if inflation ends before the space inside the bubble hits the singularity. When the Universe reheats, thermal corrections (\ref{equ:thermalpot}) stabilise the Higgs potential, preventing the collapse. The bubble then collapses into a black hole, and the primordial black holes produced in this way could potentially constitute part or all of the dark matter in the Universe~\cite{Espinosa:2017sgp}. This scenario requires fine tuning to avoid the singularity or
new heavy degrees of freedom that modify the potential at high field values~\cite{Espinosa:2018euj}. The same scenario can also produce potentially observable gravitational waves~\cite{Espinosa:2018eve}.

\makeatletter{}\section{Cosmological Constraints}
\def\Nbub{\langle {\cal N}\rangle}
\label{sec:Cosmology}
\subsection{Cosmological history}
\label{sec:CosmoHistory}
For the Universe to be currently in a metastable state rather than in its true ground state, it is not enough that the decay rate is slow today. The Universe also had to somehow end up in the metastable electroweak-scale state, and the decay rate had to be sufficiently slow in the past for the Universe to stay there through the whole history of the Universe. 
The former requirement depends on the initial conditions of the Universe, 
which are often assumed to involve Planck-scale field values, and therefore
one needs to explain how the Higgs could have relaxed into the electroweak-scale vacuum without getting trapped into the negative-energy true vacuum.
The latter condition, the survival of the current metastable state through the history of the Universe, 
requires that no bubbles of true vacuum
were nucleated in our past light cone~\cite{Espinosa:2007qp}.
This is because, once nucleated, a bubble of true vacuum expands at the speed of light and destroys everything in its way. If even a single bubble had nucleated at any time, anywhere in our past lightcone, it would have already hit us.

To describe the history of the Universe, we approximate it with the FLRW metric~(\ref{eq:le}).
The scale factor $a(t)$ satisfies the Friedmann equation (\ref{eq:e})
\begin{equation}
H^2\equiv\frac{\dot{a}^2}{a^2}=\frac{\rho}{3M_{\rm P}^2},
\label{equ:Friedmann}
\end{equation}
where $\rho$ is the energy density of the Universe.
When the dominating energy forms can be described by ideal fluids, one can write an equation of state
$p=w\rho$, which relates the pressure $p$ to the energy density $\rho$ through the equation of state parameter $w$. From the first law of thermodynamics it then follows that the energy density scales with the expansion of the Universe as
\begin{equation}
\rho\propto a^{-3(1+w)}.
\label{equ:rhoscaling}
\end{equation}

Observations indicate the the Universe currently contains three forms of energy: radiation ($w=1/3$), matter ($w=0$) and dark energy, which we assume to be a cosmological constant with $w=-1$.
The total energy density can be therefore written as a function of the scale factor as
\begin{equation}
\rho(a)=\rho_{\rm tot}^0\left(\Omega_\Lambda+\Omega_{\rm mat}\left(\frac{a_0}{a}\right)^3
+\Omega_{\rm rad}\left(\frac{a_0}{a}\right)^4\right),
\label{equ:rhoOmega}
\end{equation}
where $\Omega_\Lambda=0.69$, $\Omega_{\rm mat}=0.31$ and $\Omega_{\rm rad}=5.4\times 10^{-5}$
are the observed energy fractions of cosmological constant, matter and radiation, respectively~\cite{Tanabashi:2018oca}, $\rho_{\rm tot}^0$ is the current total energy density, and $a_0$ is the current value of the scale factor.
It is common to choose $a_0=1$ but we include it explicitly for clarity.
The Universe is therefore currently dominated by dark energy, but in past it was dominated by matter and, at even earlier times, by radiation.
Observations also show that in its very early stages, before radiation-dominated epoch, the Universe went through a period of accelerating expansion known as inflation, during which the equation of state was, again, $w\approx -1$.

To find the expected number of bubbles in the past lightcone, it is convenient to write the FLRW metric in terms of the conformal time $\eta$ as in Eq.~(\ref{eq:conft}).
In these coordinates, light satisfies $|d\vec{r}/d\eta|=1$, so if we denote the current conformal time by $\eta_0$, the comoving radius of our past light cone at conformal time $\eta$ is $r(\eta)=\eta_0-\eta$.

The dependence of the scale factor on the conformal time is determined by the Friedmann equation (\ref{eq:e}), which in terms of the conformal time is
\begin{equation}
\left(\frac{d a}{d\eta}\right)^2=\frac{\rho a^4}{3 M_{\rm P}^2}. 
\label{equ:conformalFriedmann}
\end{equation}
Using Eq.~(\ref{equ:rhoOmega})
one finds that the conformal time since the end of inflation is
\begin{equation}
\label{equ:currenteta}
\eta_0-\eta_{\rm inf}=\frac{1}{H_0}\int_0^{a_0}\frac{da}{\sqrt{\Omega_\Lambda a^4+\Omega_m a_0^3a+\Omega_r a_0^4}}
\approx 3.21(a_0H_0)^{-1}.
\end{equation}

The bubble nucleation rate $\Gamma$ may have been very different in different stages of the early evolution of the Universe. It depends on the curvature of spacetime and temperature, and also potentially on any perturbations or non-equilibrium processes that could catalyse or trigger the decay process and therefore it is function of the scale factor, $\Gamma=\Gamma(a)$.
This allows us to write an expression for the expected number of bubbles $\Nbub$ 
in our past lightcone (after some initial time $\eta_{\rm ini}$) as
\begin{eqnarray}
\Nbub &=& 
\int_{\rm past} d^4x \sqrt{-g} \Gamma(x)=
\int_{\eta_{\rm ini}}^{\eta_0} d\eta\, a(\eta)^4 \frac{4\pi r(\eta)^3}{3} \Gamma(a(\eta))
\nonumber\\
&=&\frac{4\pi}{3}\int_{\eta_{\rm ini}}^{\eta_0} d\eta a(\eta)^4 (\eta_0-\eta)^3 \Gamma(a(\eta))
=\frac{4\pi}{3}\int_0^{a_0}da\left(\eta_0-\eta\left(a\right)\right)^3\frac{a^2}{H(a)}\Gamma(a).
\label{equ:expectedbubbles}
\end{eqnarray}
If this number is much greater than one, it would be unlikely that our part of the Universe could have survived until today, and therefore our existence requires
\begin{equation}
\Nbub \lesssim 1.
\label{equ:expNbound}
\end{equation}

\subsection{Late Universe}
\label{equ:lateuni}

Let us first consider the post-inflationary Universe described by the energy density (\ref{equ:rhoOmega}) and assume that the bubble nucleation rate $\Gamma(a)$ in the past was at least as high as its current Minkowski space value $\Gamma_0$, i.e., $\Gamma(a)\ge \Gamma_0$.
In this case the expected number of bubbles is 
\begin{equation}
\Nbub_{\rm post}\ge
\Gamma_0{\cal V}_{\rm post}=\frac{4\pi}{3}\Gamma_0\int_{\eta_{\rm inf}}^{\eta_0}d\eta
(\eta_0-\eta)^3 a(\eta)^4
\approx 0.125 \Gamma_0H_0^{-4}.
\label{equ:Vpostinflation}
\end{equation}
Hence, the constraint on the nucleation rate $\Gamma_0$ from the post-inflationary era is
\begin{equation}
\Gamma_0\lesssim 8.0\,H_0^4.
\label{equ:Gamma0bound}
\end{equation}
Using Eq.~(\ref{equ:Gamma0}) and $H_0\approx70~{\rm km/s/Mpc}$, this translates to a bound
\begin{equation}
B\gtrsim 540
\end{equation}
on the bounce action.

By calculating the nucleation rate $\Gamma_0$, theories can be divided into  categories: \textit{stable}, \textit{metastable} and \textit{unstable}. 
If the rate exceeds the bound (\ref{equ:Gamma0bound}), the Universe would not have survived until the present day, and hence the vacuum is said to be unstable.
If the rate is non-zero but satisfies Eq.~(\ref{equ:Gamma0bound}), the vacuum would not have decayed by the present time but would decay in the future, and hence it is said to be metastable. Finally, if the decay rate is strictly zero, which is the case when the current vacuum state is the global minimum of the potential, then the vacuum is said to be stable.

\begin{figure}[ht!]
	\begin{center}
				\includegraphics[width=\textwidth]{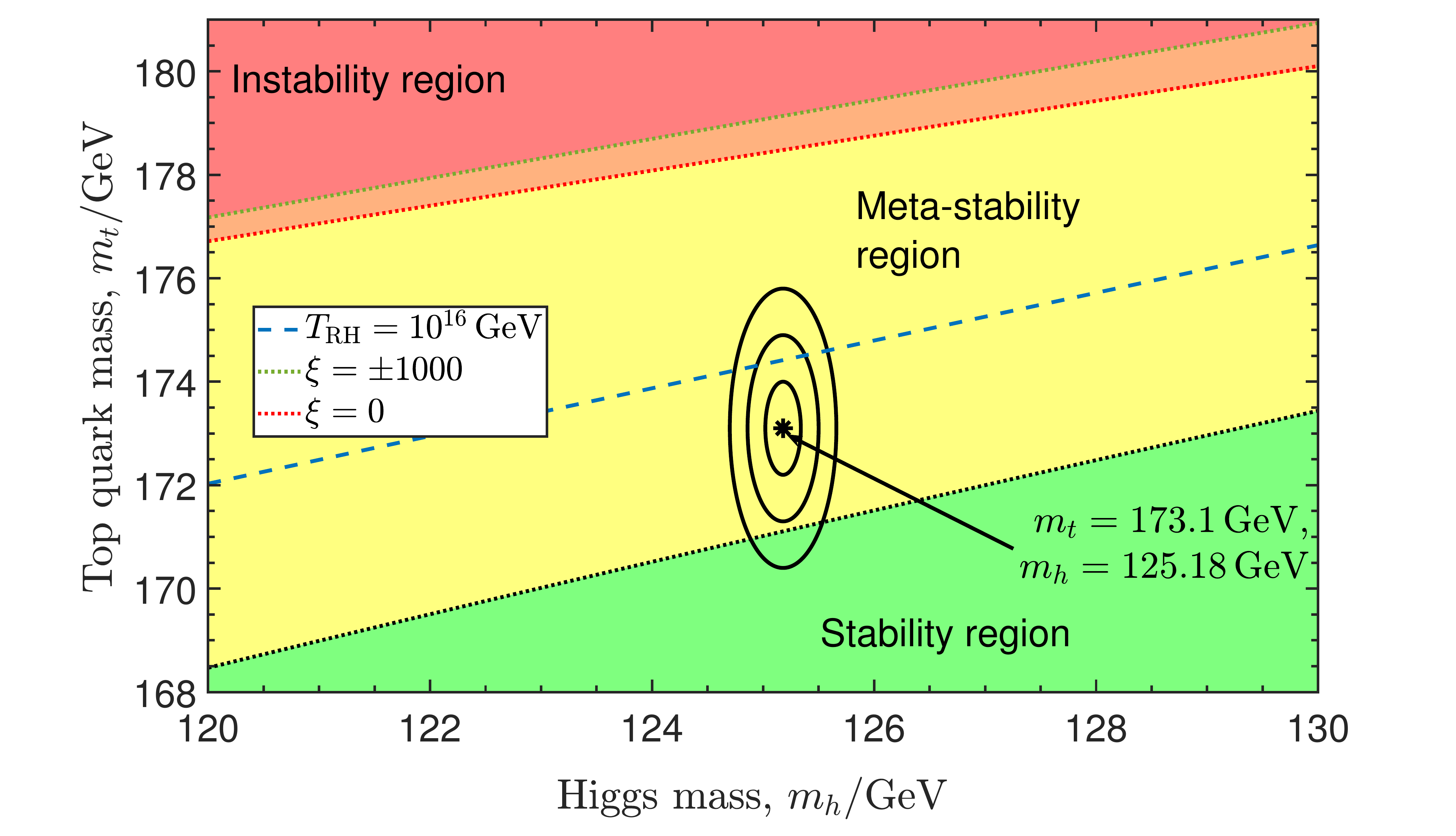}
	\end{center}
		\caption{\label{fig:dec} Stability diagram of the Standard Model vacuum state in the pole masses $M_t$, $M_h$ of the top quark and Higgs boson respectively. Ellipses show the $1\sigma,2\sigma,3\sigma$ confidence intervals for $M_t$ and $M_h$ around their central values from Ref.~\cite{Tanabashi:2018oca}. 
In the green region, the current vacuum is absolutely stable, in the yellow region it satisfies the bound (\ref{equ:Gamma0bound}), and in the red region it is so unstable that it would not have survived until the present day.
The instability boundary includes gravitational backreaction~\cite{Rajantie:2016hkj} and is shown for $\xi=0$ and $\xi=\pm 1000$ of the non-minimal curvature coupling. The blue dashed line shows the instability bound (\ref{equ:thermalbound}) obtained by taking the thermal history of the Universe into account~\cite{Rose:2015lna} and assuming a high reheat temperature $T_{\rm RH}=10^{16}~{\rm GeV}$.
For lower reheat temperatures, the instability bound becomes weaker, and approaches the red dotted line as $T_{\rm RH}\rightarrow 0$.
    }
\end{figure}

Fig.~\ref{fig:dec} shows the stability diagram of the Standard Model based on Ref.~\cite{Rajantie:2016hkj} (see Section~\ref{sec:tun:flat} for discussion), in terms of the Higgs mass $M_h$, top mass $M_t$, for three different values of the non-minimal coupling $\xi$. The ellipses show the 68\%, 95\%, and 99\% contours based on the experimental and theoretical uncertainties in the masses. 

It is worth mentioning that one could invoke the anthropic principle to evade the bound~(\ref{equ:Gamma0bound}). 
Even if the expected number of bubbles $\Nbub$ is large, there is always a non-zero probability that no bubbles were nucleated.
Life can obviously only exist in those parts of the Universe that have no bubble nucleation event in their past light cone, and therefore that is necessarily what we observe, 
no matter how low the probability is a priori. One can therefore argue that observations do not require $\Nbub\lesssim 1$.
However, the anthropic argument does not rule out bubbles hitting us in the future, and therefore, if the Universe survives for a further period of time, that imposes a bound that is not subject to the anthropic principle. For this, the quantity that matters is the time derivative of the expected number of bubbles,
\begin{equation}
\frac{d\Nbub}{dt}=\frac{4\pi}{a_0}\Gamma_0\int_{\eta_{\rm ini}}^{\eta_0} d\eta\, a(\eta)^4(\eta_0-\eta)^2.
\end{equation}
This imposes constraints that are numerically weaker but cannot be avoided by anthropic reasoning. To be concrete, one can carry out an experiment by waiting for a period of time $t_{\rm exp}$, for example one year. If, at the end of the time period, the experimenter has not been hit by a bubble wall, this gives a constraint
\begin{equation}
t_{\rm exp}\frac{d\Nbub}{dt}\lesssim 1.
\end{equation}
For the post-inflationary Universe this is
\begin{equation}
t_{\rm exp}\frac{d\Nbub}{dt}=(t_{\rm exp}H_0)\times 4.91\Gamma_0H_0^{-4},
\end{equation}
and for $t_{\rm exp}=1{\rm yr}$, one obtains the bound
\begin{equation}
\Gamma_0\lesssim 2.9\times 10^{10}\, H_0^4,\quad\mbox{or}\quad B\gtrsim 520.
\end{equation} 
This is weaker than Eq.~(\ref{equ:Gamma0bound}), but because of the very strong dependence of $\Gamma_0$ on the top and Higgs masses, it does not change the stability constraints on them significantly.

\subsection{Inflation}
\label{sec:Inflation}

Although most of the spacetime volume of our past lightcone comes from the late times, the vacuum decay rate $\Gamma(a)$ was much higher in the very early Universe. Depending on the cosmological scenario, it can be high enough to violate the bound (\ref{equ:expNbound}), and this can be used to constrain theories.

The earliest stage in the evolution of the Universe that we have evidence for is inflation, a period of accelerating expansion, which made the Universe spatially flat, homogeneous and isotropic and also generated the initial seeds for structure formation. 
In simplest models of inflation, the energy density driving it is in the form of the potential energy $V(\phi)$ of a scalar field $\phi$ known as the inflaton.
The inflaton field is nearly homogeneous, and satisfies the equation of motion
\begin{equation}
\ddot{\phi}+3H\dot{\phi}+V'(\phi)=0.
\label{equ:inflatoneom}
\end{equation}
During inflation the potential satisfies the slow-roll conditions,
\begin{equation}
\epsilon\equiv\frac{\MP ^2}{2}\left(\frac{V'}{V}\right)^2\ll 1,
\quad\mbox{and}\quad 
-1\ll 
\eta\equiv \MP ^2\left(\frac{V''}{V}\right)\ll 1.
\label{equ:slowroll}
\end{equation}
These conditions guarantee the existence of a solution in which 
the first term in Eq.~(\ref{equ:inflatoneom}) is subdominant, and
the inflaton field rolls slowly down the potential $V(\phi)$. As a consequence,
the energy density $\rho\approx V(\phi)$ and the Hubble rate are approximately constant.

The Hubble rate during inflation, $H_{\rm inf}$, is largely unknown. Observationally it is constrained from above by the limits on primordial B-mode polarisation in the cosmic microwave background radiation. This gives an upper bound $r<0.09$ on the tensor-to-scalar ratio~\cite{Array:2015xqh}, which implies $H_{\rm inf}\lesssim 3.3\times 10^{-5}\MP \approx 8.0\times 10^{13}~{\rm GeV}$ at the time when the observable scales left the horizon. 
In a realistic inflationary model, the Hubble rate decreases with time, and would therefore be lower at the end of inflation.
Although there are models in which the Hubble rate is well below the tensor bound, it is generally expected to be close to it, and in the simplest single-field inflation models it even exceeds it.
It is therefore considered to be likely that the Hubble rate was significantly higher than the Higgs mass $m_{\rm H}\approx 125~{\rm GeV}$.

The minimal inflationary model is Higgs inflation~\cite{Bezrukov:2007ep}, in which the non-minimal curvature coupling of the Higgs field is large, $\xi\sim-49000\sqrt{\lambda}$. This allows it to play the role of the inflaton, without the need for a separate inflaton field. 
During inflation, the Higgs field has a large value $\varphi\sim \MP /|\xi|$, which means that the existence of a negative-energy minimum would appear to pose a problem for the scenario, because if the Higgs field gets trapped there, it would lead to a rapid collapse of the Universe instead of inflation.
However, inclusion of higher-dimensional operators and finite temperature effects can avoid this problem~\cite{Bezrukov:2014ipa}.
Of course, if the actual top and Higgs masses lie in the stable region (see Fig.~\ref{fig:dec}), no problem arises. Furthermore, if they are just below the stability boundary, the effective Higgs potential would have an inflection point which would allow the scenario known as critical Higgs inflation~\cite{Bezrukov:2014bra,Hamada:2014wna,Hamada:2014iga}, in which the Higgs field values are significantly lower than in conventional Higgs inflation.
In the following our focus will be on the conventional scenario in which the inflaton is a separate field, and therefore we will not discuss Higgs inflation in detail.
A thorough and up-to-date review of Higgs inflation, covering also the vacuum stability issues, is given in Ref.~\cite{Rubio:2018ogq}.

Even in the scenario in which the inflaton is not the Standard Model Higgs field, 
one could expect on general grounds that the natural initial value for the Higgs field is at the Planck scale $\varphi\sim \MP $~\cite{Lebedev:2012sy}.
In that case the existence of a negative-energy true vacuum between the electroweak and Planck scales would appear to be a problem, just like in Higgs inflation. Therefore one either has to assume special initial conditions that guarantee $\varphi\ll \varphi_{\rm bar}$ everywhere, or  
find a mechanism that allows the Higgs field to roll to small values without gettting trapped in the negative energy true vacuum.

In addition, even if that problem is solved, one still needs to avoid the nucleation bubbles of true vacuum, and hence satisfy the bound~(\ref{equ:expNbound}).
Approximating inflation with a de Sitter space with constant Hubble rate $H_{\rm inf}$, the expected number of bubbles (\ref{equ:expectedbubbles}) in our past lightcone originating from inflation is
\begin{equation}
\langle{\cal N}\rangle\approx \Gamma_{\rm inf}{\cal V}_{\rm inf},
\end{equation}
where $\Gamma_{\rm inf}$ is the vacuum decay rate, and ${\cal V}_{\rm inf}$
is the volume of the inflationary part of our past light cone.
One can write this as
\begin{equation}
{\cal V}_{\rm inf}\approx 
\frac{4\pi}{9}\left[a_{\rm inf}^3H_{\rm inf}^3(\eta_0-\eta_{\rm inf})^3+3N_{\rm tot}\right]H_{\rm inf}^{-4}
\approx
\frac{4\pi}{9}\left[33.2\times\left(\frac{a_{\rm inf}H_{\rm inf}}{a_0H_0}\right)^3 +3N_{\rm tot}\right]H_{\rm inf}^{-4}
,
\label{equ:Vinf}
\end{equation}
where $a_{\rm inf}$ is the scale factor at the end of inflation, $H_{\rm inf}$ is the Hubble rate during inflation, and $N_{\rm tot}$ is the total number of e-foldings of inflation. In principle, if inflation lasted for an infinite amount of time, the volume of the inflationary past light cone would be infinite. In practice, inflation has a finite duration in most models, and the first term usually dominates in Eq.~(\ref{equ:Vinf}).

The factor $(a_{\rm inf}H_{\rm inf}/a_0H_0)$ is the ratio of the comoving Hubble lengths today and at the end of inflation. It can be expressed as
\begin{equation}
\frac{a_{\rm inf}H_{\rm inf}}{a_0H_0}=e^N,
\label{equ:efoldings}
\end{equation}
where $N$ is the number of e-foldings from the moment the largest observable scales left the horizon during inflation, to the end of inflation. It depends somewhat on the cosmological history, but is approximately~\cite{Liddle:2003as}
\begin{equation}
N\approx 60+\ln\frac{V_{\rm inf}^{1/4}}{10^{16}~{\rm GeV}}.
\end{equation}
This means that the spacetime volume of the inflationary past light cone is
\begin{equation}
{\cal V}_{\rm inf}
\approx 
46\,e^{3N}H_{\rm inf}^{-4}.
\label{equ:Vinfnum}
\end{equation}
From Eq.~(\ref{equ:expNbound}), one then obtains a bound on the decay rate during inflation
\begin{equation}
\Gamma_{\rm inf}\lesssim 0.02\,e^{-3N}H_{\rm inf}^4
\sim 10^{-80}\left(\frac{V_{\rm inf}^{1/4}}{10^{16}~{\rm GeV}}\right)^{-3}H_{\rm inf}^{4}.
\label{equ:infGammabound}
\end{equation}

In the literature, the vacuum stability during inflation is often discussed in terms of the survival probability $P_{\rm survival}$, which can be defined either as the fraction of volume that remains in the metastable vacuum at the end of inflation, or as the probability that a given Hubble volume remains in the metastable vacuum until the end of inflation. This is related to $\Nbub$ by
\begin{equation}
\Nbub\approx e^{N}(1-P_{\rm survival}),
\label{equ:Psurv}
\end{equation}
and therefore the bound (\ref{equ:Gamma0bound}) can be written as
\begin{equation}
1-P_{\rm survival}\lesssim e^{-3N}.
\label{equ:Psurvbound}
\end{equation}

One can use the bounds (\ref{equ:infGammabound}) or (\ref{equ:Psurvbound}) to constrain the Hubble rate during inflation $H_{\rm inf}$ and other parameters of the theory.
This computation can be done in two ways, either using the instanton calculation of the tunneling rate discussed in Section~\ref{sec:tunn}, or using the stochastic Starobinsky-Yokoyama approach discussed in Section~\ref{sec:StaroYoko}. The instanton calculation includes both quantum tunneling and classical excitation, and it can incorporate interactions and gravitational backreaction at short distances. Because it requires analytic continuation, it only works with constant Hubble rate $H_{\rm inf}$, but it can still be expected to be a good approximation when the Hubble rate is slowly varying. In contrast, the stochastic approach can describe a time-dependent Hubble rate and gives a more detailed picture of the time evolution, but it includes only the classical excitation process and does not include interactions on sub-Hubble scales.

In the stochastic approach, the dynamics is described by either the Langevin equation (\ref{eq:Lan}), or by the Fokker-Planck equation (\ref{eq:FP}), which gives the time evolution 
of the one-point probability distribution $P(t,\varphi)$ of the Higgs field $\varphi$.

If the Higgs field is assumed to vanish initially, $\varphi=0$, the probability distribution grows initially as
\begin{equation}
P(h,t)=\sqrt{\frac{2\pi}{H^3t}}\exp\left(-\frac{2\pi^2\varphi^2}{H^3t}\right).
\label{equ:Pearly}
\end{equation}
This is obtained by ignoring the Higgs potential $V(\varphi)$, which should be a good approximation at early times. 

After some time the potential becomes important and starts to limit this growth. If the Hubble rate $H$ is constant, the
field approaches asymptotically the equilibrium distribution (\ref{eq:p}), and it is also a good approximation if the Hubble rate is varying sufficiently slowly. Considering the tree-level potential $V(\varphi)=\lambda \varphi^4/4$ with constant $\lambda>0$, the typical (rms) value of the field is given by Eq.~(\ref{hstar}) as
\begin{equation}
\varphi_*\approx 0.363 \lambda^{-1/4}H\approx 0.605 H,
\label{equ:typicalh}
\end{equation}
where the last expression is for the experimental value of the Higgs self coupling
$\lambda\approx 0.13$. If $H\gtrsim 10^{10}~{\rm GeV}$, these field values are beyond the position (\ref{equ:hbarrier}) of the maximum of the potential.
This means that for such values of the Hubble rate, inflationary fluctuations of the Higgs field would be able to throw the Higgs field over the potential barrier, triggering the vacuum instability~\cite{Espinosa:2007qp,Lebedev:2012sy,Kobakhidze:2013tn,Gabrielli:2013hma,Kehagias:2014wza,Fairbairn:2014zia,Kobakhidze:2014xda,Hook:2014uia,Enqvist:2014bua,Bhattacharya:2014gva,Herranen:2014cua,Kamada:2014ufa,Kearney:2015vba,Shkerin:2015exa}.
This would place a rough upper bound on the Hubble rate,
\begin{equation}
H\lesssim 10^{10}~{\rm GeV}.
\label{equ:naiveHbound}
\end{equation}

To make the bound more precise, Espinosa et al.~\cite{Espinosa:2007qp} solved the equation for the initial state $P(0,\varphi)=\delta(\varphi)$,
with the boundary condition $P(\hmax,t)=0$ to account for the destruction of any Hubble volume where $\varphi>\hmax$. They then defined the survival probability of the vacuum as
\begin{equation}
P_{\rm survival}(t)=\int_{-\hmax}^{\hmax} P(h,t).
\label{equ:Psurvival}
\end{equation}
Because of the boundary conditions, the survival probability is not conserved but decreases with time, 
and from the late-time asymptotic decay,
\begin{equation}
P_{\rm survival}\sim e^{-\gamma t},
\end{equation}
one can determine the vacuum decay rate $\Gamma\approx \gamma H^3$.
This way, they found the decay rate per unit time to be
\begin{eqnarray}
\Gamma&\sim&
\frac{H^6}{32\hmax^2},\quad\mbox{if }H\gtrsim\hmax,
\label{equ:Espinosadecayratehigh}
\\
\Gamma&\sim&
\lambda^{5/4}{\hmax^3H}
\exp\left(-
\frac{8\pi^2V(\hmax)}{3H^4}\right),\quad \mbox{if }H\lesssim\hmax.
\label{equ:Espinosadecayrate}
\end{eqnarray}
One can see immediately that high Hubble rates, $H\gtrsim \hmax$, are ruled out by the bound~(\ref{equ:infGammabound}).
The relevant result is therefore Eq.~(\ref{equ:Espinosadecayrate}). Comparing with Eq.~(\ref{equ:infGammabound}) one obtains the constraint
\begin{equation}
H\lesssim \left(\frac{8\pi^2}{9N}V(\hmax)\right)^{1/4}.
\label{equ:Hbound}
\end{equation}

The numerical value of this constraint depends on the number of e-foldings $N$ and, in particular, the height of the potential barrier, which is highly dependent on the precise Higgs and top masses. 
The bound on the ratio $H/\hmax$ is much less sensitive to the mass values, and therefore also quote the bounds in units of $\hmax$ rather than ${\rm GeV}$.
To obtain indicative bounds in physical units, one can use the central estimate for $\hmax$ in Eq.~(\ref{equ:hbarrier}).
Using $N=60$, the bound (\ref{equ:Hbound}) becomes
\begin{equation}
H\lesssim 0.067\hmax
.\label{equ:Hboundnum}
\end{equation}

The same result be also obtained using the instanton approach~\cite{Kobakhidze:2013tn}, which gives the decay rate ~(\ref{equ:GammaHM}),
\begin{equation}
\Gamma_{\rm inf}\sim H_{\rm inf}^4e^{-B(H_{\rm inf})},
\end{equation}
where $B(H_{\rm inf})$ is the relevant instanton action in de Sitter space with Hubble rate $H_{\rm inf}$.
The bound (\ref{equ:infGammabound}) can therefore be expressed as
\begin{equation}
B(H_{\rm inf})\gtrsim 3N+4\approx 180.
\end{equation}

Fig.~\ref{fig:SM_bounce_action} shows that for Hubble rates near $\hmax$, the relevant instanton solution is the Hawking-Moss instanton, whose action (\ref{eq:HMactionFixedBackground}) agrees with the exponent in Eq.~(\ref{equ:Espinosadecayrate}) in the limit where the barrier height is much less than the false vacuum energy. The instanton and Fokker-Planck calculations are therefore in good agreement in this case.

As discussed in Section~\ref{sec:HMtransition}, the relevant instanton for lower Hubble rates, $H<H_{\rm cross}$, is the Coleman-de Luccia solution~\cite{Rajantie:2017ajw}. However, this is below the bound (\ref{equ:Hbound}) and the Coleman-de Luccia action is very high, $B\sim 1800$, so that it gives a negligible decay rate, and therefore this does not change the bound~(\ref{equ:Hbound}).

There has been some debate about the correct field value used for the boundary condition (\ref{equ:Psurvival}) in the Fokker-Planck calculation.
Hook et al.~\cite{Hook:2014uia} applied the boundary condition $P(t,\varphi_{\rm cl})=0$ at $\varphi=\varphi_{\rm cl}$, determined from the condition
\begin{equation}
-V'(\varphi_{\rm cl})=\frac{3H^3}{2\pi}.
\end{equation}
This condition means that at $h>h_{\rm cl}$ the classical motion of the field due to the potential gradient dominates over the quantum noise. Therefore it allows field trajectories that cross the top of the barrier but return to the metastable side because of the quantum fluctuations. This leads to a slower decay rate in the case of the high Hubble rate,
\begin{equation}
\Gamma\approx \frac{H^6}{32 \varphi_{\rm cl}^2},~\mbox{for}~H\gtrsim \hmax.
\end{equation}
East et al.~\cite{East:2016anr} considered the cutoff point the value $\varphi_{\rm sr}$, where
\begin{equation}
\varphi_{\rm sr}=-\frac{V'(\varphi_{\rm sr})}{3H^2}.
\end{equation}
This is the value above which the Higgs field
no longer satisfies the slow roll condition and therefore the stochastic approach fails.
The choice of the boundary condition becomes less important when $H\ll \hmax$, and therefore it does not affect the bound Eq.~(\ref{equ:Hbound}) very much.
By solving the Fokker-Planck equation numerically, the authors obtained the bound
\begin{equation}
H\lesssim 0.067 \hmax,
\end{equation}
which coincides numerically with Eq.~(\ref{equ:Hbound}).

There are aspects of physics that are not included in the approximations leading to the bound (\ref{equ:Hbound}), and which can therefore provide a way to evade the bound.
First, the high spacetime curvature $R=12H^2$ during inflation modifies the effective potential both at the tree level through the non-minimal coupling $\xi$ and through the curvature-dependence of the loop corrections. 
The non-minimal coupling gives rise to an effective curvature-dependent mass term
(\ref{equ:curvaturemass}),
\begin{equation}
{\cal M}^2=m^2+12\left(\xi-\frac{1}{6}\right) H^2.
\end{equation}
If $\xi$ is positive, it increases the potential height between the electroweak and true vacua and helps to stabilise the electroweak vacuum even if the Hubble rate is well above the bound (\ref{equ:Hbound})~\cite{Espinosa:2007qp,Kehagias:2014wza,Herranen:2014cua}. On the other hand, negative values of $\xi$ make the vacuum less stable. For $\xi<0$, Joti et al.~\cite{Joti:2017fwe} obtained the bound
\begin{equation}
H\lesssim\frac{0.005}{\sqrt{-\xi}}\hmax.
\label{equ:Jotibound}
\end{equation}
The stabilising effects of the non-minimal coupling have also been discussed in Refs.~\cite{Kamada:2014ufa,Espinosa:2015qea,Shkerin:2015exa,Kohri:2016qqv,Kohri:2016wof,Kawasaki:2016ijp,Calmet:2017hja,Markkanen:2018bfx}

\begin{figure}
\begin{center}
			\includegraphics[width=10cm]{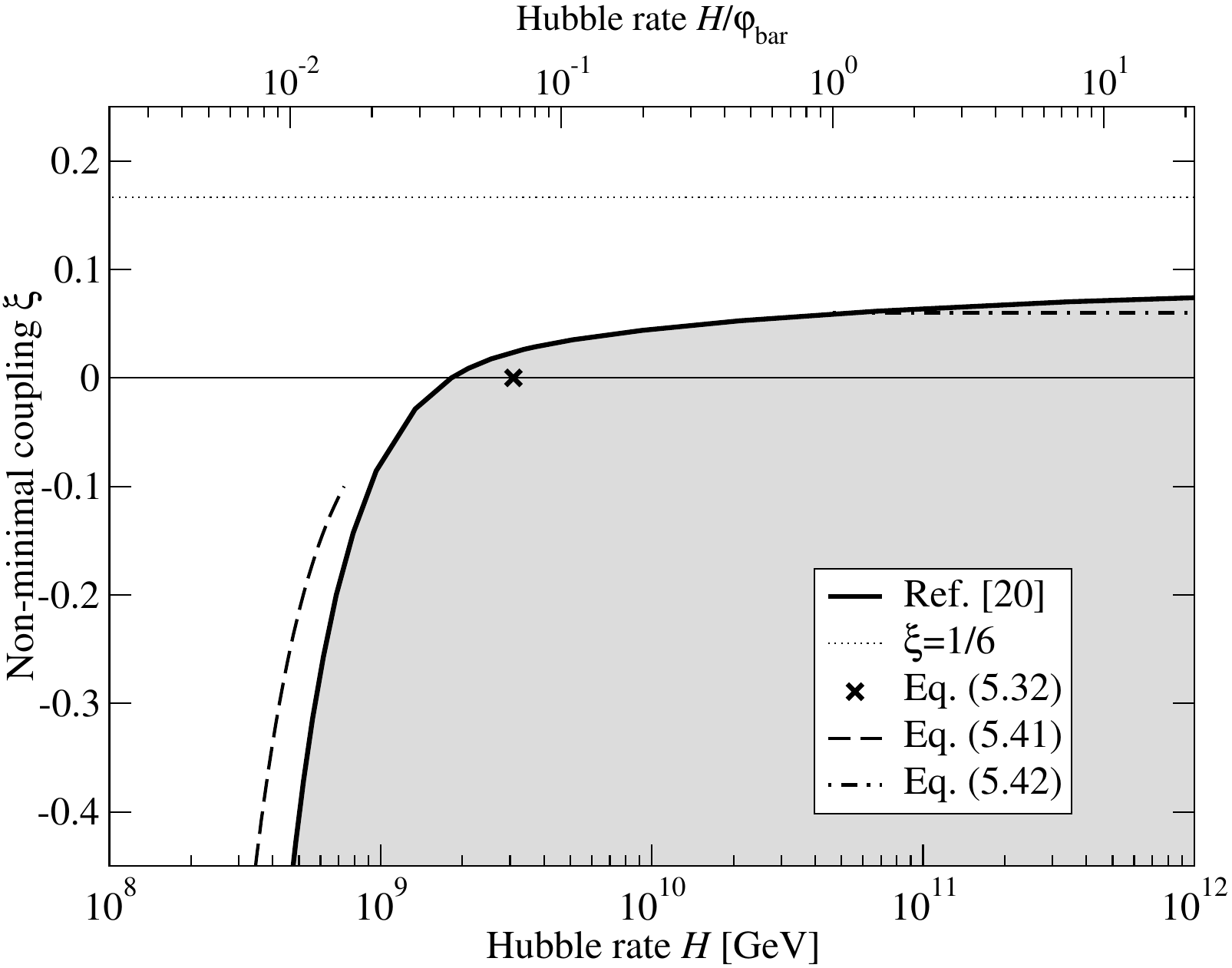}
\end{center}
\caption{\label{fig:xibounds}Stability bounds on the non-minimal coupling $\xi$ (renormalised at the electroweak scale) and the Hubble rate during inflation $H_{\rm inf}$. 
The coloured area shows the unstable region based on the numerical results from Markkanen et al.~\cite{Markkanen:2018bfx}, 
the cross corresponds to Eq.~(\ref{equ:Hbound}),
the dashed line to Eq.~(\ref{equ:Jotibound}) from Ref.~\cite{Joti:2017fwe},
and
the dash-dotted line to Eq.~(\ref{equ:Herranenbound}) from Ref.~\cite{Herranen:2014cua}.
The bottom axis refers to units calculated using the barrier position from Eq.~(\ref{equ:hbarrier}).
}
\end{figure}

The curvature dependent loop corrections mean that the non-minimal coupling $\xi$ runs with the renormalisation scale, and even it is zero at low energies, it runs to a negative value $\xi\approx -0.03$ at the relevant scales for the instability $\mu_\Lambda\sim 10^{10}~{\rm GeV}$
~\cite{Herranen:2014cua}.
Curvature contributions to the loop corrections to the rest of the effective potential can be approximated using renormalisation group improvement~\cite{Herranen:2014cua}, by choosing the 
renormalisation scale as $\mu_*\approx H$ when $H\gtrsim \varphi$, rather than $\mu_*\approx \varphi$ which had been used previously. Using the curvature-dependent renormalisation scale, such as Eq.~(\ref{eq:ccrun}), has become the norm in the more recent literature~\cite{Kearney:2015vba,East:2016anr,Rodriguez-Roman:2018swn}. Having $\mu_*\sim H$  means that for sufficiently high Hubble rates the effective coupling becomes negative, $\lambda(\mu_*)<0$, and the potential barrier disappears completely, unless $\xi$ is sufficiently large. Both of these effects, running $\xi$ and the curvature-dependent renormalisation scale, tend to de-stabilise the vacuum.
Taking them into account gives the bound~\cite{Herranen:2014cua}
\begin{equation}
\xi\gtrsim 0.06 \quad\mbox{for}\quad H\gtrsim \hmax.
\label{equ:Herranenbound}
\end{equation}
The full curvature-dependent effective potential was computed at one-loop order in Ref.~\cite{Markkanen:2018bfx}, and confirms this expectation. 
The stability bounds as a function of the Hubble rate $H$ and the non-minimal coupling $\xi$ are shown in Fig.~\ref{fig:xibounds}.
For comparison, the bound from particle collider experiments is $|\xi|\lesssim 2.6\times 10^{15}$~\cite{Atkins:2012yn}.

A sufficiently large positive non-minimal coupling $\xi$ can also avoid the Higgs field initial condition problem. It was found in Ref.~\cite{Calmet:2017hja} that if
\begin{equation}
\xi\gtrsim H/10^{-4}\MP ,
\end{equation}
the positive curvature contribution to the effective potential allows
the Higgs field to roll from Planck-scale values to its electroweak minimum during inflation without getting trapped into the negative-energy true vacuum.

The bound (\ref{equ:Hbound}) also does not take into account any direct coupling between the Higgs and the inflaton field $\phi$. Although a direct coupling is not radiatively generated, in general it is possible and the precise form it would have and its effects on vacuum stability depend on the details of the inflaton sector. 
The simplest example is a coupling of the form $\lambda_{\phi h}\phi^2 h^2$ in chaotic inflation with a quadratic potential. During inflation, the inflaton field has a high value $\phi\gtrsim \MP $, and therefore the coupling produces an effective mass term for the Higgs field,
\begin{equation}
{\cal M}^2=m^2+\lambda_{\phi h}\phi^2.
\end{equation}
Coupling values $\lambda_{\phi h}\lesssim 10^{-6}$ would not spoil the flatness of the inflaton potential~\cite{Lebedev:2012sy,Gross:2015bea}, and if $\lambda_{\phi h}\gtrsim 10^{-10}$, it would 
stabilise the vacuum during inflation and allow the Higgs field to roll to its current small field values even if starts from a Planck-scale value at the beginning of inflation~\cite{Lebedev:2012sy,Fairbairn:2014zia,Gross:2015bea}.
This coupling has also been discussed in Refs.~\cite{Kamada:2014ufa}.
Considering the non-minimal curvature coupling $\xi$ and the direct Higgs-inflaton coupling $\lambda_{\phi h}$ together, Ref.~\cite{Ema:2017loe} finds the constraint
\begin{equation}
10^{-10}\lesssim \lambda_{\phi h}+10^{-10}\xi\lesssim 10^{-6},
\end{equation}
in the quadratic chaotic inflation model.

Other forms of the Higgs-inflaton coupling have been considered in Refs.~\cite{Bhattacharya:2014gva,Hook:2014uia,Ballesteros:2015iua,Cline:2018ebc}.
There are also other effects that could potentially stabilise the vacuum state during inflation.
Non-zero temperature $T\gtrsim 6\times 10^{13}~{\rm GeV}$ during inflation~\cite{Fairbairn:2014zia}, moduli fields~\cite{Ema:2016ehh}, coupling to a spectator scalar field~\cite{Gong:2017mwt,Han:2018yrk}, or top quark production~\cite{Rodriguez-Roman:2018swn} could all generate an effective stabilising term in the effective potential.

\makeatletter{}\subsection{Reheating}
The end of inflation can be defined as the point at which the Universe no longer undergoes accelerated expansion, which occurs when $w=-1/3$. This marks the beginning of the so-called  \textit{reheating} phase during which the energy density stored as potential energy gets converted into the hot thermal plasma of the Big Bang. If the acceleration is sourced by a slowly rolling inflaton $\phi$, during reheating the slow-roll conditions seize to hold and the inflaton will begin a phase where its (average) kinetic energy is comparable to its potential energy. This usually manifests as coherent oscillations around the minimum of the potential. Reheating is said to be completed when the energy density of the hot Big Bang overtakes that of the inflaton sector, which often proceeds via direct couplings allowing the inflaton to decay into SM constituents. It is however worth pointing out that it is perfectly possible to have successful reheating without any couplings between the inflaton and the SM sector, for examples of such models see Refs.~\cite{Figueroa:2016dsc,Tenkanen:2016jic,Dimopoulos:2018wfg,Haro:2018jtb}.

An inflaton field coherently oscillating  around the minimum of its potential may source a very potent non-perturbative amplification of quantum modes, which takes place during the early stages of reheating and is hence often referred to as \textit{preheating} \cite{Kofman:1994rk,Kofman:1997yn}.
If a phase of preheating occurs, it does not lead to the completion of reheating as the created particles tend to shut off any non-perturbative behaviour through backreaction and a perturbative decay channel is often required to ensure the complete decay of the inflaton. 

From the point of view of a possible vacuum destabilization, preheating is a crucial epoch because vacuum decay is potentially induced by a large amplification of the Higgs field~\cite{Herranen:2015ima}.
It is important to note that at the time of preheating, the Universe has not 
yet reheated to a high temperature, and therefore the thermal effects discussed in Section~\ref{sec:finiteTdecay}
cannot stabilise the vacuum state. 

Let us proceed to consider the familiar Lagrangian appropriate for the Higgs doublet in curved space (\ref{eq:higg1}). 
We consider Hubble rates well above the electroweak scale, $H\gg M_h$, and therefore we can neglect the tree-level mass parameter, and use the action
\ee{S=\int d^4x\,\sqrt{|g|}\bigg[\f{1}{2}\nabla_\mu \varphi\nabla^\mu \varphi-\f{\xi}{2}R \varphi^2-\f{\lambda}{4} \varphi^4\bigg]\label{eq:act6}\,.}
We also assume a single-field model of inflation with a canonical kinetic term and the potential $U(\phi)$. The inflaton $\phi$ is taken to dominate the energy density of the Universe completely and because of this the Higgs field may be considered as a subdominant spectator that can be neglected in the Einstein equation. Using then
\ee{\rho = \f{1}{2}\dot{\phi}^2+U(\phi)\,;\qquad p = \f{1}{2}\dot{\phi}^2-U(\phi)\,,}
in the Friedmann equations (\ref{eq:e}), we can solve for the Ricci scalar $R$ \ee{R=6\bigg[\bigg(\f{\dot{a}}{a}\bigg)^2+\f{\ddot{a}}{a}\bigg]=\f{1}{M_{\rm P}^{2}}\left[4 U(\phi)-\dot{\phi}^2\right]\,.\label{R}}

After inflation ends, the inflaton field $\phi$ rolls down its potential, and initially oscillates coherently about its minimum $\phi_{\rm min}$, until it eventually decays. We assume that the 
inflaton potential vanishes at the minimum, $U(\phi_{\rm min})$, as is usually the case.
We can see from Eq.~(\ref{R}) that during every oscillation, when $\phi\approx\phi_{\rm min}$, the Ricci scalar becomes negative, $R<0$. This, in turn, means that the non-minimal term $\sim \xi R \varphi^2$ gives rise to a tachyonic mass term (\ref{equ:curvaturemass}) for the Higgs field. As already discussed in Section \ref{sec:mod},
this gives rise to significant excitation of the field. 
The fact that the non-minimal term can lead to extremely efficient particle creation during preheating was first discussed in Refs.~\cite{Bassett:1997az,Tsujikawa:1999jh}. 

Particle creation from a periodically tachyonic effective mass was analyzed in detail in Ref.~\cite{Dufaux:2006ee} where it was named \textit{tachyonic resonance}. It is much more extreme than the resonant effects usually taking place during preheating. Hence a dangerous fluctuation of the Higgs field can be generated during  a single oscillation of the inflaton.

For concreteness, we now focus on the case of a quadratic inflaton potential
\begin{equation}
U(\phi)=\frac{1}{2}m^2\phi^2.
\end{equation}
Although as a complete model of inflation, this is not compatible with observations~\cite{Akrami:2018odb}, it approximates the shape of the potential around the minimum in general single-field models.
The behaviour of the inflaton field during its coherent oscillations can be approximately written as $\phi=\phi_0(t)\cos(mt)$ where $\phi_0$ is a slowly changing amplitude $\phi_0(t)=\sqrt{6}H(t) M_{\rm P}/m$ \cite{Kofman:1994rk,Kofman:1997yn}. 

We will focus only on a very brief time period immediately after inflation, when no thermalization has yet taken place. In cosmic time the properly normalized mode is obtained from Eq.~(\ref{eq:adsol2}) as $f(\eta)\rightarrow f(t)/\sqrt{a}$ giving the mode equation  
\ee{
\label{eq:omg}
\ddot{f}(t)+3H\dot{f}(t)+\bigg[\f{\mathbf{k}^2}{a^2}-\f{9}{4}H^2-\f{3}{2}\dot{H}+\xi R\bigg]f(t)=0\,.
}
By using the Friedmann equations (\ref{eq:e}) in this approximation the mode equation can be cast in the Mathieu form \cite{Bassett:1997az,Herranen:2015ima}
\ee{\f{d^2f(t)}{dz^2}+\bigg[A_k-2q\cos(2z)\bigg]f(t)=0,\qquad z=m t\,,\label{eq:mathieu}}
\ee{\nonumber A_k= \f{\mathbf{k}^2}{a^2m^2}+\xi \f{\phi_0^2}{2 M_{\rm P}^2},\qquad q=\f{3\phi_0^2}{4 M_{\rm P}^2}\bigg(\f{1}{4}-\xi\bigg)\, .}
Making use of the analysis in Ref.~\cite{Dufaux:2006ee} we can derive an analytical result for the occupation number of the Higgs field $n_\mathbf{k}$ after the first oscillation
\ee{n_\mathbf{k}=e^{2X_\mathbf{k}}\,,\qquad X_\mathbf{k}=\int_{\Delta z}\Omega_\mathbf{k}\,dz \approx \sqrt{\xi}\f{\phi_0}{M_{\rm P}}\approx\sqrt{\xi}\, \label{eq:occapp},}
where $\Omega^2\equiv-\omega^2$. The $\omega^2$ is the term in the square brackets in Eq.~(\ref{eq:mathieu}) and $\Delta z$ covers the time period when $\omega^2<0$.
Including only the IR modes $k<aH$, neglecting the expansion of space, the self-interaction and furthermore assuming $\xi\gtrsim 1$ we can estimate the generated Higgs fluctuations at horizon scale, $\langle\hat{\varphi}^2\rangle_{a H}$, after the first oscillation of the inflaton as \cite{Herranen:2015ima} 
\ee{\langle\hat{\varphi}^2\rangle_{a H}\approx \int_0^{a H} \f{dk\,{k}^2}{2\pi^2 a^3}\,2|f(t)|^2 n_\mathbf{k}
\approx \left(\f{H}{2\pi}\right)^2\f{2\exp\left\{\sqrt{\xi}\f{2\phi_0}{M_{\rm P}}\right\}}{3\sqrt{3\xi}}\,.\label{eq:tacfluc}}

If $\phi_0\sim \MP$, as in chaotic inflation, one can see from Eq.~(\ref{eq:tacfluc}), that the Higgs fluctuations are exponentially amplified if $\xi\gtrsim1$. The fluctuation  $\Delta \varphi\sim $ can become larger than the position of the potential barrier in the SM
\ee{\Delta \varphi \equiv \sqrt{\langle|\hat{\Phi}|^2\rangle_{a H}} =\sqrt{4 \langle\hat{\varphi}^2\rangle_{a H}} \gtrsim  \varphi_{\rm bar}\,.} 
Note that a large and positive $\xi$ gives rise to a destabilizing effect after inflation. This is opposite to what happens during inflation when it suppresses fluctuations by effectively making the field heavy (see Section~\ref{sec:Inflation}).

In general, once a significant particle density is produced it tends to work against any further particle production \cite{Kofman:1997yn}. For the Higgs the main backreaction comes from the self-interaction term, which contributes to the effective mass (\ref{equ:curvaturemass}), along with the curvature terms visible in (\ref{eq:omg}), as 
\ee{{\cal M}^2=-\f{9}{4}H^2-\f{3}{2}\dot{H}+\xi R+6\lambda\langle\hat{\varphi}^2\rangle\,,}
very similarly as we derived in the 1-loop approximation for a scalar singlet in Eq. (\ref{eq:effm2}) of Section \ref{eq:effpotc}.
In order for tachyonic particle creation to take place one must have $\xi |R|\gtrsim 6\lambda \langle\hat{\varphi}^2\rangle$ for $\xi\gtrsim 1$. However, in Section (\ref{sec:crun}) it was shown that the Hubble rate contributes to the RG scale through curvature induced running (see Eq. (\ref{eq:ccrun})). If $H\gtrsim\hmax$, the four-point coupling is negative, implying that the backreaction in fact enhances the instability, and will not suppress tachyonic particle creation even if a large variance is generated.

Backreaction also arises from the gravitational disturbance of the generated particle density. In order to reach this threshold one must create enough particles such that their energy density approaches $3H^2 M_{\rm P}^2$. The relevance of gravitation backreaction we can estimate from the approximate energy density for the Higgs \cite{Herranen:2015ima}
\ee{\rho_{\rm Higgs}\approx 24\xi H^2 \langle\hat{\varphi}^2\rangle +6\lambda \langle\hat{\varphi}^2\rangle^2\,.}
When $\rho_{\rm Higgs}\sim 3H^2 M_{\rm P}^2$ the Higgs starts to influence the dynamics of spacetime requiring a non-linear analysis. Below we will assume that when the gravitational backreaction threshold is reached the particle production will seize.

More detailed calculations of the process have been carried out using linearised approximation~\cite{Kohri:2016wof} and lattice field theory simulations~\cite{Ema:2016kpf,Figueroa:2017slm}.
The most detailed analysis, carried out in Ref. \cite{Figueroa:2017slm}, used the tree-level RGI effective potential with three-loop running, and considered different top quark masses. The main conclusion was that the instability is triggered with high probability for $\xi\gtrsim 4-5$, for a top quark mass $M_{\rm t}\approx 173.3$GeV.
This implies an upper bound on $\xi$ in the context of quadratic chaotic inflation.

\begin{figure}[t]
\begin{center}
			\includegraphics[width=0.8\textwidth]{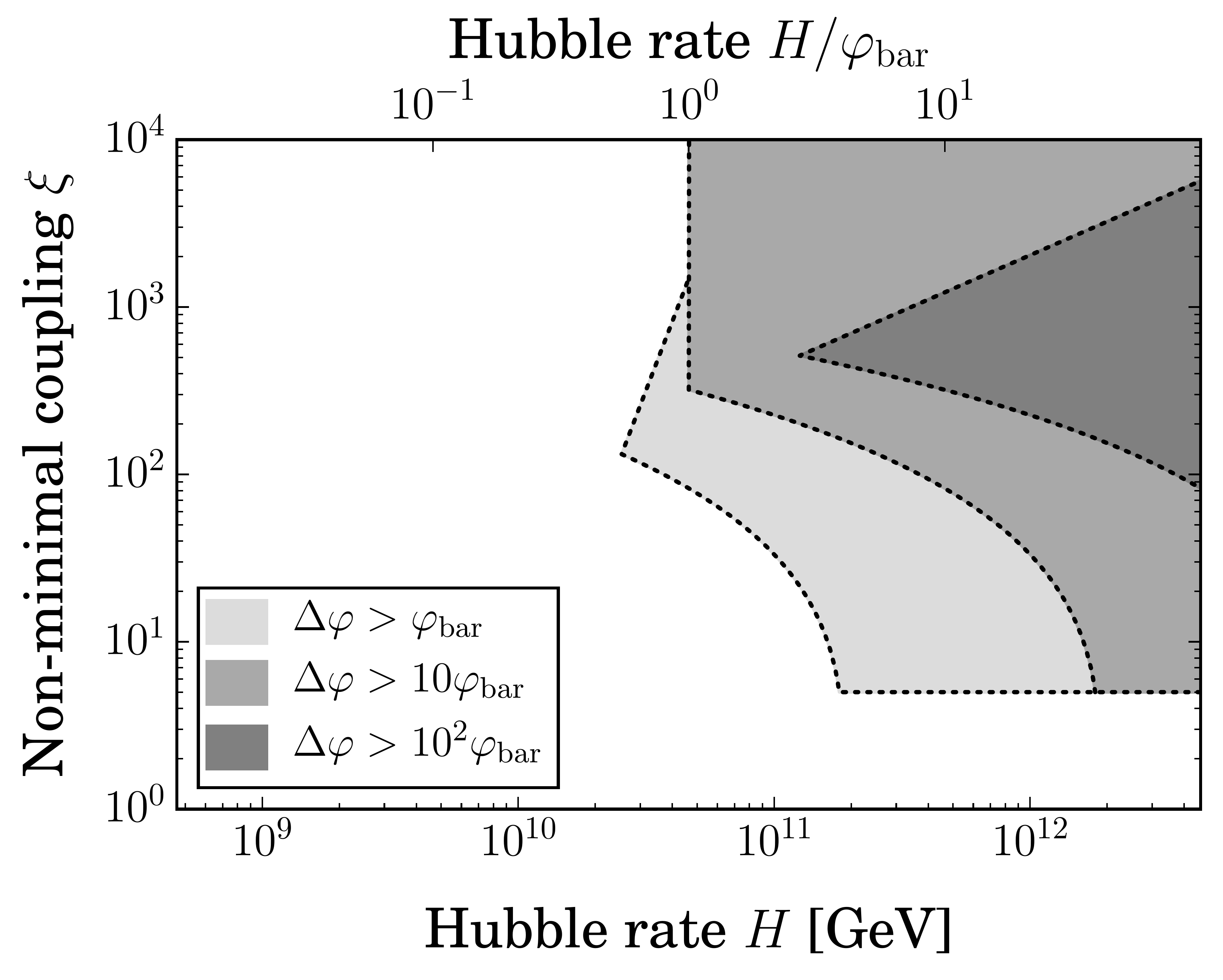}
\end{center}
\caption{Regions where the Higgs fluctuations $\Delta \varphi$ generated after inflation by a single inflaton oscillation are greater than the barrier height $\varphi_{\rm bar}$ for a model with no direct couplings between the Higgs and the inflaton. The Higgs fluctuations are given by Eq.~(\ref{eq:tacfluc}) while taking into account backreaction effects from self-interactions and gravity. The amplitude of the inflaton at the end of inflation is assumed to satisfy $\phi_0=0.3 M_{\rm P}$ and we have used the value $\varphi_{\rm bar}=4.64\times 10^{10}$GeV from Section~\ref{sec:effpSM}.
	Regions below $\xi=5$ have been cut according to the bound obtained in Ref.~\cite{Figueroa:2017slm}. \label{fig:xiboundspre}
}
\end{figure}

The regions where a dangerously large fluctuation of the Higgs is generated after a single oscillation of the inflaton are shown in Fig.\,\ref{fig:xiboundspre}. The rather complicated shapes are the result of the interplay of the variance (\ref{eq:tacfluc}) and the constraints coming from self interactions and gravitational backreaction. In Fig.\,\ref{fig:xiboundspre} we have assumed that the amplitude of the inflaton at the end of inflation satisfies $\phi_0=0.3 M_{\rm P}$. While this is true for the quadratic (chaotic) model of inflation, it is not true generically. Since Eq.~(\ref{eq:tacfluc}) is exponentially dependent on $\phi_0$ the predictions are very sensitive to the specifics of inflation. Similarly, the duration of reheating plays a crucial role and for prolonged reheating a possible instability may be further enhanced. The derivation of Eq.~(\ref{eq:tacfluc}) is based on the adiabatic approximation~\cite{Dufaux:2006ee}, which can be shown to break down for small $\xi$ \cite{Postma:2017hbk}. Furthermore, very little particle creation is expected when close to the (approximately) conformally invariant point $\xi={1}/{6}$. For these reasons and the lattice results of Ref.~\cite{Figueroa:2017slm} we have conservatively cut out regions with $\xi\leq5$.

As discussed in the previous section, the set-up in Eq.~(\ref{eq:act6}) assuming that the inflaton is decoupled from the SM is in many ways the minimal one. Couplings between the inflaton and the SM sector may of course be introduced or even required by a specific reheating model. Vacuum stability during preheating in models with no non-minimal coupling but with direct couplings between the inflaton and the Higgs was investigated in Refs.~\cite{Ema:2017rkk,Gross:2015bea,Enqvist:2016mqj}.
In particular, in Refs.~\cite{Ema:2017rkk} it was shown that in some cases vacuum decay during preheating may take place also for low-scale inflation. 

In Refs.~\cite{Kohri:2016wof,Ema:2017ckf, Ema:2017loe,Ema:2016kpf} both the non-minimal coupling and direct Higgs-inflaton couplings were considered. In a sense in this case the Higgs fluctuations are sourced in a complicated manner by the interplay of tachyonic resonance \cite{Dufaux:2006ee} and the (usual) parametric resonance \cite{Kofman:1997yn}. For the precise coupling ranges where significant particle production takes place and possible implications for instability, see Ref.~\cite{Ema:2017loe}. 
We also point out that particle creation resulting from the non-adiabatic change in the background curvature when inflation ends, already shown in Ref.~\cite{Ford:1986sy}, can be enough to probe the unstable region of the effective Higgs potential \cite{Herranen:2015ima}.

%As the main conclusion one may state that  an instability of the electroweak vacuum arising from the non-perturbative (resonant) dynamics during preheating can generically occur in a variety of models and set-ups.  

\subsection{Hot Big Bang}
\label{sec:HBB}

After reheating, the Universe entered a thermal radiation-dominated state, in
which
vacuum decay rate can be approximated by the thermal rate (\ref{equ:Gammathermal}) at the relevant temperature,
and the Hubble rate was given by the equation
\begin{equation}
\label{equ:Hubblethermal}
H(T)^2=g_*(T)\frac{\pi^2}{90}\frac{T^4}{\MP ^2},
\end{equation}
where $g_*(T)$ is the effective number of degrees of freedom and has the value 
$g_*(T)=106.75$ in the Standard Model at high temperatures.

Using Eq.~(\ref{equ:expectedbubbles})
one can write the expected number of true vacuum bubbles in our past light cone from this era as~\cite{Espinosa:2007qp,Salvio:2016mvj}
\begin{equation}
\frac{d\langle{\cal N}\rangle}{d\ln T}
=
\frac{4\pi}{3}\left(\frac{g_{*S}^0}{g_*(T)}\right)
\left(\frac{T_0}{T}\right)^3
\frac{\left(\eta_0-\eta\left(T\right)\right)^3}{H(T)}
\Gamma(T).
\end{equation}
where $g_{*S}^0=3.94$ is the effective number of entropy degrees of freedom today.
Using Eqs.~(\ref{equ:currenteta}) and (\ref{equ:Hubblethermal}) this becomes
\begin{equation}
\frac{d\langle{\cal N}\rangle}{d\ln T}
\approx 1.49 \frac{\MP }{H_0^3}
\left(\frac{T_0}{T}\right)^3\frac{\Gamma(T)}{T^2},
\label{equ:expNthermal}
\end{equation}

If the Universe reheated instantaneously after inflation, the reheat temperature $T_{\rm RH}$ to which the Universe equilibrate, is related to the Hubble rate at the end of inflation $H_{\rm inf}$ through
where $g_*\ge 106.75$ is the effective number of degrees of freedom.
Because the rate decreases when the temperature decreases, Eq.~(\ref{equ:expNthermal}) is dominated by high temperatures $T\sim T_{\rm RH}$.
Therefore one can approximate
\begin{equation}
\langle{\cal N}\rangle\approx
\frac{\MP T_0^3}{H_0^3}\frac{\Gamma(T_{\rm RH})}{T_{\rm RH}^5}
\approx 
\frac{\MP T_0^3}{H_0^3 T_{\rm RH}}e^{-B(T_{\rm RH})}
.
\end{equation}
Requiring that $\langle{\cal N}\rangle\ll 1$ leads to the bound
\begin{equation}
B(T_{\rm RH})\gtrsim 3\ln\frac{T_0}{H_0}+\ln\frac{\MP }{T_{\rm RH}}
\approx
202+\ln\frac{\MP }{T_{\rm RH}},
\end{equation}
which is satisfied by the numerical result (\ref{equ:thermalBnumerical}) for the current central Higgs and top mass values, and therefore it does not imply a bound on the reheat temperature.

As at zero temperature, the vacuum stability depends sensitively on the top and Higgs masses.
A detailed analysis~\cite{Rose:2015lna} based on integrating Eq.~(\ref{equ:expNthermal}) gives an upper bound on the top quark mass,
\begin{equation}
\frac{M_t}{\rm GeV}
<
0.283\left(\frac{\alpha_s-0.1184}{0.0007}\right)
+0.4612\frac{M_h}{\rm GeV}
+1.907\log_{10}\frac{T_{\rm RH}}{\rm GeV}
+\frac{1.2\times 10^3}{0.323\log_{10}\frac{T_{\rm RH}}{\rm GeV}+8.738}.
\label{equ:thermalbound}
\end{equation}

In practice, reheating is not instantaneous, and there may have been a period when the Standard Model degrees of freedom were in thermal equilibrium but were not the dominant energy component.
In the scenario in which the inflaton field decays slowly and dominates the energy density of the Universe for an extented period, the maximum temperature is~\cite{Espinosa:2007qp,EliasMiro:2011aa,Rose:2015lna}
\begin{equation}
T_{\rm max}=\left(\frac{3}{8}\right)^{2/5}
\left(\frac{40}{\pi^2}\right)^{1/8}
\frac{g_*^{1/8}(T_{\rm RH})}{g_*^{1/4}(T_{\rm max})}
\left(\MP  H_{\rm inf}T^2_{\rm RH}\right)^{1/4}.
\end{equation}
Because the Universe was not radiation-dominated Eq.~(\ref{equ:expNthermal}) does not describe the period when $T\gtrsim T_{\rm RH}$. Instead, one has
\begin{equation}
\frac{\langle{\cal N}\rangle}{d\ln T}
\approx \frac{\MP }{T^2} H_0^3 \left(\frac{T_0}{T_{\rm RH}}\right)^3
\left(\frac{T_{\rm RH}}{T}\right)^{10}\Gamma(T).
\end{equation}
However, this has only a small effect on the numerical bounds~\cite{Rose:2015lna}.

In Fig.~\ref{fig:dec}, the blue dashed line shows the bound (\ref{equ:thermalbound}) calculated with a reheat temperature $T_{\rm RH}\sim 10^{16}~{\rm GeV}$. As can be seen, the inclusion of the thermal history of the Universe reduces the allowed mass range compared with the zero-temperature bounds. The central experimental values are still allowed, but the instability boundary lies within two standard deviations from them.

\makeatletter{}\section{Concluding Remarks}
\label{sec:con}
The current experimental data 
show that with a high likelihood, the electroweak vacuum state of the Standard Model is metastable.
Even though 
the vacuum state could
be stabilised by new physics beyond the Standard Model, 
and even in the Standard Model
parameters corresponding to a stable vacuum are still allowed by experimental errors, 
it is important to study the implications of the possible metastability. That allows one to understand whether the metastability is compatible with observations, and if so, what constraints it places on the parameters of the theory.

If the electroweak vacuum really is metastable, then bubbles of the true, negative-energy vacuum can be nucleated by quantum tunneling or classical excitation, as discussed in Section~\ref{sec:tunn}. Once a bubble has formed, it expands at the speed of light, destroying everything in its way. This clearly has not happened yet in our part of the Universe, which means there has not been a single bubble nucleation event in our whole past light cone. In Section~\ref{sec:Cosmology}, we showed how the likelihood of this can be estimated by computing the nucleation rate and integrating it over the past light cone. Because the past light cone includes all of the different cosmological eras, and the nucleation rate and its dependence on theory parameters is different in each era, this provides a rich set of constraints on both the cosmological history and on the Standard Model parameters. 

In this review, we have focussed on four different cosmological eras: inflation, preheating, hot radiation-dominated phase, and the late Universe. Vacuum stability in the late Universe one obtains constraints on the Higgs and top masses, and they are made tighter by considering the hot radiation-dominated phase, as summarised in Fig.~\ref{fig:dec}. Survival of the vacuum through inflation and the subsequent preheating phase constrains the Hubble rate during inflation and the Higgs-curvature coupling $\xi$ (Figs.~\ref{fig:xibounds} and \ref{fig:xiboundspre}), as well as other aspects of inflationary models. A demonstration of the power of these considerations is that for quadratic chaotic inflation, the non-minimal coupling is constrained to be within the range $0.06\lesssim \xi\lesssim 5$, which is 15 orders of magnitude stronger than the experimental bounds from the Large Hadron Collider~\cite{Atkins:2012yn}.
  Cosmological vacuum decay has a unique connection to gravity via the early Universe, which
  opens up an observational window to particle physics well beyond what colliders can achieve. 

In this work we have reviewed the, already rather significant, body of work investigating the cosmological consequences of the SM Higgs possessing a metastable potential. We have also discussed the relevant theoretical frameworks required for such studies. The multidisciplinary nature of the problem is perhaps one of the reasons behind the ongoing significant interest as particle physics, quantum field theory and gravity all play a prominent role.  Although  the specifics of the theory behind early Universe dynamics are not currently known what has become quite apparent is that a metastable Higgs potential generically leads to non-trivial constraints, which are completely invisible to colliders.
  
On the other hand, despite the large number of existing studies, much remains to be explored. For example, at the moment very few  works exist that go beyond the simple quadratic model of inflation. This is equally true for the inflationary and reheating epochs. There is also a great deal of scope for improving calculation techniques in order to obtain more precise and robust constraints, for example by going beyond the semiclassical approximation or fully including gravitational effects. The work on the cosmological aspects of Higgs vacuum metastability is only starting.

%It also remains to be conclusively answered whether vacuum decay is suppressed for the thermal plasma of the hot Big Bang, especially if the reheating temperature is high. Finally, we point out a potentially very fruitful research direction that still remains entirely unexplored: vacuum transitions induced by the violent dynamics present in the early Universe are in principle possible in the vast landscape of beyond the SM theories. As for the SM it may then be possible to constrain the allowed parameter spaces by invoking cosmological consistency, thus providing novel tool for determining the viability of beyond the SM theories. 

\section*{Acknowledgments}
The authors are grateful to numerous individuals for discussion and collaboration on this topic. 
AR and TM were funded by the STFC grant ST/P000762/1, and SS was funded by the Imperial College President's PhD Scholarship and the UCL Cosmoparticle Initiative.

\makeatletter{}\providecommand{\href}[2]{#2}\begingroup\raggedright\endgroup

\end{document}